\documentclass[10pt,preprint,a4paper]{aastex}

\usepackage{amsmath}                % American Mathematical Society package
\usepackage{amsfonts}               % American Mathematical Society fonts
\usepackage{amssymb}                % American Mathematical Society symbol
\usepackage{epsfig}                 % EPS figures

\usepackage{subfigure}

\usepackage{graphicx}
\usepackage{grffile}

\newcommand{\cm}{{~\rm cm}}
\newcommand{\s}{{~\rm s}}
\newcommand{\km}{{~\rm km}}
\newcommand{\g}{{~\rm g}}
\newcommand{\K}{{~\rm K}}
\newcommand{\erg}{{~\rm erg}}
\newcommand{\yr}{{~\rm yr}}
\newcommand{\Myr}{{~\rm Myr}}
\newcommand{\Gyr}{{~\rm Gyr}}
\newcommand{\pc}{{~\rm pc}}
\newcommand{\kpc}{{~\rm kpc}}

\begin{document}

\title{HEATING COLD CLUMPS BY JET-INFLATED BUBBLES IN COOLING FLOW CLUSTERS}

\author{Shlomi Hillel\altaffilmark{1} and Noam Soker\altaffilmark{1}}

\altaffiltext{1}{Department of Physics, Technion -- Israel Institute of Technology, Haifa
32000, Israel; shlomihi@tx.technion.ac.il, soker@physics.technion.ac.il}

\begin{abstract}
We simulate the evolution of dense-cool clumps embedded in the intra-cluster medium (ICM) of cooling flow clusters of galaxies
in response to multiple jet-activity cycles, and find that the main heating process of the clumps is mixing with the hot shocked jets' gas, the bubbles,
while shocks have a limited role.
We use the PLUTO hydrodynamical code in {{{{two dimensions with imposed axisymmetry}}}}, to follow the thermal evolution of the clumps.
We find that the inflation process of hot bubbles, that appear as X-ray deficient cavities in observations,
is accompanied by complicated induced vortices inside and around the bubbles.
The vorticity induces efficient mixing of the hot bubbles' gas with the ICM and cool clumps, resulting in a substantial
increase of the temperature and entropy of the clumps.
For the parameters used by us heating by shocks barely competes with radiative cooling, even after 25 consecutive shocks excited during
0.5~Gyr of simulation.
Some clumps are shaped to filamentary structure that can turn to observed optical filaments.
We find that not all clumps are heated.
Those that cool to very low temperatures will fall in and feed the central supermassive black hole (SMBH), hence closing the feedback cycle
in what is termed the cold feedback mechanism.
\end{abstract}

%%% \keywords{ }

% ==========================================================
\section{INTRODUCTION}
\label{s-intro}
% ==========================================================

A negative feedback mechanism determines the thermal evolution of the intra-cluster medium (ICM) in the inner regions of cooling flow (CF) clusters and groups of
galaxies (e.g., \citealt{Binney1995, Farage2012, Pfrommer2013}).
This feedback mechanism is driven by active galactic nucleus (AGN) jets that inflate X-ray deficient cavities (bubbles; e.g.,
\citealt{Dong2010, OSullivan2011, Gaspari2012a, Gaspari2012b, Birzan2011, Gitti2012, GilkisSoker2012, RefaelovichSoker2012}).
Examples of bubbles in cooling flows include Abell 2052 \citep{Blanton2011}, NGC 6338 \citep{Pandge2012}, NGC 5044 \citep{David2009}, HCG 62 \citep{Gitti2010},
Hydra A \citep{Wise2007}, NGC 5846 \citep{Machacek2011} NGC 5813 \citep{Randall2011}, A 2597 \citep{McNamara2001}, Abell 4059 \citep{Heinz2002}, NGC 4636 \citep{Baldi2009},
NGC 5044 \citep{Gastaldello2009, David2011}, and RBS 797 \citep{Schindler2001, Cavagnolo2011, Doria2012}.
A relevant feature is that in most cases the two opposite bubbles of a bubble pair depart from exact axisymmetrical morphology.
This implies a relative motion of the ICM and the source of the jets, and in some cases a change in direction of the jets' axis.
We will not be able to simulate these flow patterns with our numerical grid on which we impose axial symmetry.

The process of bubble inflation is at the heart of the feedback mechanism, as it is related to other processes such as vortex shedding in the ICM
(e.g., \citealt{RefaelovichSoker2012, Walgetal2013}), sound wave excitation \citep{Sternberg2009, GilkisSoker2012},
and mixing of the ICM with hot shocked jets' material \citep{GilkisSoker2012, Sokeretal2013}.
The bubbles seem to be a key ingredient in the feedback mechanism not only in cooling flows, but also in other astrophysical objects \citep{Sokeretal2013}, such as
core collapse supernovae \citep{PapishSoker2014}.

Vortices inside the bubbles and in their surroundings play major roles in the formation of bubbles, their evolution, and their interaction with the ICM
(e.g. \citealt{Heinz2005, Sternberg2008b, Sternberg2009, RefaelovichSoker2012}).
\cite{Omma2004} find that a turbulent vortex trails each cavity, and that this vortex contains a significant quantity of entrained and uplifted material (also \citealt{Roediger2007}),
and \cite{GilkisSoker2012} find that vigorous mixing caused by vortices implies that the region within few$\times 10 \kpc$ is multi-phase.
These processes lead to the formation of small cool regions, that if are not heated by another jet-activity episode cool and flow inward to feed the AGN.
The process of feeding the AGN with cold clumps in the feedback mechanism cycle is termed the \emph{cold feedback mechanism},
and was suggested by \cite{Pizzolato2005}.
The cold feedback mechanism has been later strengthened by observations of cold gas and by more detailed studies (e.g.,
\citealt{Revaz2008, Pope2009, Wilman2009, Pizzolato2010, Wilman2011, Nesvadba2011, Cavagnolo2011, Gaspari2012a, Gaspari2012b,
McCourt2012, Sharma2012, Farage2012, GilkisSoker2012, Waghetal2013, BanerjeeSharma2014, McNamaraetal2014, Li2014, VoitDonahue2014, Voitetal2014}).

To inflate the wide bubbles very close to the origin of the jets, termed `fat bubbles,' either slow (sub-relativistic) massive wide (SMW) jets (bipolar outflows)
\citep{Sternberg2007}, precessing jets \citep{Sternberg2008a, Falceta-Goncalves2010}, or a relative motion of the jets to the medium
\citep{Bruggen2007, Soker2009, Morsony2010, Mendygral2012} are required.
No fat bubbles are formed when the jets penetrate to too large a distance, while in intermediate cases elongated and/or detached from the center bubbles
are formed (e.g., \citealt{Basson2003, Omma2004, Heinz2006, VernaleoReynolds2006, AlouaniBibi2007, Sternberg2007, ONeill2010, Mendygral2011, Mendygral2012}).
In the present study we will inflate bubbles by SMW jets but our results hold for bubbles inflated by precessing jets or a relative motion of the
ICM as well.
Our demonstration that bubbles in cooling flow clusters are inflated by SMW outflows \citep{Sternberg2007}, and our suggestion that such SMW
could also have been active during galaxy formation \citep{Sokeretal2009} require many AGN to form SMW bipolar outflows.
Such common SMW bipolar outflows are supported by recent observations (e.g., \citealt{Moe2009, Dunn2010, Tombesi2012, Aravetal2013, Harrisonetal2014}).

In our setting, two opposite jets are launched along a common axis. The heating of the gas perpendicular to the jets' axis need not be $100\%$
efficient, as observations show that heating does not completely offset cooling (e.g., \citealp{Wis04, McN04, Cla04, Hic05, Bre06, Sal08, Wilman2009}),
and a \emph{moderate CF} exists \citep{Soker2001}.
\textit{Moderate} implies here that the mass cooling rate to low temperatures is much lower than the
cooling rate expected without heating, but it is much larger than the accretion rate onto the supermassive black hole (SMBH) at the center of the cluster.
The cooling gas is either forming stars (e.g., \citealp{Odea08, Raf08}), forming cold clouds (e.g., \citealt{Edge2010}), accreted
by the SMBH to maintain the cold feedback mechanism \citep{Pizzolato2010}, or is
expelled back to the ICM and heated when it is shocked or mixed with the hot jets' material.
The mixing of cold ICM clumps with the hot shocked jets' material is the focus of our present study.

In section \ref{s-numerical-setup} we describe the numerical code and setup.
In section \ref{s-global_flow_structure} we describe the global flow structure, and in section \ref{s-clump_bubble_interaction} we turn to study the interaction of
jet-inflated bubble with the ICM.
Our study of multiple jet-launching episodes, up to 25 episodes, is described in section \ref{multiepisodes} where we follow the entropy of the cold clumps.
In section \ref{s-summary} we summarize our main findings and their implications.

% ==========================================================
\section{NUMERICAL SETUP}
\label{s-numerical-setup}
% ==========================================================

We use the PLUTO code \citep{Mignone2007} for the hydrodynamic simulations.
The simulations were carried out in a {{{{two-dimensional grid with imposed azimuthal symmetry}}}}.
The computational grid is in the quadrant where the two coordinates $x$ and $z$ are positive.
The $z$ coordinate is chosen along the azimuthal symmetry axis that coincides with the axis of the jet.
In reality two opposite jets are lunched simultaneously,
such that the flow here is assumed to be symmetric with respect to the $z = 0$ plane,
amounting to reflective boundary conditions at $z = 0$.
A polar grid is used, where the azimuthal coordinate runs from $\theta = 0^\circ$ (symmetry axis) to $\theta=90^\circ$ (equator) and the radial coordinate runs from $r = 0.5 \kpc$ to $r = 600 \kpc$.
{{{{The grid has $256$ angular divisions with a constant angular size of $\Delta \theta=0.35^\circ$.
The radial cell size is such that $\Delta r= r \Delta \theta$. Namely, the radial cell size increases with radius.
At a radius of $r=10 \kpc$, for example, the cell size is $(\Delta r, r \Delta \theta) = (60 \pc, 60 \pc)$. }}}}

On the inner boundary of $r = 0.5 \kpc$ we inject a jet into the grid within the angle range of $\theta = 0^\circ - 70^\circ$ for a wide jet \citep{Sternberg2007}, and a reflective boundary condition is imposed for $\theta = 70^\circ - 90^\circ$.
The power of the two jets together is $P_{2j} = 2 \times 10^{45} \erg \s^{-1}$ (half of it in each direction), with a sub-relativistic jet velocity of $v_j = 9500 \km \s^{-1}$.
The mass deposition rate is thus
\begin{equation}
\dot{M}_{2j} = \frac{2 P_{2j}}{v_j^2} \simeq 70 M_{\odot}~\yr^{-1}.
\end{equation}
When the jet is turned off reflecting boundary conditions apply for the entire inner sphere at $r = 0.5 \kpc$.

The boundary of the computational domain at $r = 600 \kpc$ was pushed out far enough so that in our region of interest at $r \la 50 \kpc$ there are no boundary effects during the time of the simulation.
We use a grid stretched in the $r$ direction in order to reduce the computational cost of the spurious domains.
At the initial temperature of $T = 4 \times 10^7 \K$ of the ambient gas the sound speed is
\begin{equation}
c_s = \left( \frac{ \gamma k T}{\mu m_H} \right) ^ {1 / 2} = 950 \km~\s^{-1},
\end{equation}
where we use $\gamma = 5 / 3$ and $\mu = 0.61$.
Thus, our region of interest remains without interference from the outer boundary for a period of time of approximately
\begin{equation}
\frac{550 \kpc}{950 \km~\s^{-1}} \simeq 570 \Myr.
\end{equation}

The simulation begins with an isothermal sphere of gas with a density profile of (e.g., \citealt{VernaleoReynolds2006})
\begin{equation}
\rho(r) = \frac{\rho_0}{\left[ 1 + \left( r / a \right) ^ 2 \right] ^ {3 / 4}},
\end{equation}
with $a = 100 \kpc$ and $\rho_0 = 10^{-25} \g \cm^{-3}$.
A gravity field is added to maintain hydrostatic equilibrium,
{{{{
\begin{equation}
g(r) = \frac{1}{\rho} \frac{\mathrm{d}p}{\mathrm{d}r},
\end{equation}
where the pressure is found from the given $\rho(r)$ and the constant temperature $T$.
The gravity field is kept constant in time.
}}}}
Hydrostatic equilibrium has been verified numerically in a sterile simulation, i.e., with no jets.
{{{{Radiative cooling is included using one of the PLUTO modules, where we insert
the tabulated cooling function from Table 6 in \cite{SutherlandDopita1993}.
Not only the dense clumps cool, but so does the ICM itself during the course of our simulations.
The ICM cooling time near the center of our cluster model is $\tau_{\rm cool} = 2 \Gyr$, compared with
the $500 \Myr$ of our longest simulated case.
This amounts to significant, but partial, cooling of the ICM, and can be seen, e.g., in the temperature maps over time
of figure \ref{RunM20D0.3Temp}.
However, by the time our simulation is over the jets are expected to encounter new regions of the ICM, either due
to precession of the jets to other directions (as simulated, e.g., by \citealt{Li2014}), and/or due to a non-radial ICM motion relative to the AGN.
These effects cannot be modelled in our 2D simulations, and are postponed to a future 3D study. }}}}

In order to follow certain regions of the simulations, we mark them with `tracers' in the PLUTO code.
Tracers are artificial flow scalars which are frozen-in to the hydrodynamic flow.
They are given a value of $\xi = 1$ for a traced region, and $\xi = 0$ elsewhere.
We verified that the sum $\Sigma \xi_i M_i$ is constant in time, where $\xi_i$ and $M_i$ are the tracer value and mass in each numerical cell $i$, respectively,
and the sum is over all cells.

At the beginning of each run, $t=0$, we position dense clumps at several locations in the ICM to study their interaction with the jet.
The initial cross section of each clump is a circle in the meridional $zx$ plane, such that in 3D it is actually a torus.
The dynamics of toroidal clumps is expected to differ from the dynamics of spherical clumps, in terms of the drag forces and
interaction with their surroundings, the development of instabilities, and so on.
However, {{{{ the 2D setting is adequate to present the dominate role of mixing in heating the clumps. We expect
that in 3D the vortices, which now have one more degree of freedom, will be even more efficient in mixing cooling clumps with hot bubble gas. }}}}
{{{{{ In 3D the vortices will cascade to small scales (a process impossible in 2D).
We will therefore use the term `vorticity' in referring to the shown vortices.  }}}}}
Thus, in 2D, features near the symmetry axis must be treated very carefully and with caution.
The full 3D problem is computationally expensive, and will be examined in a future work.
{{{{ In addition to vorticity induced directly by the jet, we observe in simulations other instabilities such as Rayleigh-Taylor
and Kelvin-Helmholtz (e.g., \citealt{RefaelovichSoker2012}).
The jet-induced vorticity, however, is necessarily present and is geometrically different from these instabilities.
As well, vortices are formed near the jets but can reach large distances from the center (see figure \ref{figure: 3clumps_r45_rhofactor1.3_t305}).
}}}}

{{{{The simulations performed in this study are summarized in table \ref{table: simulations}.
% TTTTTTTTTTTTTTTTTTTTTTTTTTTTTTTTTTTTTTTTTTTTTTT
\begin{table}[htb]
%\tabletypesize{\tiny}
\small
\centering
%\begin{tabular}{l | c | c | c | c | c | c}
\begin{tabular}{lcccccl}
\hline
Simulation      &  $t_{\rm jet}$ (Myr)    &  $\delta$  &  R (kpc)  &  $N_{\rm cl}$  & Figure  &  Main feature  \\
\hline
%\hline
M20$\delta$0.3  &  $10$ ($t_{\rm q}=10$)  &  $0.3$     &  $1$      &  $3$           &  \ref{figure: 3clumps_r45_rhofactor1.3_t50}, \ref{figure: 3clumps_r45_rhofactor1.3_t305}, \ref{RunM20D0.3Temp}, \ref{RunM20D0.3Trace}  &  Global vorticity; Multiple shocks  \\
S20$\delta$1    &  $20$                   &  $1$       &  $1$      &  $1$           &  \ref{figure: rhofactor2_6panels}, \ref{figure: rhofactor2_zoomed}, \ref{Tracers}  &  Mixing via vortices  \\
S20$\delta$2    &  $20$                   &  $2$       &  $1$      &  $1$           &  \ref{figure: rhofactor3}, \ref{Tracers}  &  Vortices; Filaments  \\
S20$\delta$3    &  $20$                   &  $3$       &  $1$      &  $1$           &  \ref{Tracers}, \ref{figure: rhofactor4_size123}  &  \\
S20$\delta$3R2  &  $20$                   &  $3$       &  $2$      &  $1$           &  \ref{figure: rhofactor4_size123}  &  \\
S20$\delta$3R3  &  $20$                   &  $3$       &  $3$      &  $1$           &  \ref{figure: rhofactor4_size123}  &  \\
S20$\delta$1C5  &  $20$                   &  $1$       &  $1$      &  $5$           &  \ref{figure: confA_rhofactor2}, \ref{figure: confA_rhofactor2_tracer}  &  Filaments  \\
\hline
\end{tabular}
\caption{
{{{{Simulations performed in this work.
The name of each simulation is built as follows: `M' denotes multiple jet-activity episodes; `S' denotes a single jet episode;
`$\delta\#$' denotes the mass density contrast $\delta$ (eq.~\ref{delta}) in the clumps (given in the third column of the table);
`R$\#$' denotes the radius of the clumps in kpc (given in the fourth column);
`C$\#$' denotes the number of clumps in the simulation (given in the fifth column).
The second column shows the duration of the jet $t_{\rm jet}$.
In Run~M20$\delta$0.3 the jet is periodic with quiescence time of $t_{\rm q}=10 \Myr$.
The last two columns list the figures that present the results and the main features of the flow.}}}}
}
\label{table: simulations}
\end{table}
% TTTTTTTTTTTTTTTTTTTTTTTTTTTTTTTTTTTTTTTTTTTTTTT
}}}

% ==========================================================
\section{GLOBAL FLOW STRUCTURE}
\label{s-global_flow_structure}
% ==========================================================

The global flow structure influences the mixing of the clump with the shocked jets' material.
For that we start by presenting in figure \ref{figure: 3clumps_r45_rhofactor1.3_t50} and \ref{figure: 3clumps_r45_rhofactor1.3_t305}
the global flow structure of {{{{Run~M20$\delta$0.3}}}}, where a periodic jet is injected into the grid.
Later we will follow in greater details the evolution of the clumps in this run.
The clumps, which have an initial 3D torus structure in our {{{{2D}}}} grid, start with an initial constant density having a contrast of
\begin{equation}
\delta =\frac {\rho_{\rm clump}-\rho_{\rm ICM}}{\rho_{\rm ICM}},
\label{delta}
\end{equation}
relative to the density of the ICM, $\rho_{\rm ICM}$.
The jet is active for $t_{\rm jet} = 10 \Myr$ and with a quiescence period of $t_{\rm q} = 10 \Myr$ between active phases,
i.e., a time period of $t_{\rm jp} = 20 \Myr$.
%FFFFFFFFFFFFFFFFFFFFFFFFFFFFFFFFFFFFFFFFFFFFFFFFFFF
\begin{figure}[htb]
\centering
\subfigure{\includegraphics[width=0.49\textwidth]{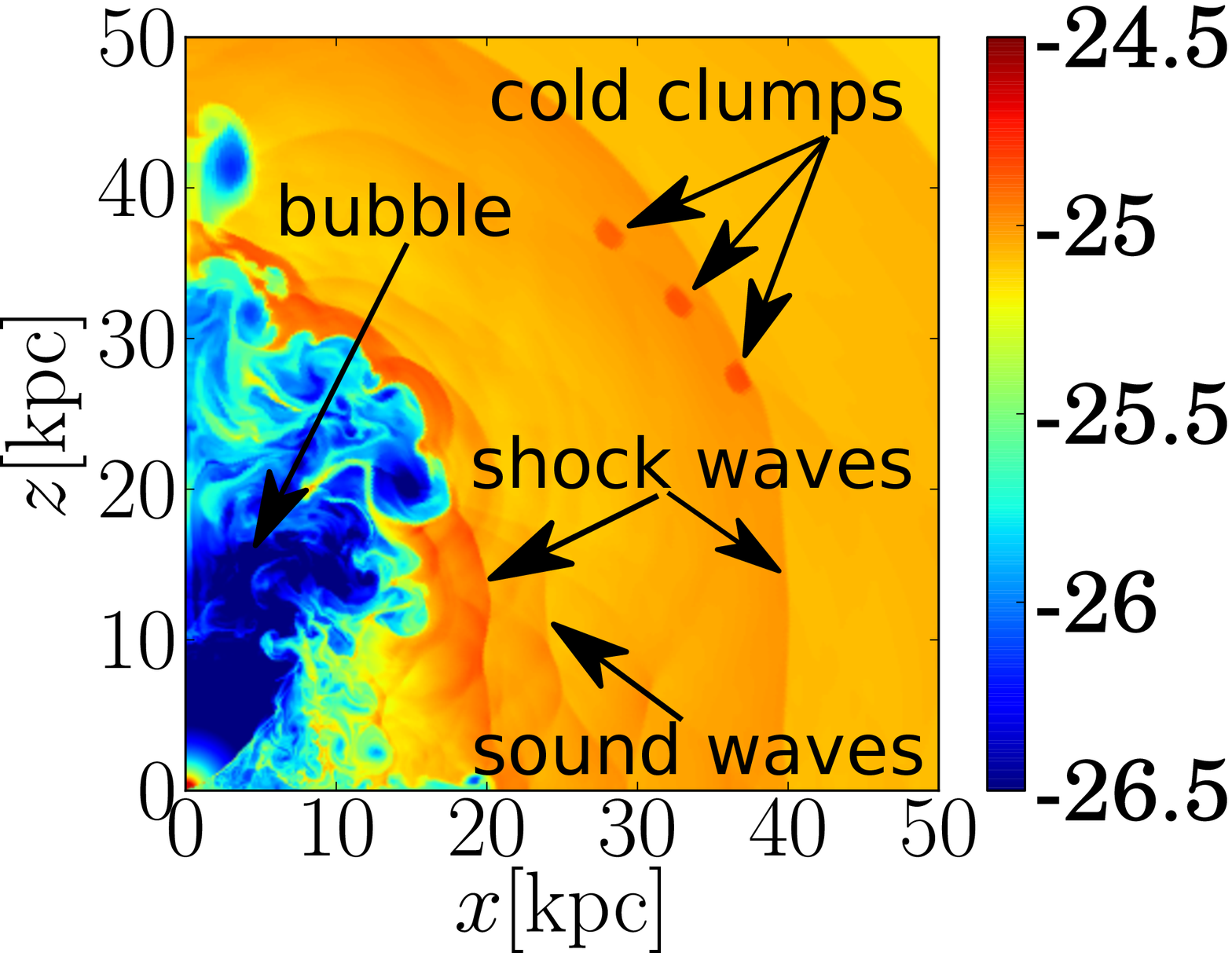}}
\subfigure{\includegraphics[width=0.49\textwidth]{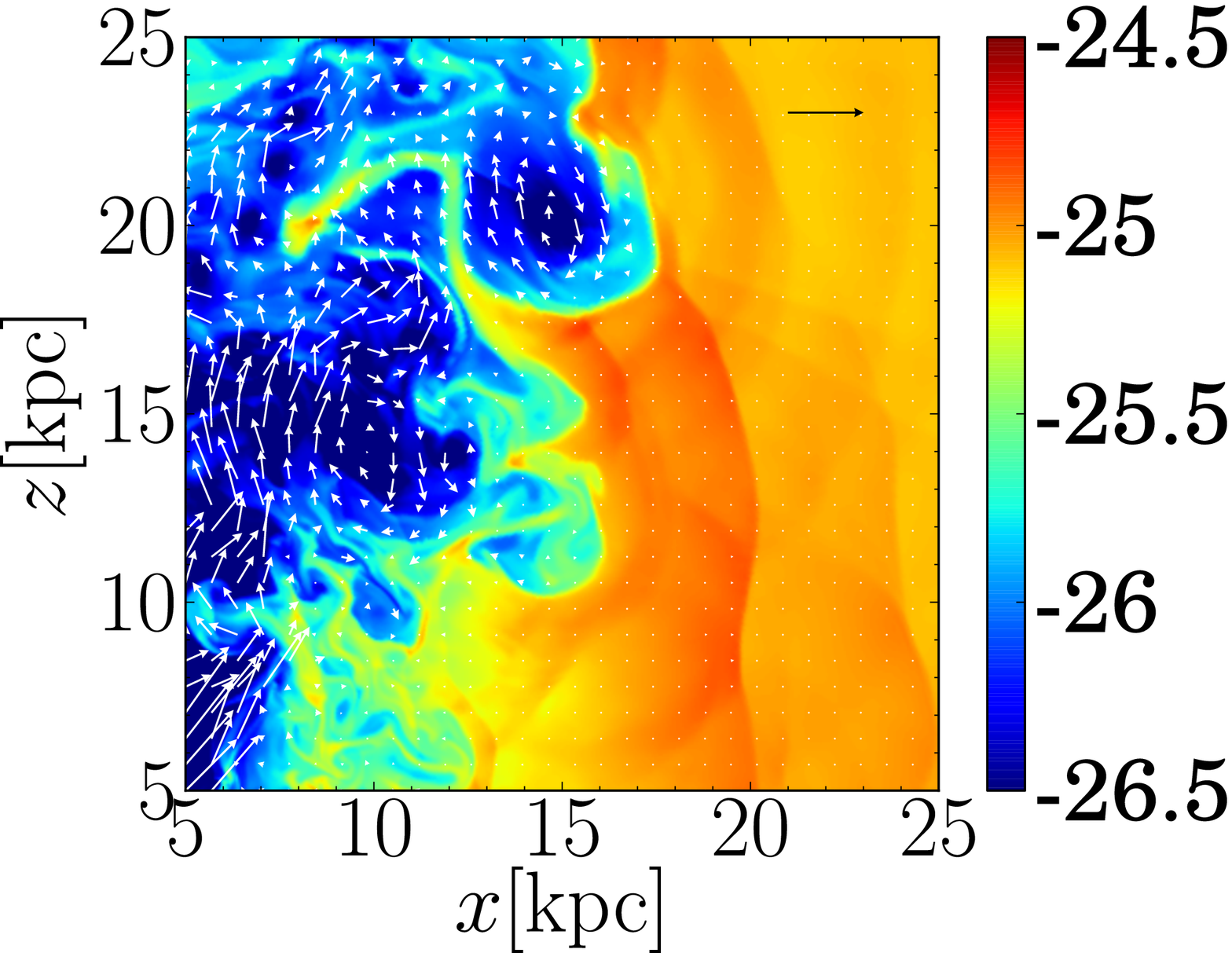}}
\caption{The global flow structure of {{{{Run~M20$\delta$0.3}}}} at $t=50 \Myr$.
In {{{{this simulation}}}} the jet is active for $t_{\rm jet} = 10 \Myr$ in each episode, with a quiescence period of $t_{\rm q} = 10 \Myr$
between active episodes.
The color coding of density is in $\g \cm^{-3}$ and logarithmic scale.
In the left panel we show the density map at $t=50 \Myr$ of the regions influenced by the jets.
Marked are three dense-cold clumps with an initial over-density of $\delta =(\rho_{\rm clump}-\rho_{\rm ICM})/\rho_{\rm ICM}=0.3$,
whose evolution we follow in later sections, as well as shock waves and sound waves.
Each clump has an initial shape of a torus in 3D with a cross section of radius $R=1 \kpc$, and the centers of the cross sections at $t=0$
are at a distance of $r=45 \kpc$ from the center of the grid.
In the right panel we zoom on one region at the same time, and also present velocity arrows, emphasizing vorticity and flow
toward the jet's axis in some ICM regions. Velocity is proportional to the arrow length, with inset showing an arrow
for $10,000 \km \s^{-1}$. Note the different scaling of the panels.
}
\label{figure: 3clumps_r45_rhofactor1.3_t50}
\end{figure}
%FFFFFFFFFFFFFFFFFFFFFFFFFFFFFFFFFFFFFFFFFFFFFFFFFFF
%FFFFFFFFFFFFFFFFFFFFFFFFFFFFFFFFFFFFFFFFFFFFFFFFFFF
\begin{figure}[htb]
\centering
\subfigure{\includegraphics[width=0.217\textwidth]{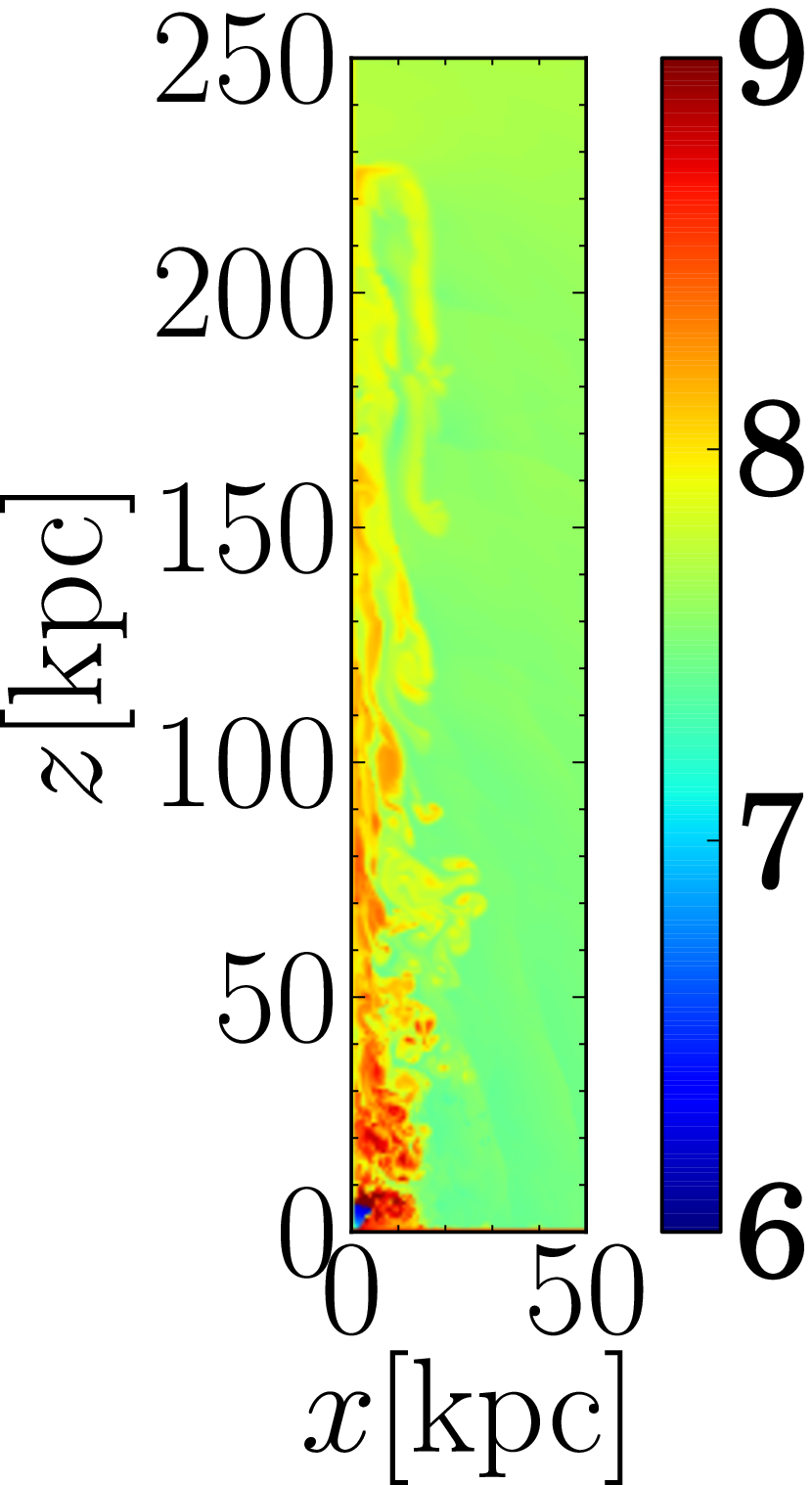}}
\subfigure{\includegraphics[width=0.278\textwidth]{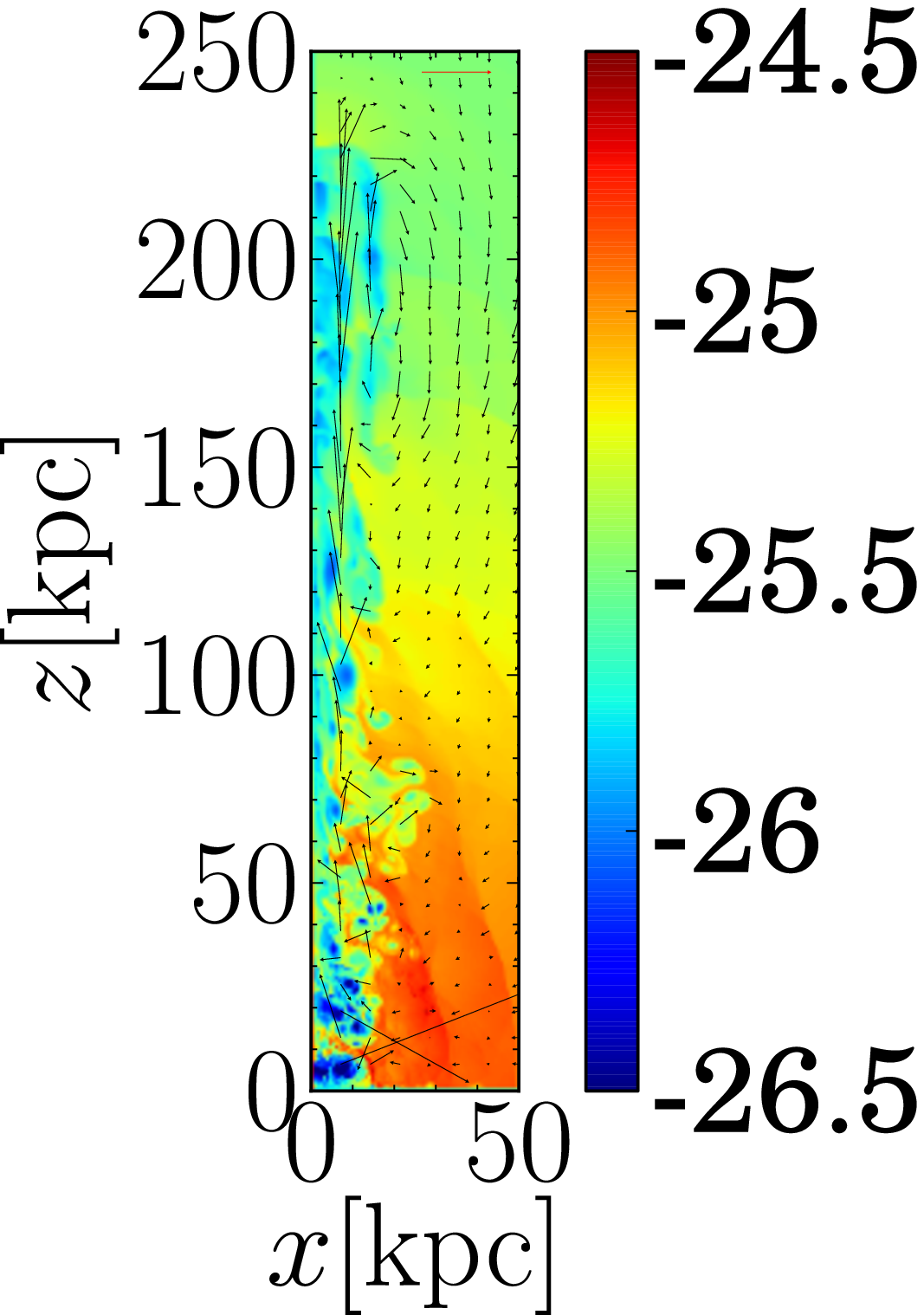}}
\subfigure{\includegraphics[width=0.485\textwidth]{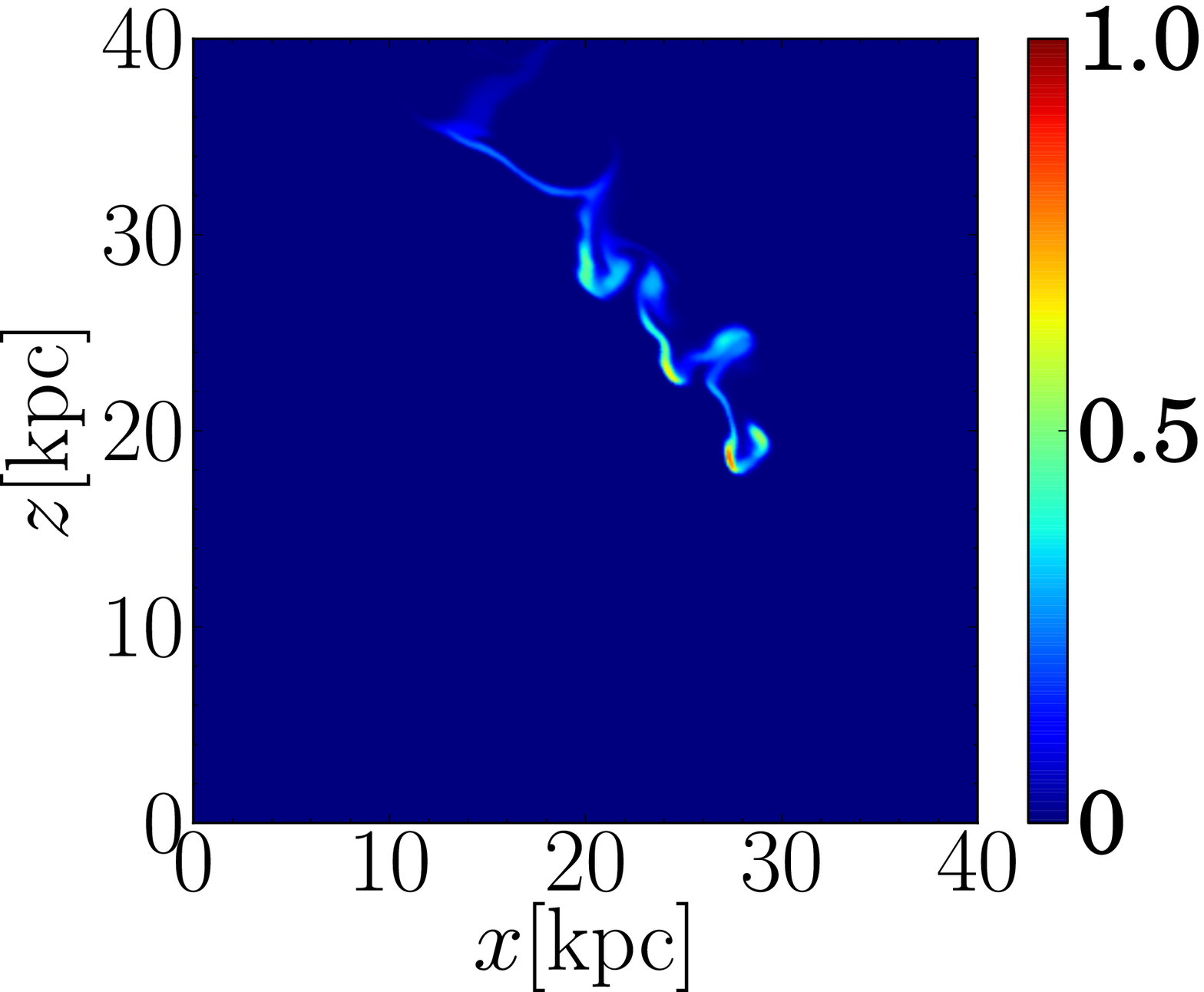}}
\caption{The global flow structure of {{{{Run~M20$\delta$0.3}}}} at $t=305 \Myr$.
The panels, from left to right, show the temperature map, density map with velocity arrows, and tracers of the material that
started in the three cold clumps, respectively.
Velocity is proportional to the arrow length, with red horizontal arrow scaled to $1,000 \km \s^{-1}$.
Color coding of tracers stands for the concentration of material originated in the clumps.
The density scale is as in Fig.~\ref{figure: 3clumps_r45_rhofactor1.3_t50}, and the temperature scale is
in $\K$ and logarithmic scale. Note different scaling of panels.
}
\label{figure: 3clumps_r45_rhofactor1.3_t305}
\end{figure}
%FFFFFFFFFFFFFFFFFFFFFFFFFFFFFFFFFFFFFFFFFFFFFFFFFFF

Due to numerical limitations of the {{{{2D}}}} code we launch jets along the same direction and they expand along the symmetry axis.
In reality, for such a long duration we expect that a relative motion of the central SMBH and the ICM, as well as different
directions of jets launching, will cause a spreading of the shocked jets' gas in the cluster.
For that, the global morphology of the jets at distances of $r \ga 40 \kpc$, which is the location of the first jets after two episodes,
does not represent reality well. Nonetheless, we continue the run for a long time in order to study the influence of the global flow
on the dense clumps evolving at distances $r \la 45 \kpc$.

We note the following properties of the global flow that can be seen in the different panels of Fig.
\ref{figure: 3clumps_r45_rhofactor1.3_t50} and \ref{figure: 3clumps_r45_rhofactor1.3_t305}, and
teach us about the physics and relevance to observations.
\begin{enumerate}
\item {\it Bubbles.} In most resolved CF clusters X-ray deficient bubbles are observed close to the center.
Therefore, it is crucial to obtain such bubbles in simulations that aim at reproducing the heating processes in clusters.
Our simulations show that we can indeed form large low-density bubbles near the center, termed `fat bubbles,'
as previously done, e.g., by \cite{Sternberg2007} and \cite{GilkisSoker2012}, and are similar to observed bubbles.
Due to the limitation of constant jet launching direction mentioned above, the bubbles are seen only at early time near
the center (Fig. \ref{figure: 3clumps_r45_rhofactor1.3_t50}).
\item {\it Vorticity.} Inflation of bubbles by jets introduces vortices inside the bubble and on its boundaries.
These vortices mix material in their surroundings, and enable mixing of cold gas with hot shocked jet material {{{{ \citep{Li2014} }}}}.
As we show in this study, this mixing has a significant role in the heating process of the ICM gas.
\item {\it Shock waves.} The high velocity of the jet material generates shock waves propagating outward in the ICM,
as marked on Fig. \ref{figure: 3clumps_r45_rhofactor1.3_t50}.
These forward shocks are almost spherical and propagate outward.
Each jet is shocked in a reverse shock that leads to the formation of a hot bubble.
The shock waves heat the ICM and raise its entropy, but as we show later, this process is much less efficient than mixing \citep{GilkisSoker2012}.
\item {\it Sound waves.} Multiple sound waves are observed in some CF clusters, e.g., Perseus \citep{Fabian2012}.
As shown before by \cite{Sternberg2009} and \cite{GilkisSoker2012} each bubble inflation episode excites multiple sound waves
in the ICM. We find here the same, but as we inflate many bubbles, the sound wave structure becomes very rich.
\item{\it Mixing of cold clumps.} As can be seen in the right panel of Fig. \ref{figure: 3clumps_r45_rhofactor1.3_t305},
the material that started in the upper left cold clump starts to mix with hot gas that is the shocked jets' material.
As we show later, this efficiently heats the cold gas originated in this clump.
{{{{ \item{\it Mixing within the ICM.} The large vortex seen in Fig. \ref{figure: 3clumps_r45_rhofactor1.3_t305} induces mixing
within the ICM medium itself. In 3D we expect the mixing to be more vigorous even.
Such mixing is very important in determining the evolution of the ICM, including dust evolution \citep{VoitDonahue2014} } }}}
{{{{ \item{\it Turbulence.} The large vortex discussed above, which in 3D will most likely be composed of several/many vortices,
will induce turbulence in the ICM. Namely, part of the injected energy ends up as kinetic energy of the ICM.  }}}}
\end{enumerate}

That we reproduce bubbles, shock waves, and sound waves similar to those observed in many CF clusters, gives us confidence that we are able to
simulate the heating processes of the ICM and the cold clumps embedded in it.
The gas flow, vorticity and global flow, are very hard to observe.
As we show below, they play a crucial role in facilitating the heating of the gas by the jets.

We emphasize again that in reality, namely, as observed \citep{Bogdan2013}, the jets' axes change direction, making mixing with scattered cold clumps and the ICM
much more efficient. We do not simulate this in our {{{{2D}}}} code, but we show that even in our `pessimistic' setting we can obtain efficient heating.
We will not study mixing of metals in the ICM, as it requires a more accurate treatment of the inflow and cooling.
We rather concentrate on the heating processes.

% ==========================================================
\section{FLOW STRUCTURE OF CLUMP-BUBBLE INTERACTION}
\label{s-clump_bubble_interaction}
% ==========================================================

% ==========================================================
\subsection{Mixing}
\label{s-mixing}
% ==========================================================

In this section we study single jet-launching episode and its flow structure.
The thermal evolution of the clumps is discussed in section \ref{s-heating-the-cold-gas}.
We place clumps of different density contrasts (eq.~\ref{delta}), of different
radii of their cross section $R = 1, 2, 3 \kpc$, at different azimuthal angles from the $z$ axis $\theta$
and different distances $r$ from the source of the jet (center).
In our {{{{2D}}}} numerical grid, each clump has a 3D shape of a torus.
In the simulations in this section we inject a single jet, as described in section \ref{s-numerical-setup}, along the $z$ axis (vertical direction in all figures) at $t=0$,
and turn it off at $t=20 \Myr$.

In {{{{Run~S20$\delta$1}}}} we place a single clump having a density contrast of $\delta = 1$, initial radius of cross section of
$R = 1 \kpc$, a distance from the center of $r = 20 \kpc$, and at $\theta = 45^\circ$ from the $z$-axis.
The results of this run are shown in Fig.~\ref{figure: rhofactor2_6panels}.
A `fat-bubble' is formed as in \cite{Sternberg2007},
and an almost spherical shock wave propagates outward (see section \ref{s-global_flow_structure}).
The clump itself goes through several phases of evolution,
as clearly seen in the different panels of Fig.~\ref{figure: rhofactor2_6panels}, and the zoomed plot in Fig.~\ref{figure: rhofactor2_zoomed}.
First it experiences {\it shock compression}, at $t \simeq 10 \Myr$.
The shock heats the clump, but on the other hand compresses it and shortens its radiative cooling time.
Here the clump survives, and at $t \simeq 30-50 \Myr$, when it is highly distorted to a horseshoe shape, starts to mix with
the hot shocked jet material (colored blue in the Fig.~\ref{figure: rhofactor2_6panels}).
The evolution in the {\it mixing phase} is dictated mainly by the many
vortices that are formed by the jet-ICM interaction \citep{GilkisSoker2012}.
In the {\it entraining phase}, starting at $t \sim 100 \Myr$, the dense part of the clump is entrained and dragged outward
(last panel of Fig.~\ref{figure: rhofactor2_6panels}).
%FFFFFFFFFFFFFFFFFFFFFFFFFFFFFFFFFFFFFFFFFFFFFFFFFFF
\begin{figure}[htb]
\centering
\subfigure{\includegraphics[width=0.42\textwidth]{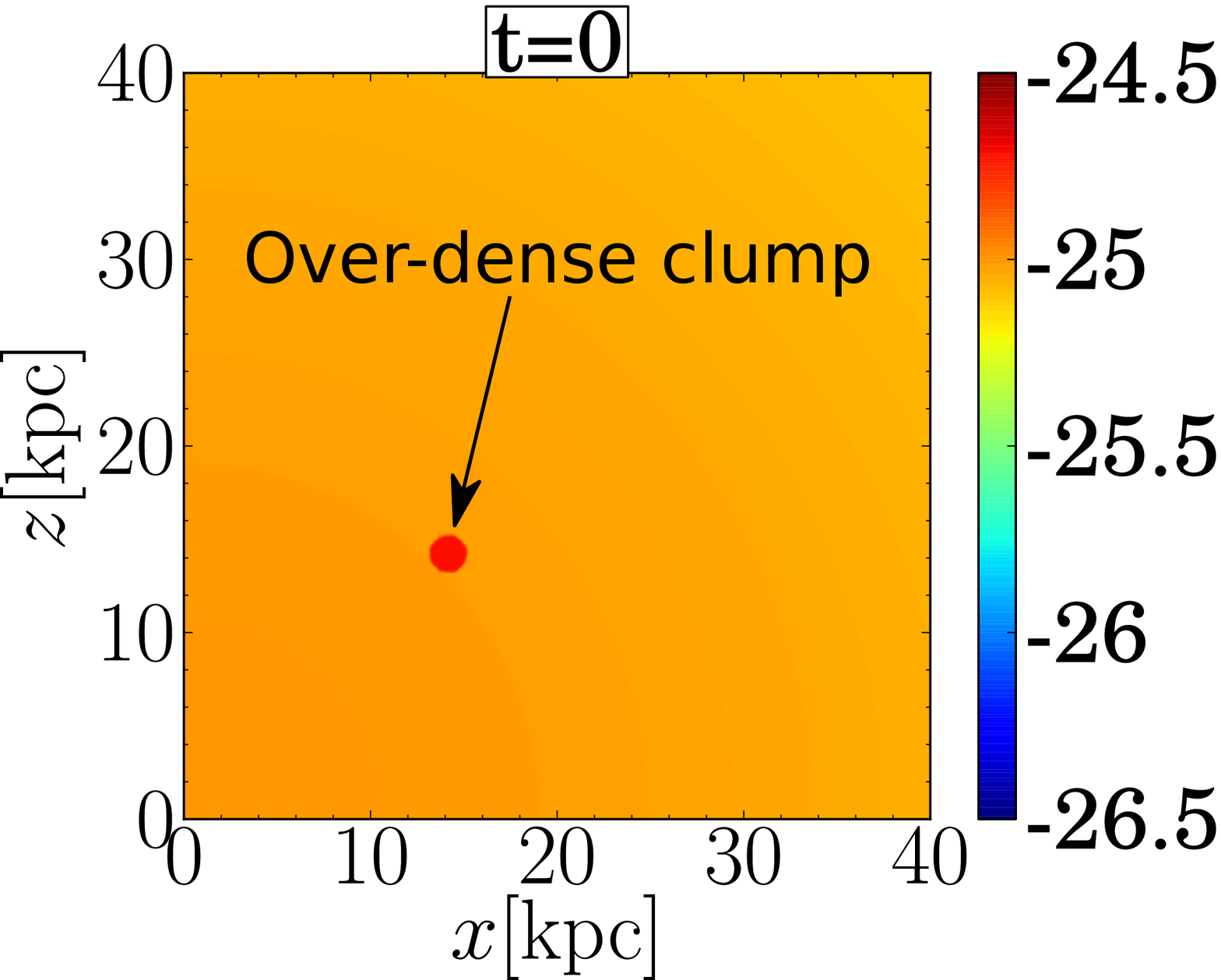}}
\subfigure{\includegraphics[width=0.42\textwidth]{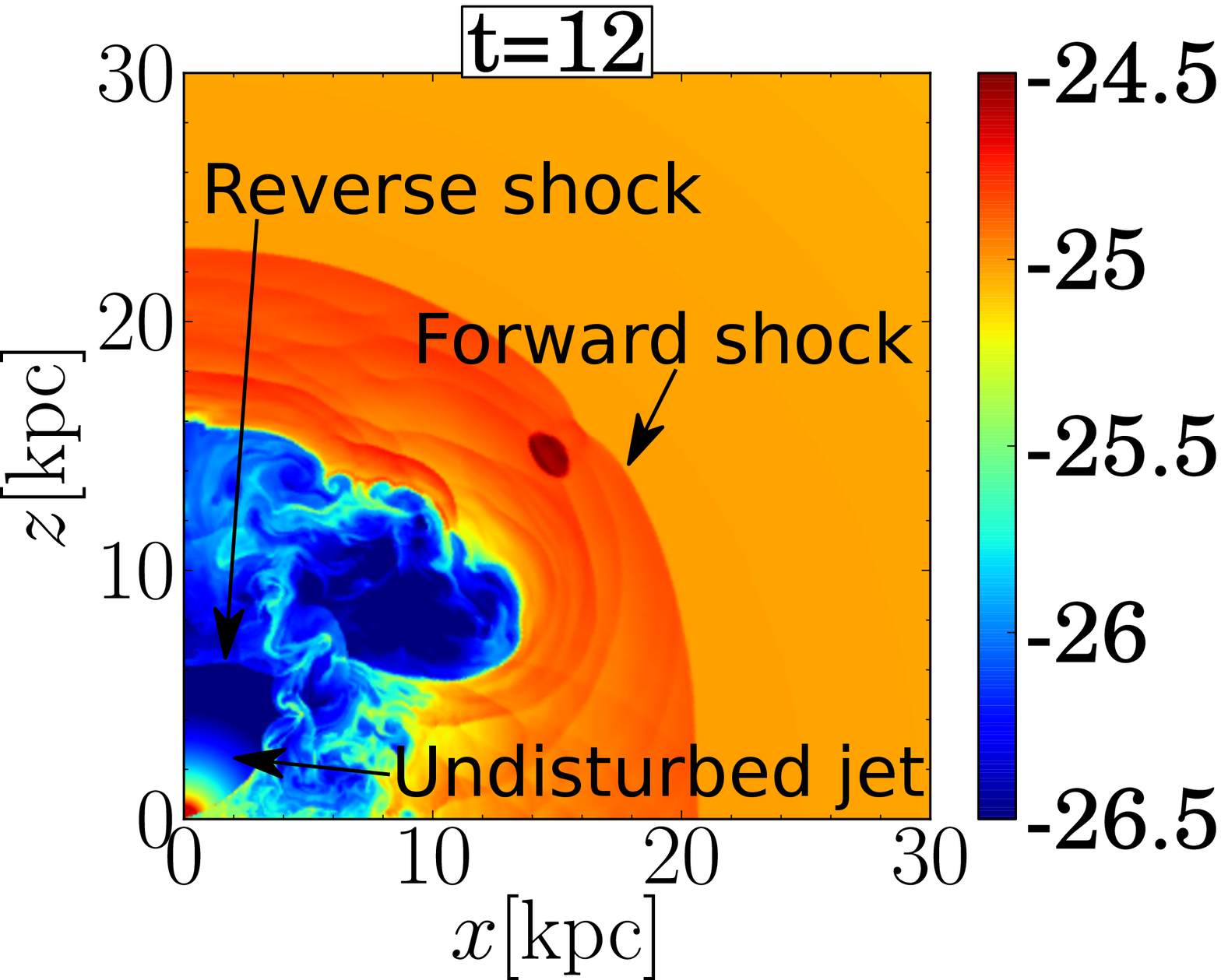}} \\
\subfigure{\includegraphics[width=0.42\textwidth]{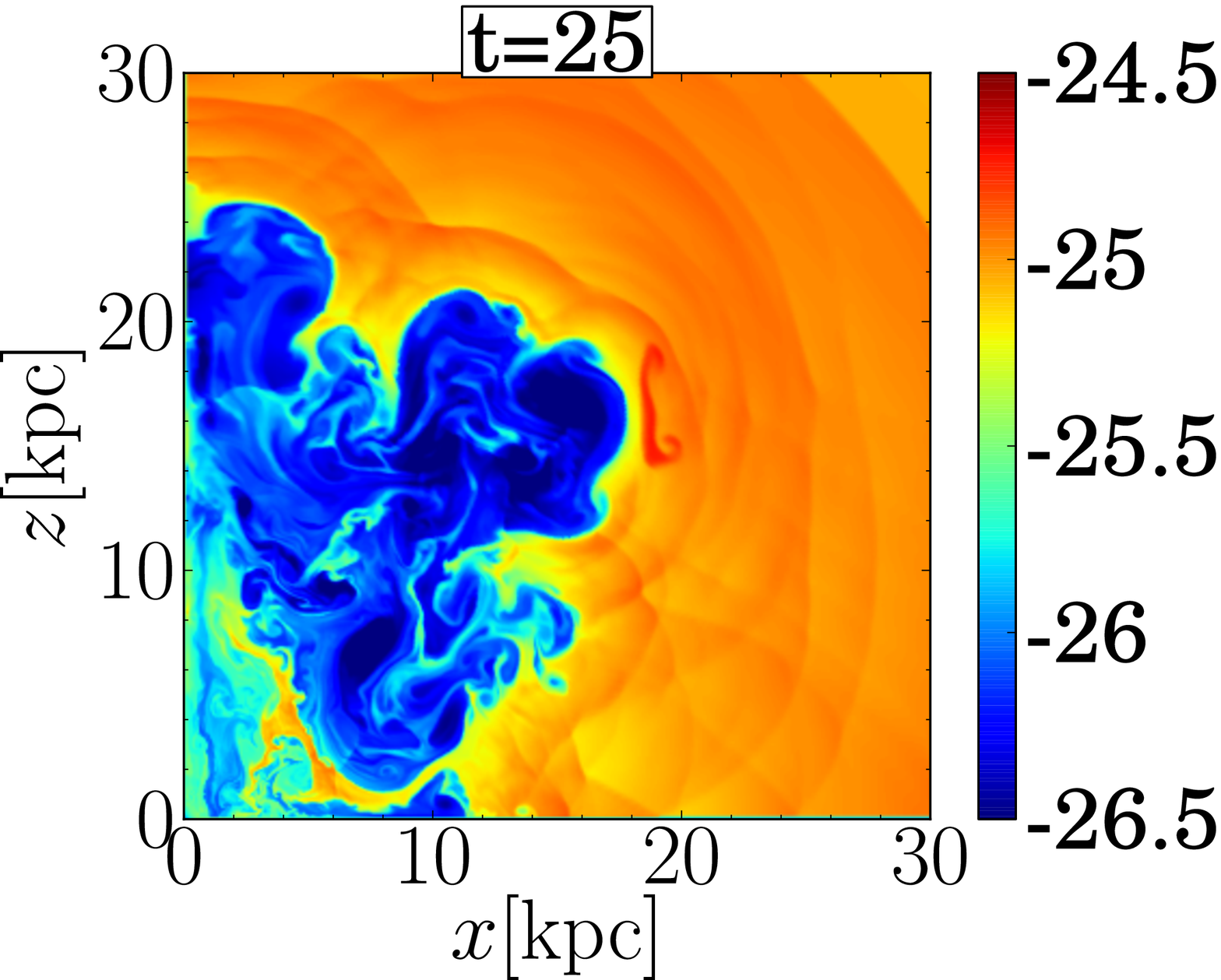}}
\subfigure{\includegraphics[width=0.42\textwidth]{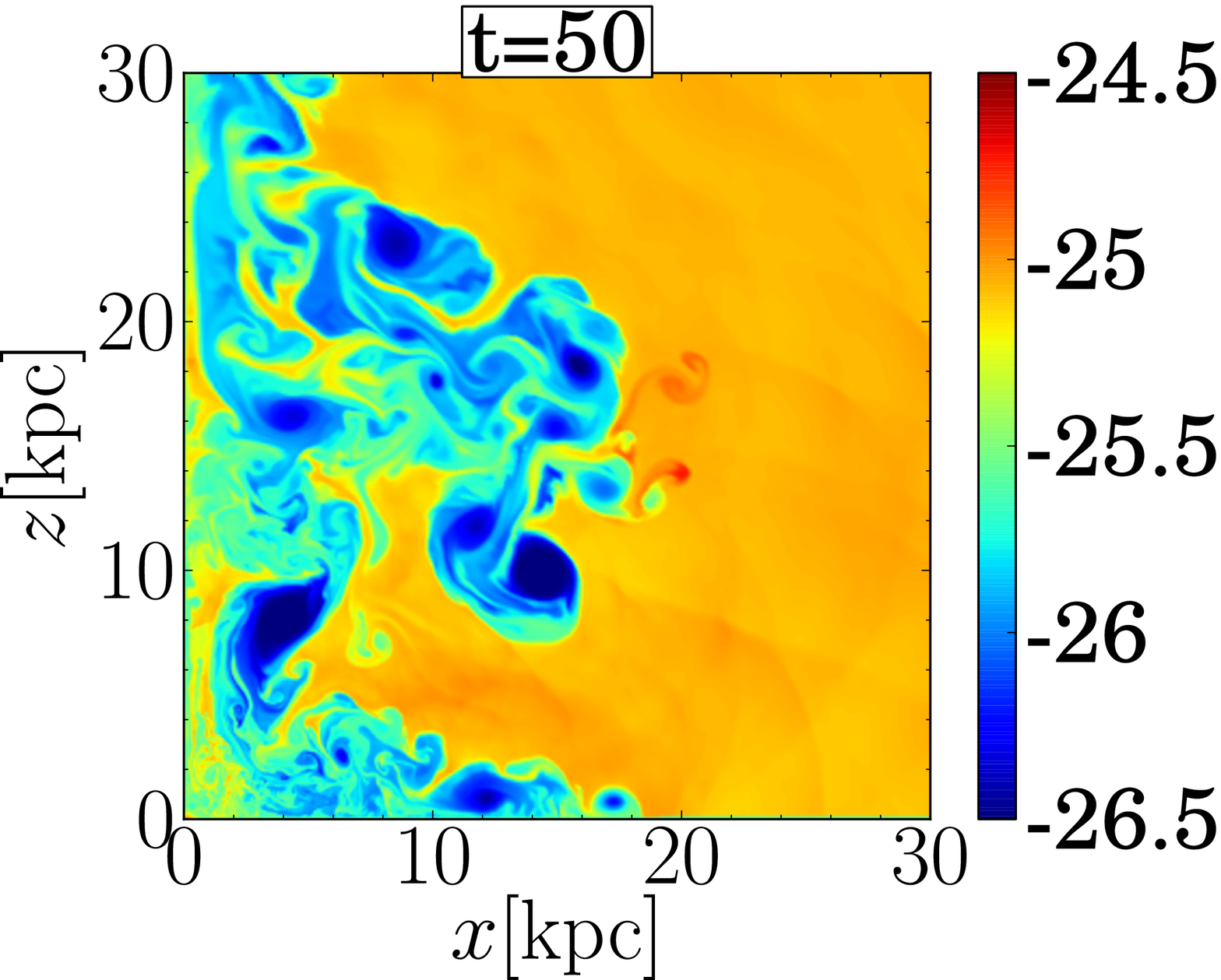}} \\
\subfigure{\includegraphics[width=0.42\textwidth]{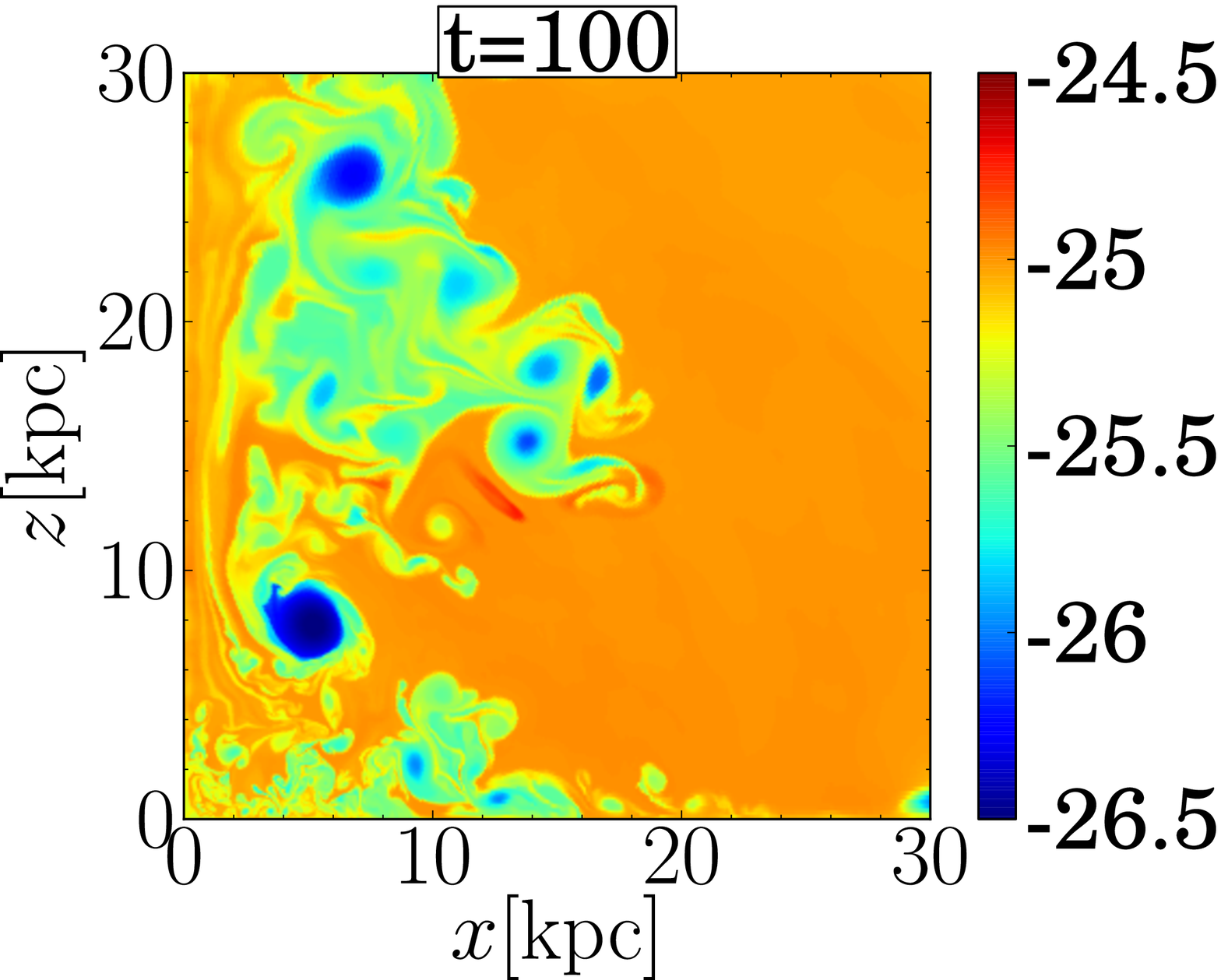}}
\subfigure{\includegraphics[width=0.42\textwidth]{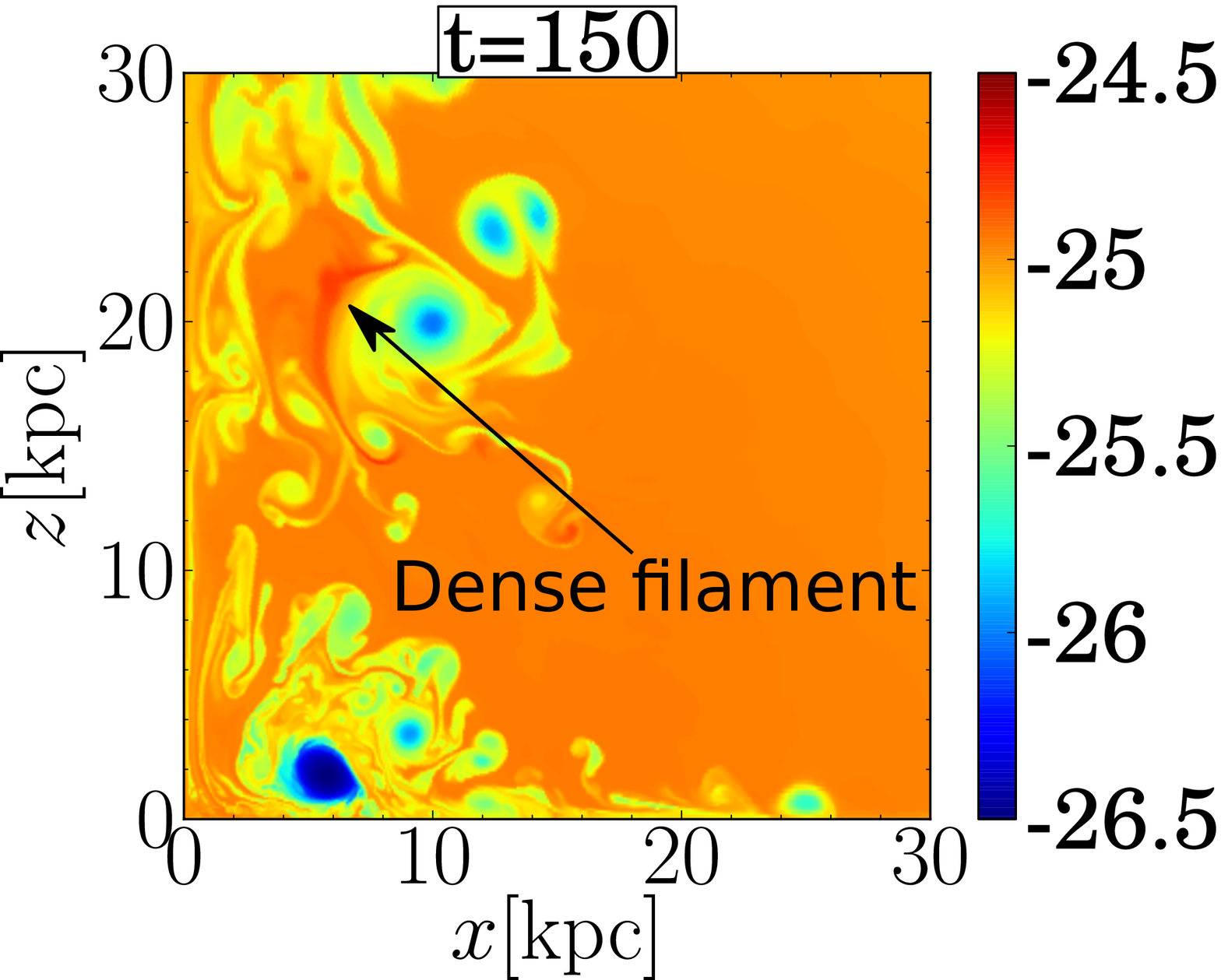}}
\caption{
Several quantities in the meridional plane for {{{{Run~S20$\delta$1}}}}. The clump has a shape of a torus in 3D with a cross section
of radius $R=1 \kpc$, the center of the cross section is at a distance $r= 20 \kpc$ and at an angle
$\theta=45^\circ$ from the $z$ axis.
The clump's density contrast is $\delta = 1$.
The jet operates for $t_{\rm jet}= 20 \Myr$ starting at $t=0$, and is then switched off.
The color bar gives the density in log scale in units of $\log{\rho} (\g \cm^{-3})$.
The darkest red regions in each panel are clump's material.
The time of each plot is indicated in units of $\Myr$.
As we show later, the vigorous mixing occurring around $t \simeq 35 \Myr$ leads to a substantial heating of the clump's material.
}
\label{figure: rhofactor2_6panels}
\end{figure}
%FFFFFFFFFFFFFFFFFFFFFFFFFFFFFFFFFFFFFFFFFFFFFFFFFFF
%FFFFFFFFFFFFFFFFFFFFFFFFFFFFFFFFFFFFFFFFFFFFFFFFFFF
\begin{figure}[htb]
\centering
\subfigure{\includegraphics[width=0.49\textwidth]{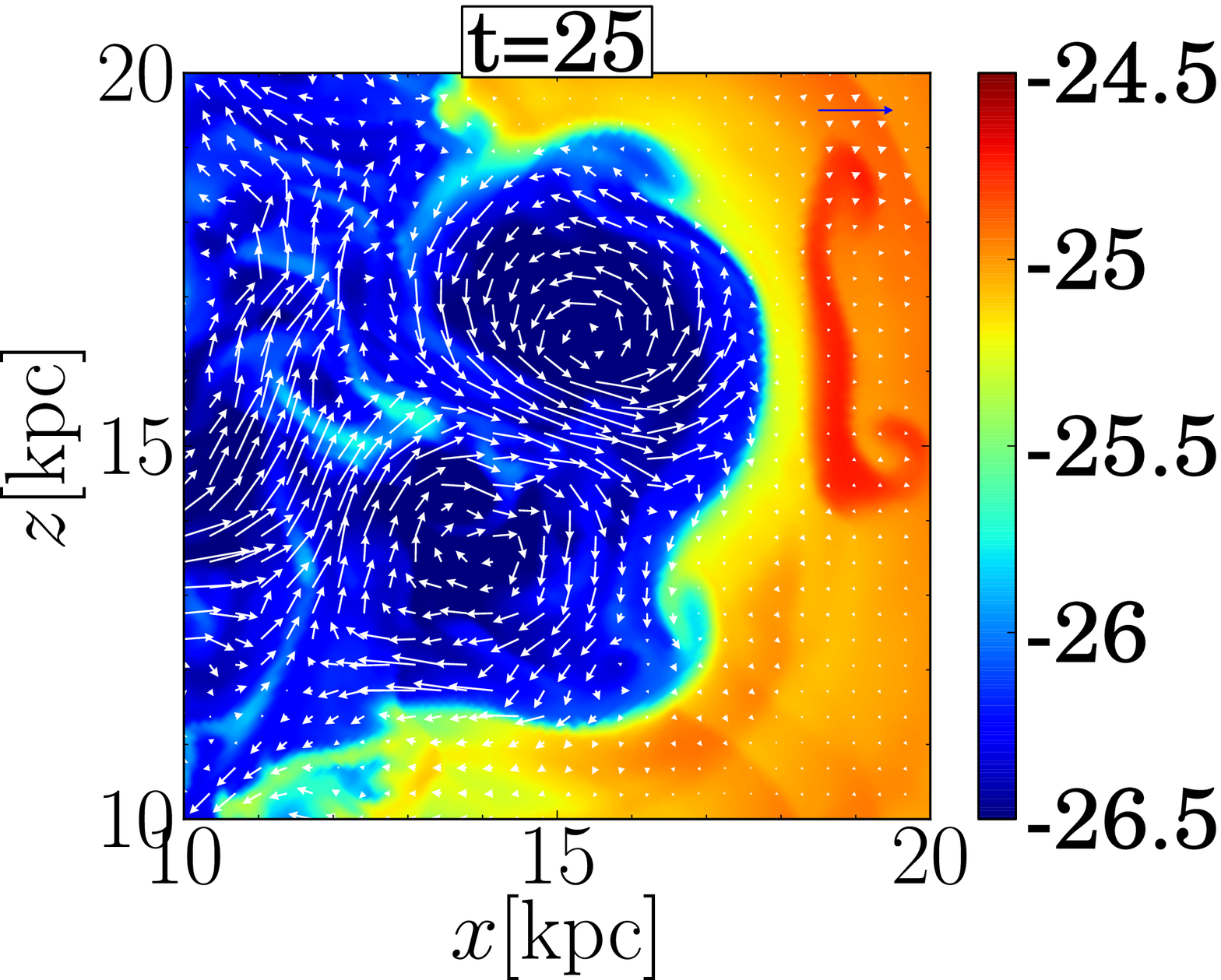}}
\subfigure{\includegraphics[width=0.49\textwidth]{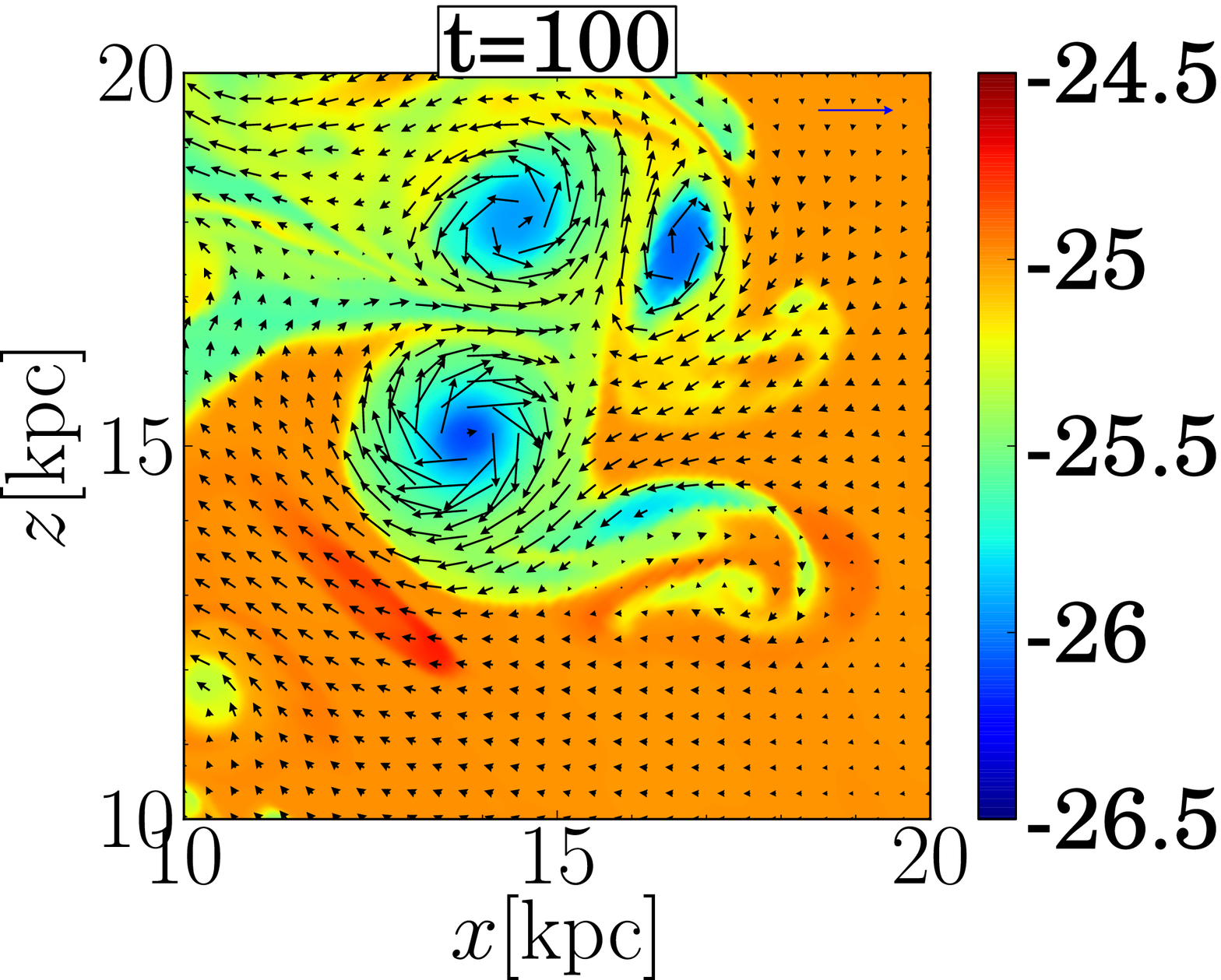}}
\caption{
Density and velocity plots for {{{{Run~S20$\delta$1}}}} presented in figure \ref{figure: rhofactor2_6panels}, emphasizing the vorticity
near the destructed clump (red area) at $t = 25$ (left) and $t = 100$~Myr (right).
The color bar gives the density in log scale in units of $\log{\rho} (\g \cm^{-3})$.
Three clear vortices are seen at $t=100 \Myr$.
Such vortices are behind the vigorous mixing of the dense clump with the shocked jet's gas.
Velocity is proportional to the arrow length,
with inset showing an arrow for $5,000 \km \s^{-1}$ for the $t = 25\Myr$ plot (left),
and $1,000 \km \s^{-1}$ for the $t = 100\Myr$ plot (right).
}
\label{figure: rhofactor2_zoomed}
\end{figure}
%FFFFFFFFFFFFFFFFFFFFFFFFFFFFFFFFFFFFFFFFFFFFFFFFFFF

For comparison in Fig.~\ref{figure: rhofactor3}
we present {{{{Run~S20$\delta$2}}}} of the evolution of a denser clump having an initial density contrast of $\delta = 2$,
but otherwise identical to the clump in {{{{Run~S20$\delta$1}}}}.
In Fig. \ref{Tracers} we show the location of the gas (`tracers') originated in the clump of {{{{Run~S20$\delta$1}}}}, {{{{Run~S20$\delta$2}}}}, and {{{{Run~S20$\delta$3}}}} for
initial density contrasts of $\delta = 1,2,3$, {{{{respectively}}}}.
The differences between the clumps is mainly due to the stochastic flow of vorticity. More mass in the denser clumps makes its tracers more
prominent in the figure as it spreads.
%FFFFFFFFFFFFFFFFFFFFFFFFFFFFFFFFFFFFFFFFFFFFFFFFFFF
\begin{figure}[htb]
\centering
\subfigure{\includegraphics[width=0.49\textwidth]{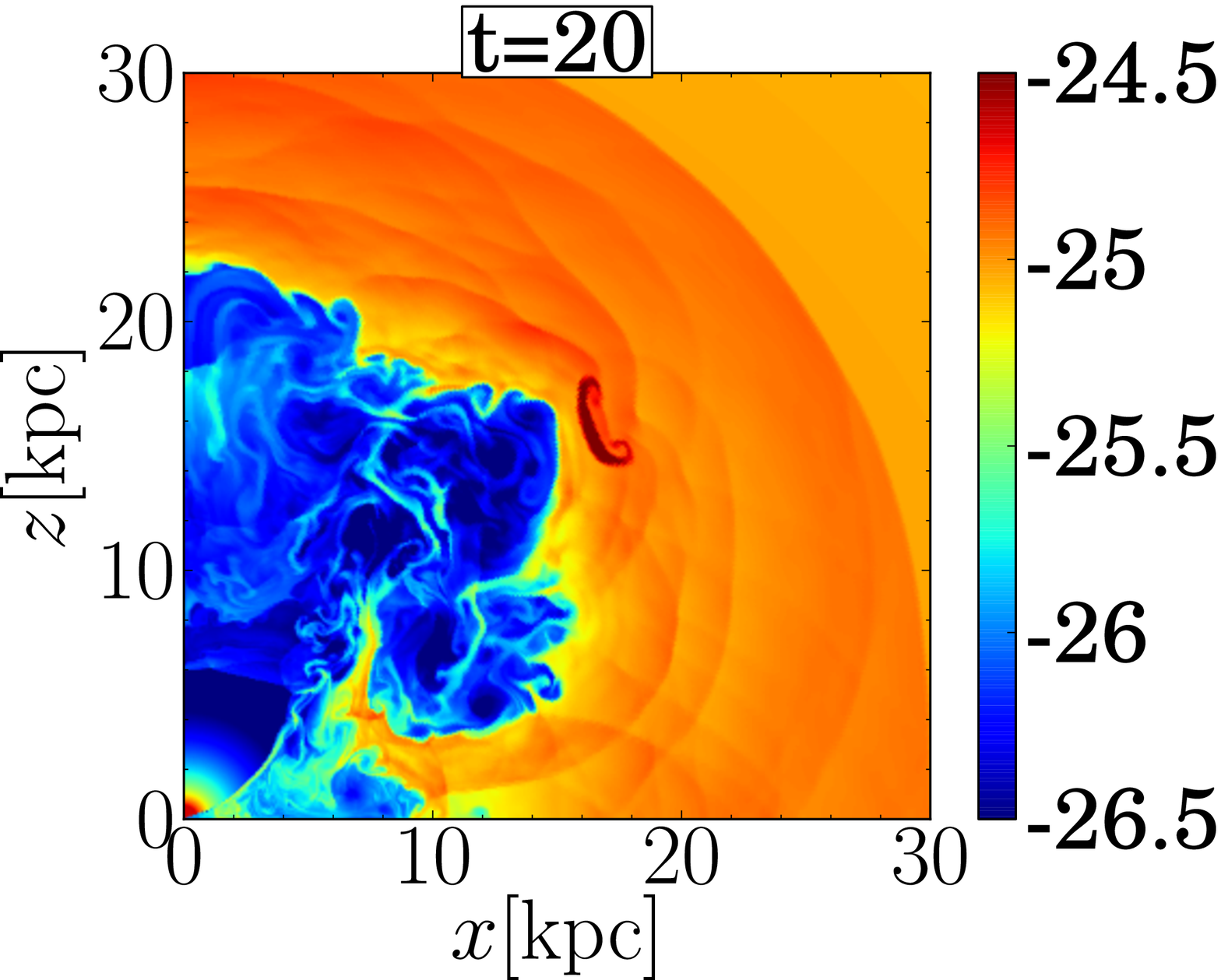}}
\subfigure{\includegraphics[width=0.49\textwidth]{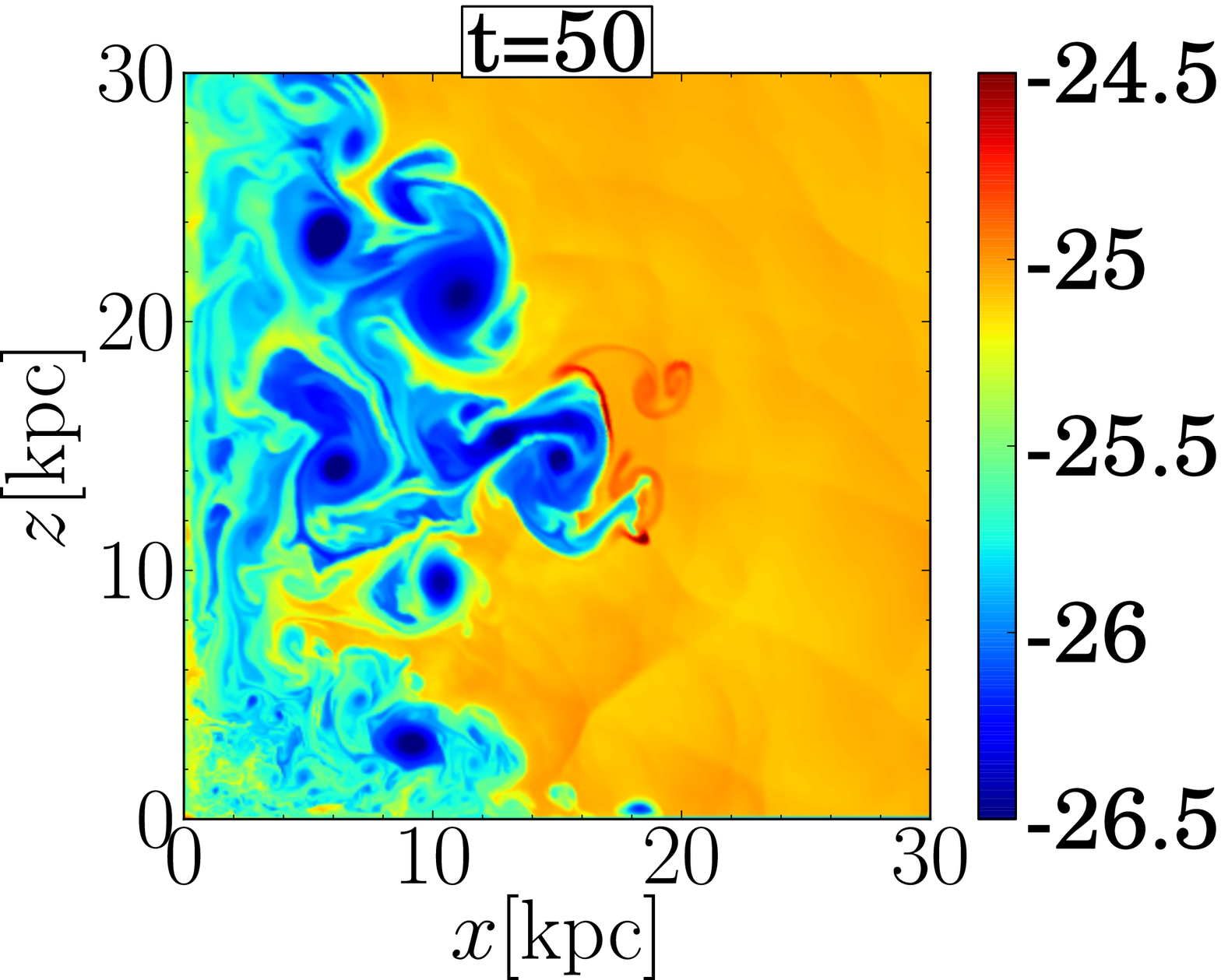}} \\
\subfigure{\includegraphics[width=0.49\textwidth]{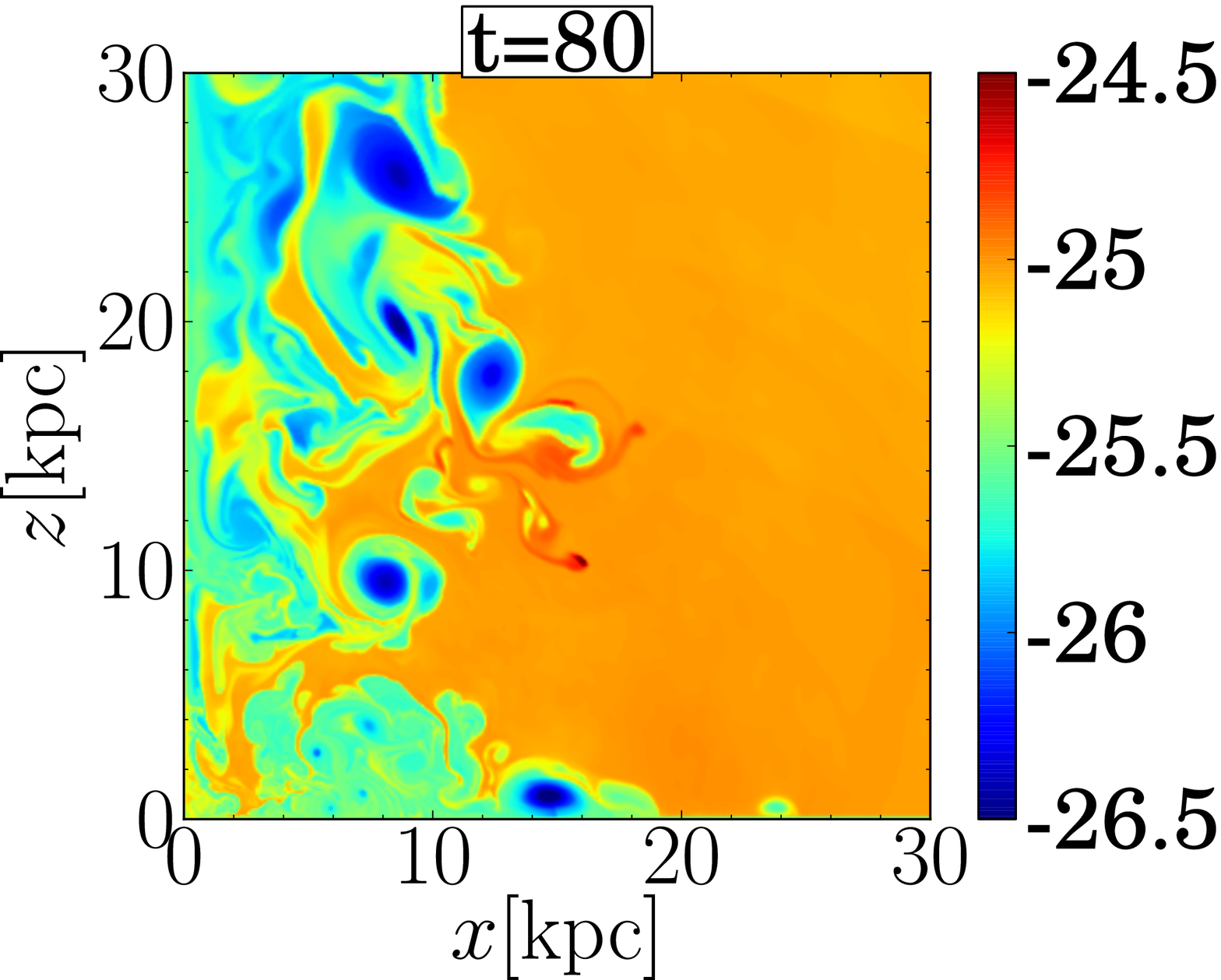}}
\subfigure{\includegraphics[width=0.49\textwidth]{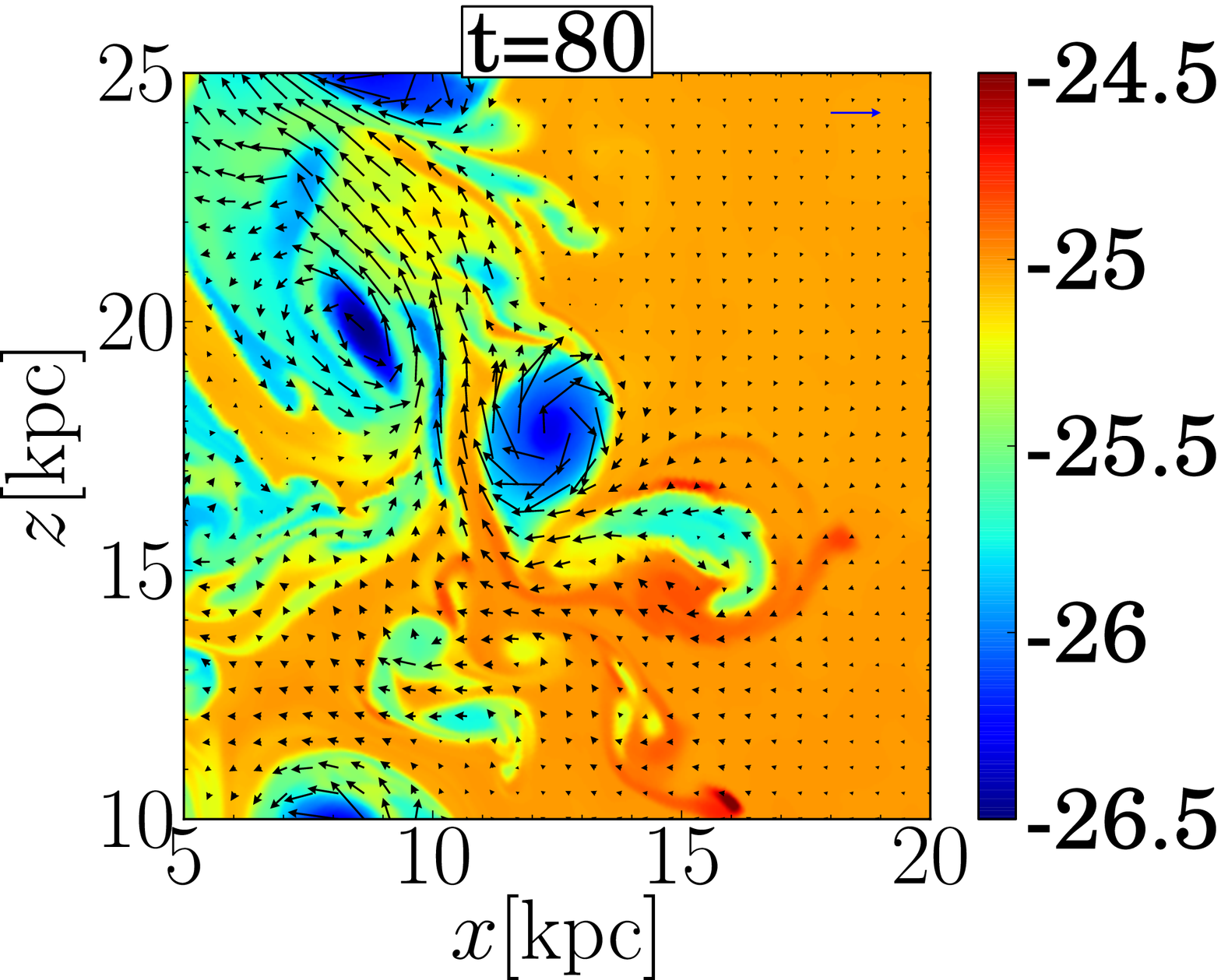}}
\caption{
Results for {{{{Run~S20$\delta$2}}}}, where the clump has an initial density contrast of $\delta  = 2$, but otherwise the
initial setting is as in {{{{Run~S20$\delta$1}}}} presented in Fig. \ref{figure: rhofactor2_6panels}.
Plots are shown at times of $t = 20, 50, 80 \Myr$ as indicated, with a zoomed plot on the right bottom panel.
Velocity is proportional to the arrow length, with inset showing an arrow for $1,000 \km \s^{-1}$.
The spreading of the clump's gas and the formation of dense filaments is evident in the bottom two panels.
}
\label{figure: rhofactor3}
\end{figure}
%FFFFFFFFFFFFFFFFFFFFFFFFFFFFFFFFFFFFFFFFFFFFFFFFFFF
%FFFFFFFFFFFFFFFFFFFFFFFFFFFFFFFFFFFFFFFFFFFFFFFFFFF
\begin{figure}[htb]
\centering
\subfigure{\includegraphics[width=0.32\textwidth]{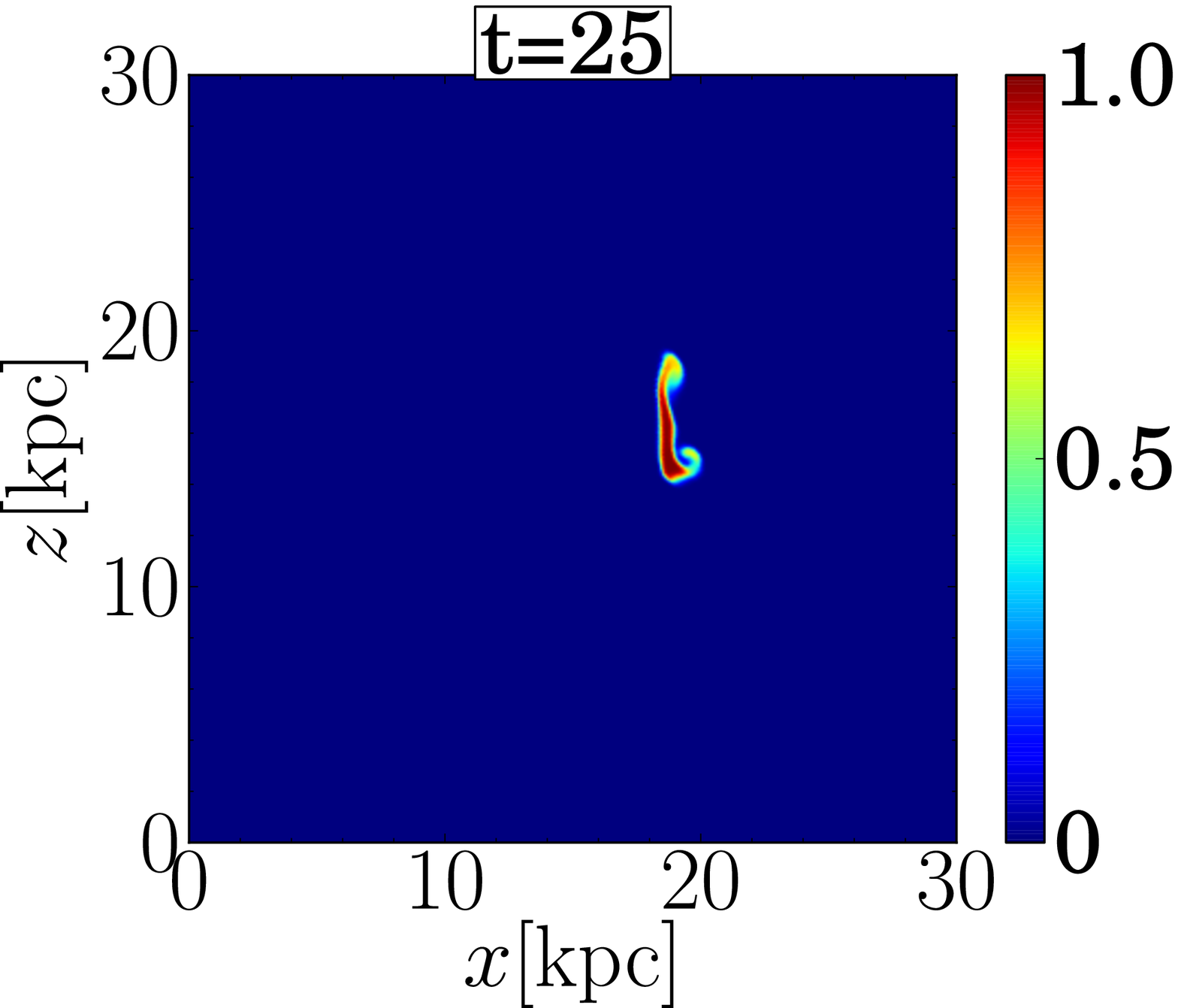}}
\subfigure{\includegraphics[width=0.32\textwidth]{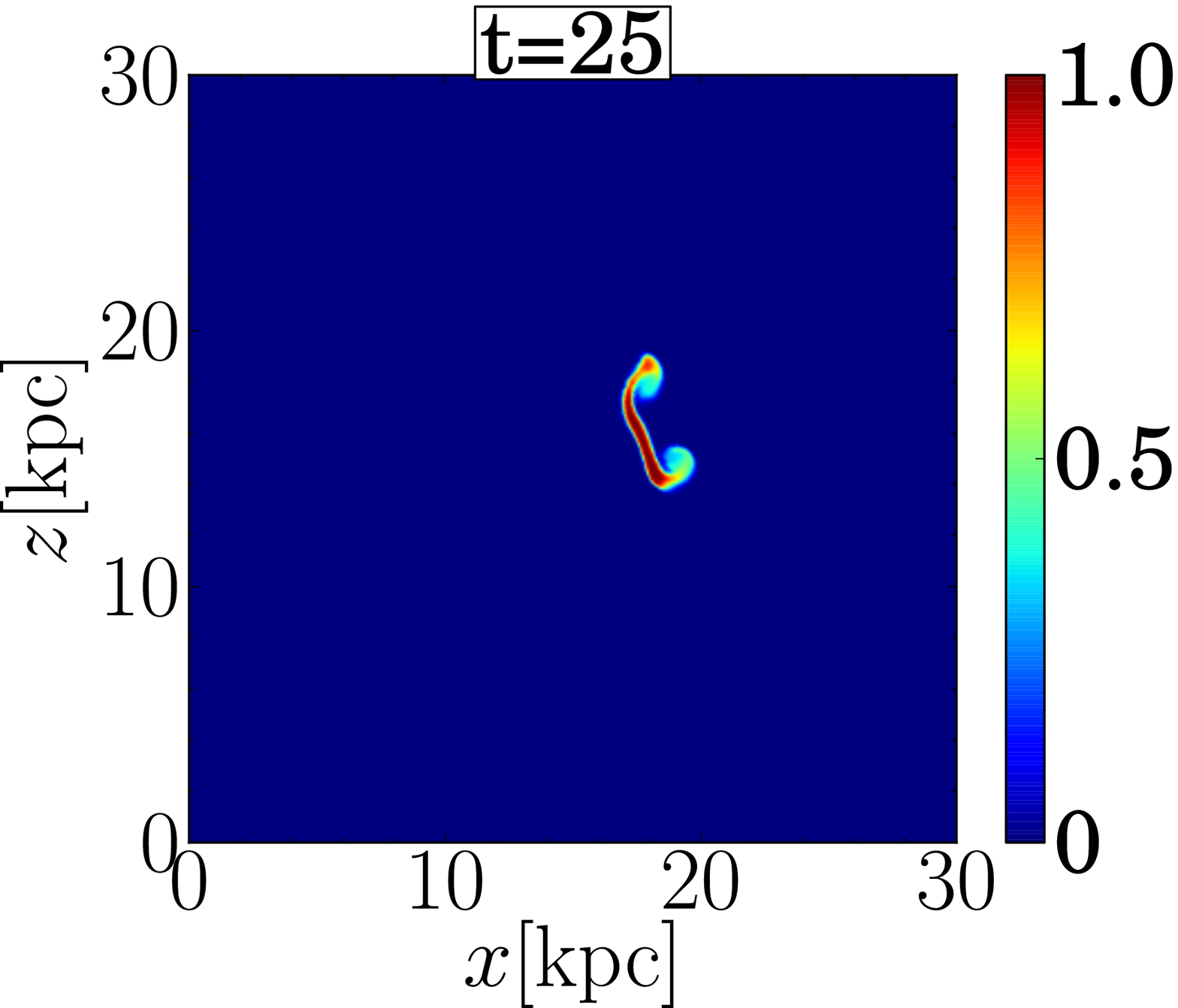}}
\subfigure{\includegraphics[width=0.32\textwidth]{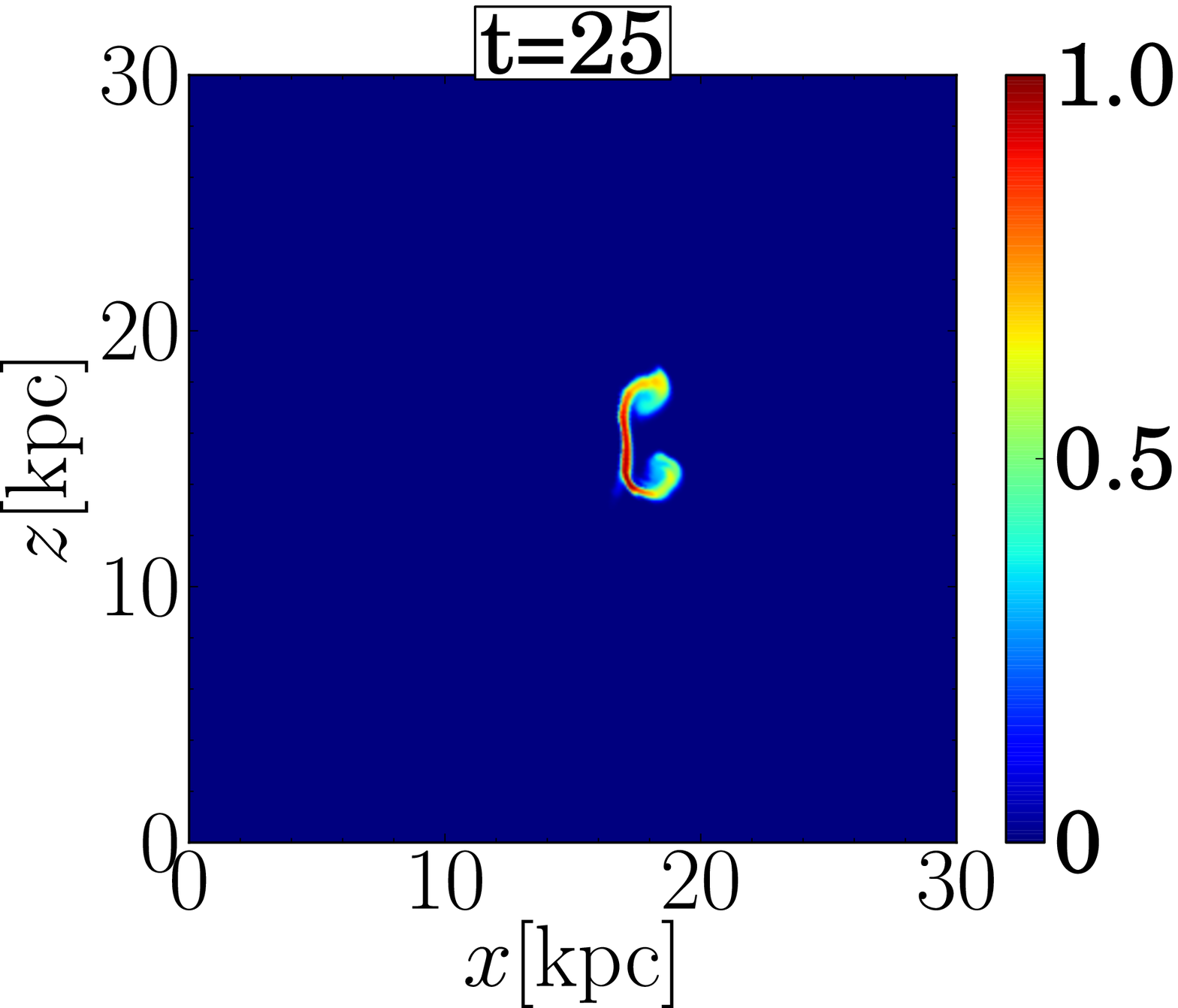}} \\
\subfigure{\includegraphics[width=0.32\textwidth]{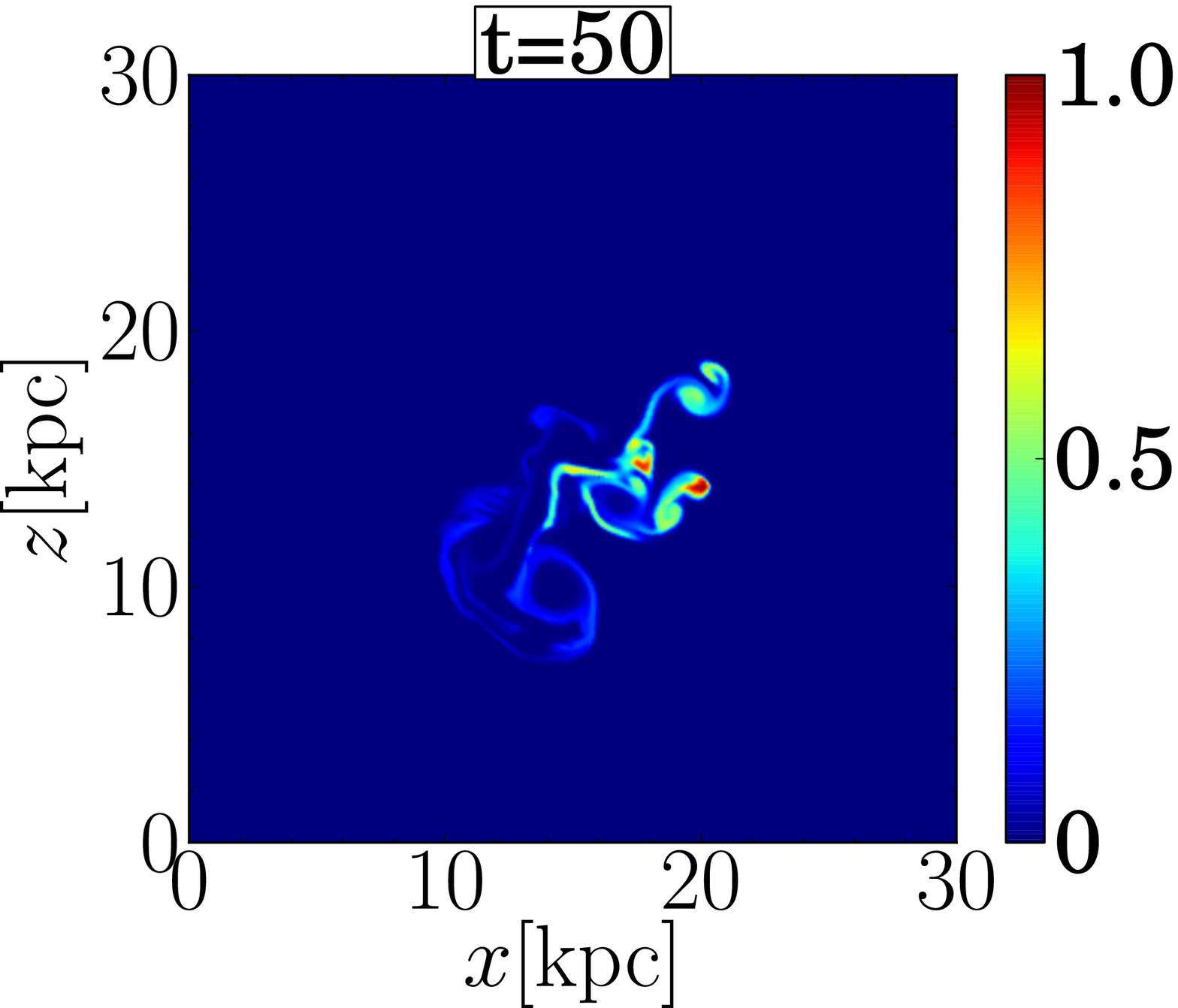}}
\subfigure{\includegraphics[width=0.32\textwidth]{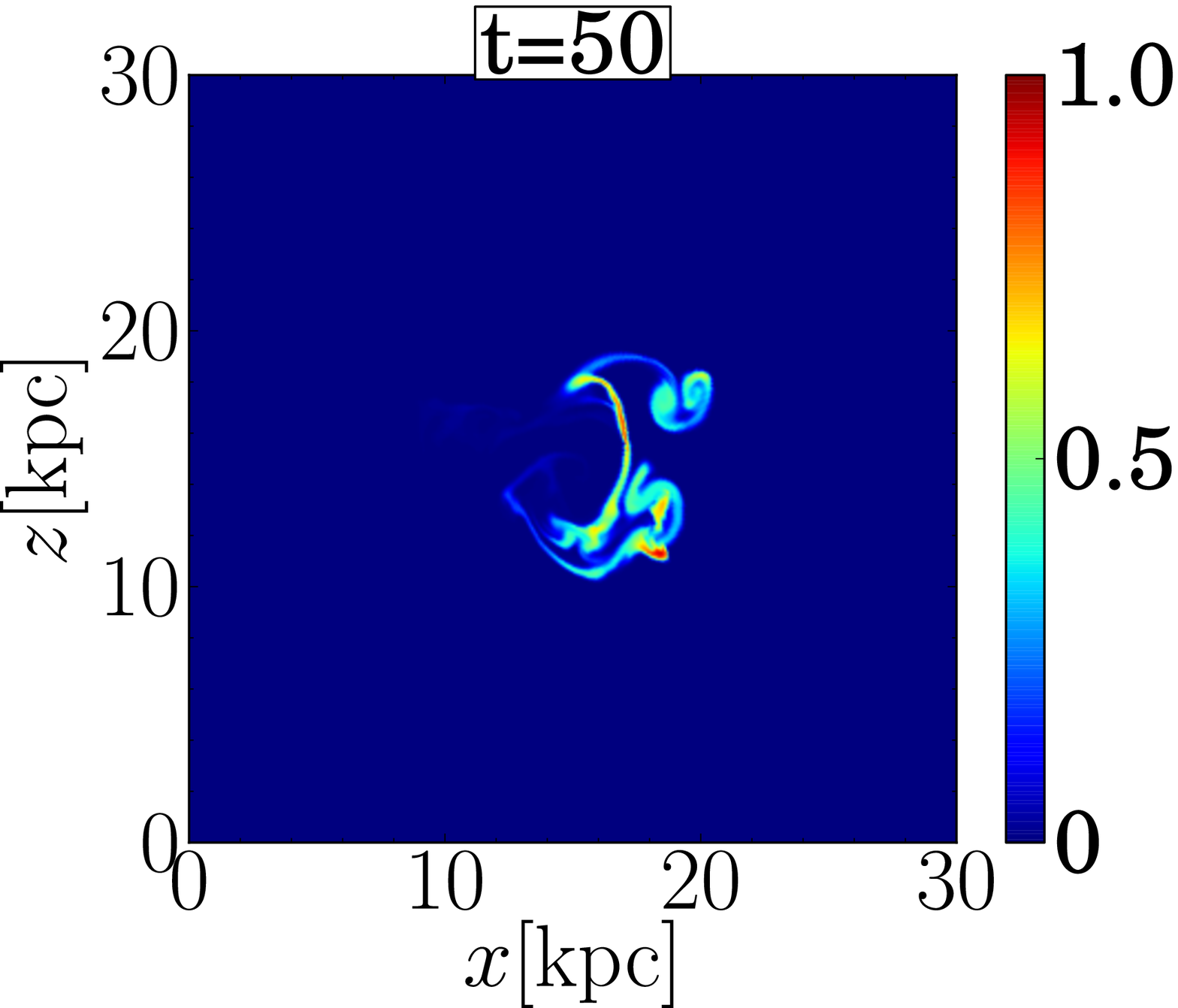}}
\subfigure{\includegraphics[width=0.32\textwidth]{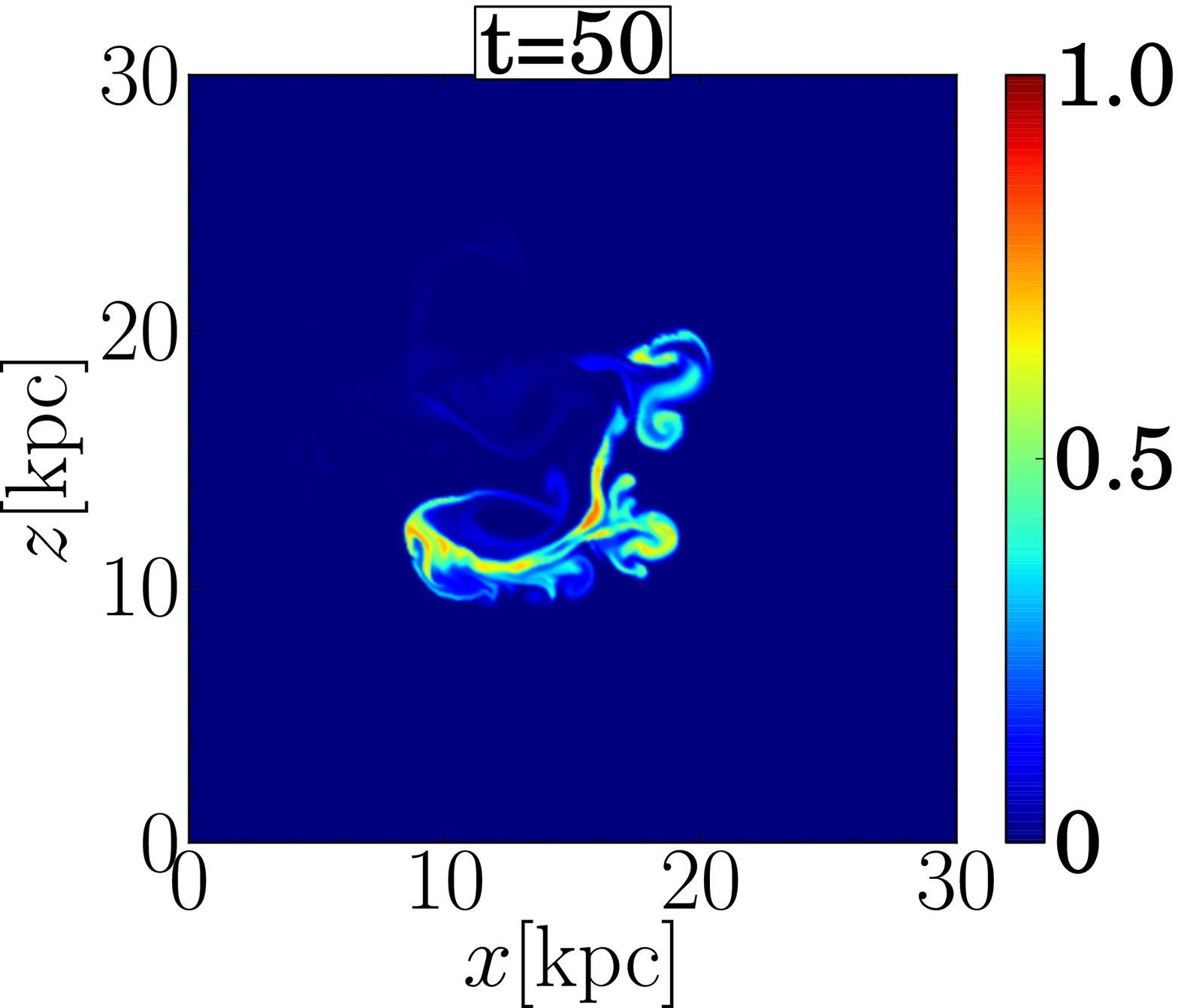}} \\
\subfigure{\includegraphics[width=0.32\textwidth]{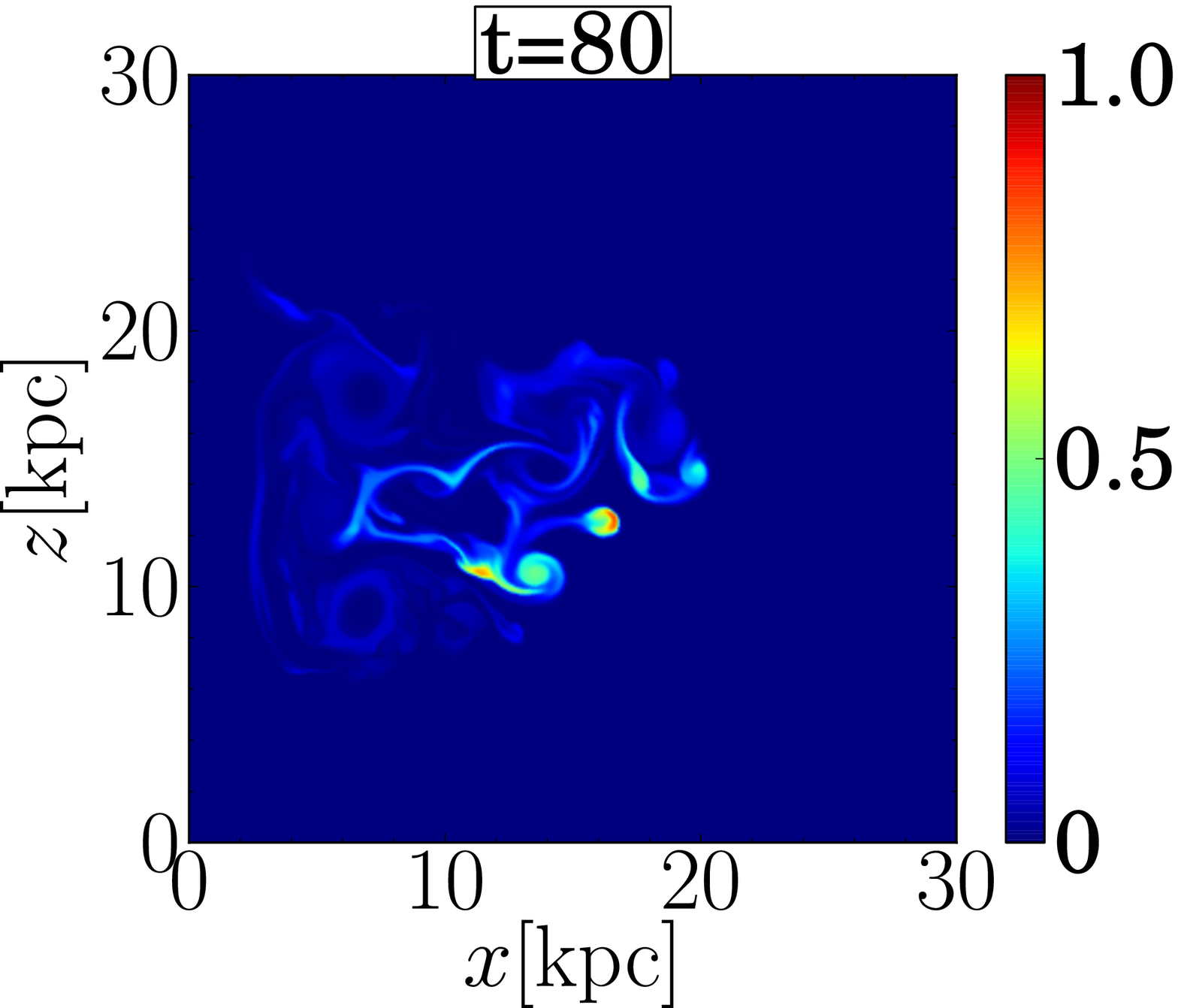}}
\subfigure{\includegraphics[width=0.32\textwidth]{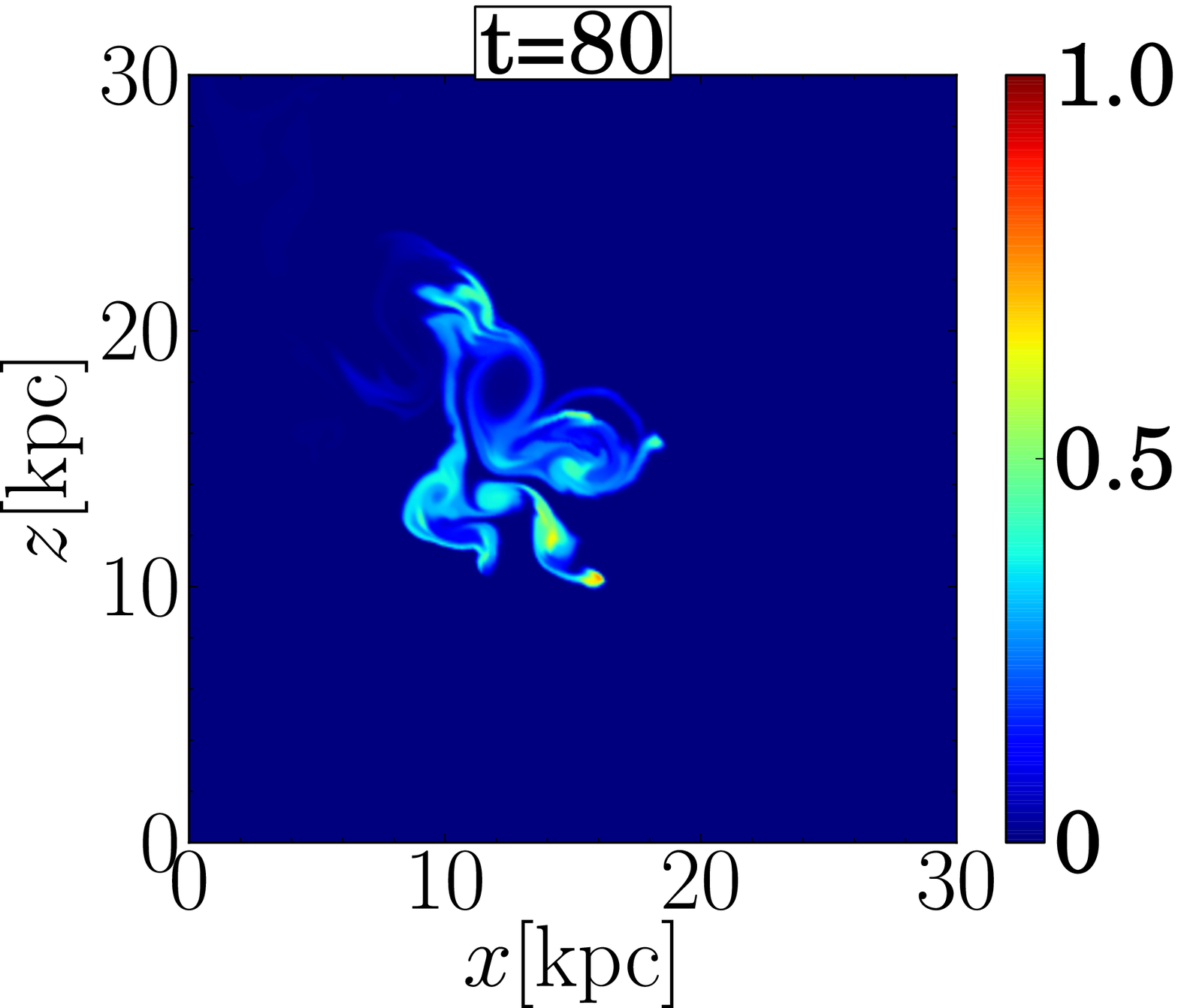}}
\subfigure{\includegraphics[width=0.32\textwidth]{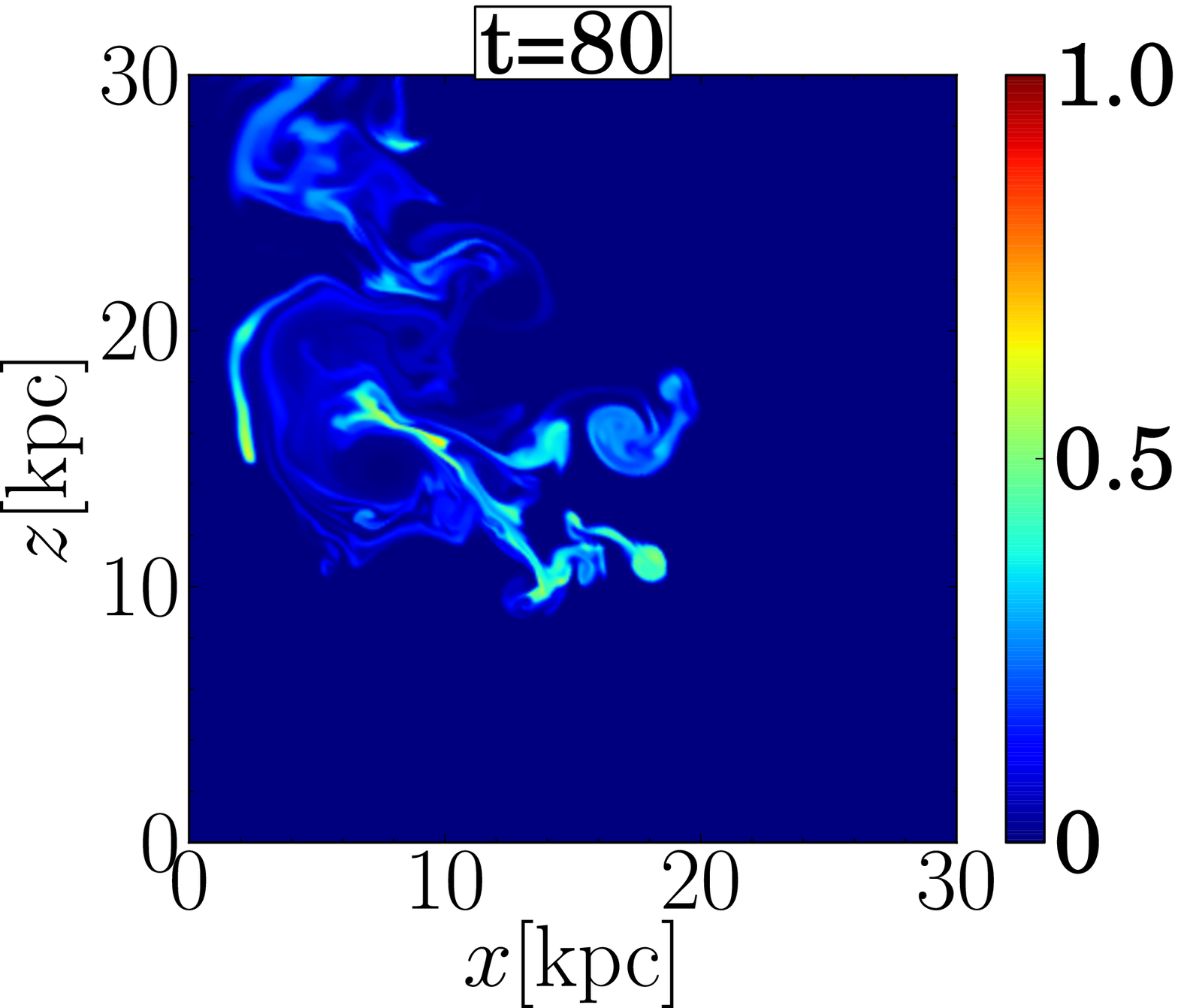}}
\caption{Comparing the location of gas originated in the dense clump of {{{{Run~S20$\delta$1}}}} (density contrast of $1$; left column),
{{{{Run~S20$\delta$2}}}} (density contrast of $2$; middle column), and {{{{Run~S20$\delta$3}}}} (density contrast of $3$; right column).
The plots from top to bottom are at times of $t = 25, 50, 80 \Myr$.
The plots in the top row, at $t = 25 \Myr$, show the initial spreading due to the shock wave.
After the initial spreading, further spreading depends on local stochastic flow structure and vorticity.
}
\label{Tracers}
\end{figure}
%FFFFFFFFFFFFFFFFFFFFFFFFFFFFFFFFFFFFFFFFFFFFFFFFFFF

To close this section that emphasizes the evolution of dense clumps toward mixing with shocked jet's material,
we follow the evolution of clumps starting with a density contrast of  $\delta = 3$, but having an initial different
cross section radii of $R=1, 2, 3 \kpc$ (runs {{{{Run~S20$\delta$3}}}}, {{{{Run~S20$\delta$3R2}}}}, and {{{{Run~S20$\delta$3R3}}}}, respectively).
The properties of the jet are as in the other runs in this section.
The results of the evolution of each clump are shown in the temperature maps of figure \ref{figure: rhofactor4_size123}.
It is evident that these high density clumps suffer catastrophic cooling, down to the numerical floor temperature of $10^4 \K$.
These cold regions are compressed and cannot be resolved in the figure. They might form optical filaments.
%FFFFFFFFFFFFFFFFFFFFFFFFFFFFFFFFFFFFFFFFFFFFFFFFFFF
\begin{figure}[htb]
\centering
\subfigure{\includegraphics[width=0.32\textwidth]{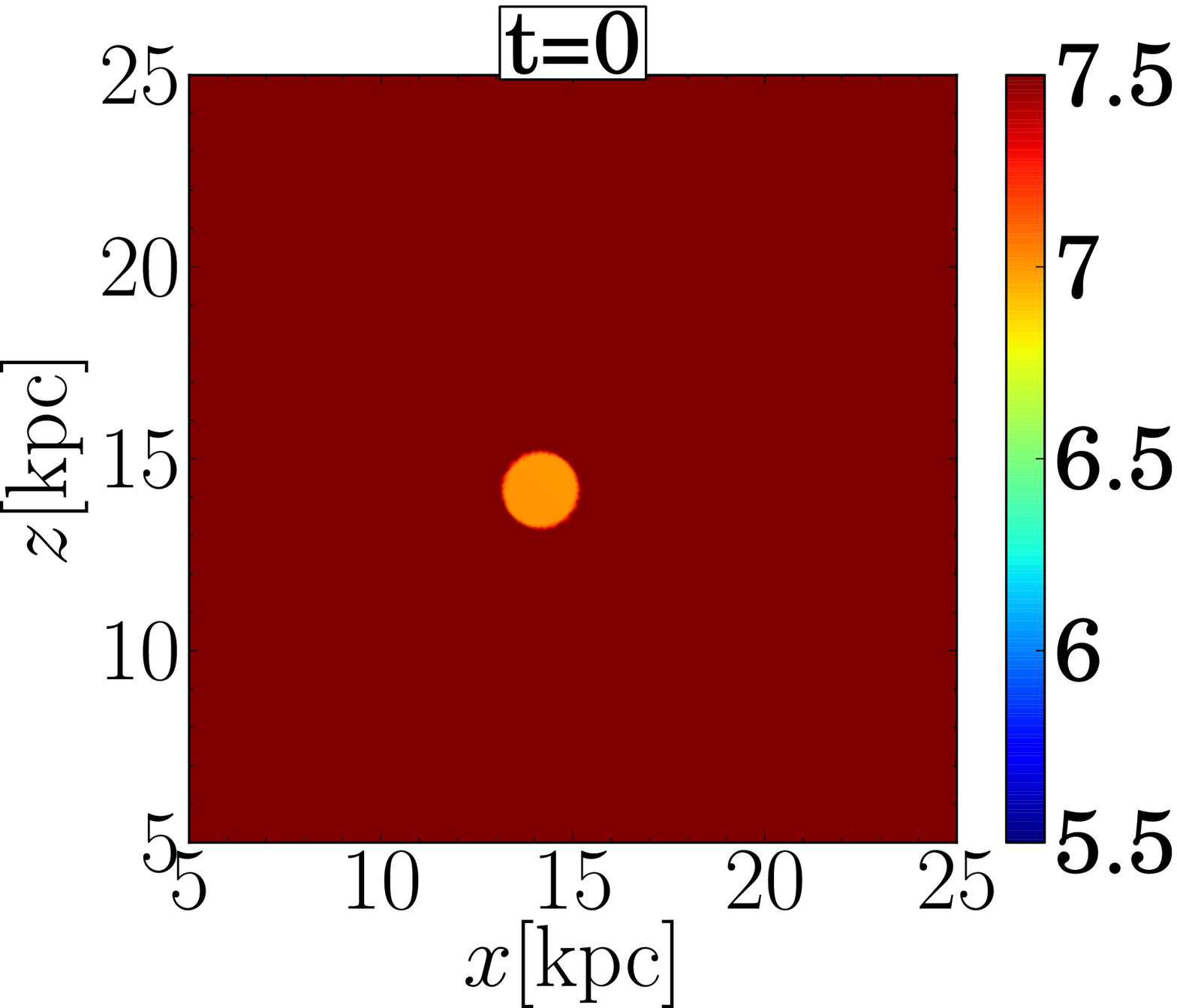}}
\subfigure{\includegraphics[width=0.32\textwidth]{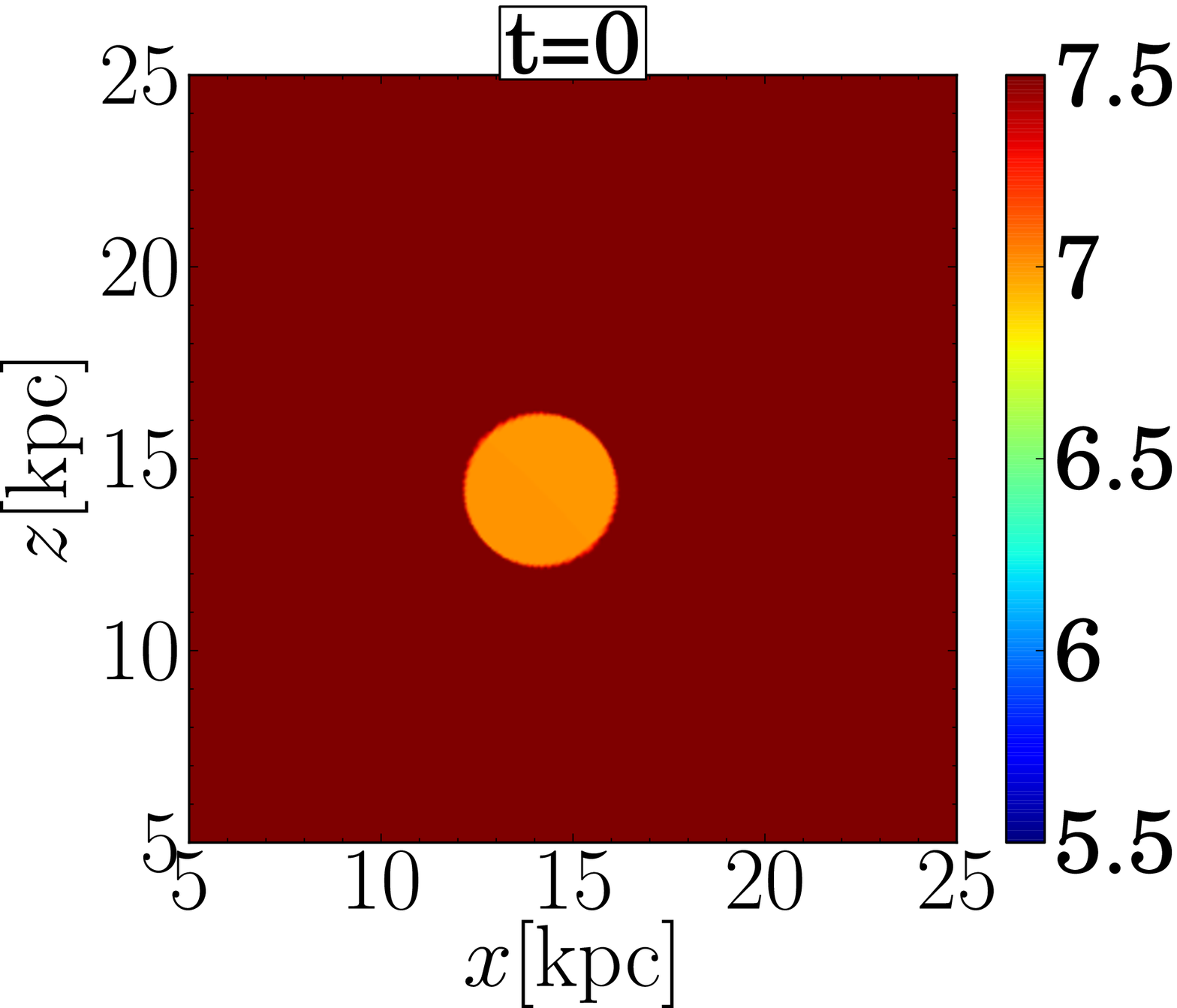}}
\subfigure{\includegraphics[width=0.32\textwidth]{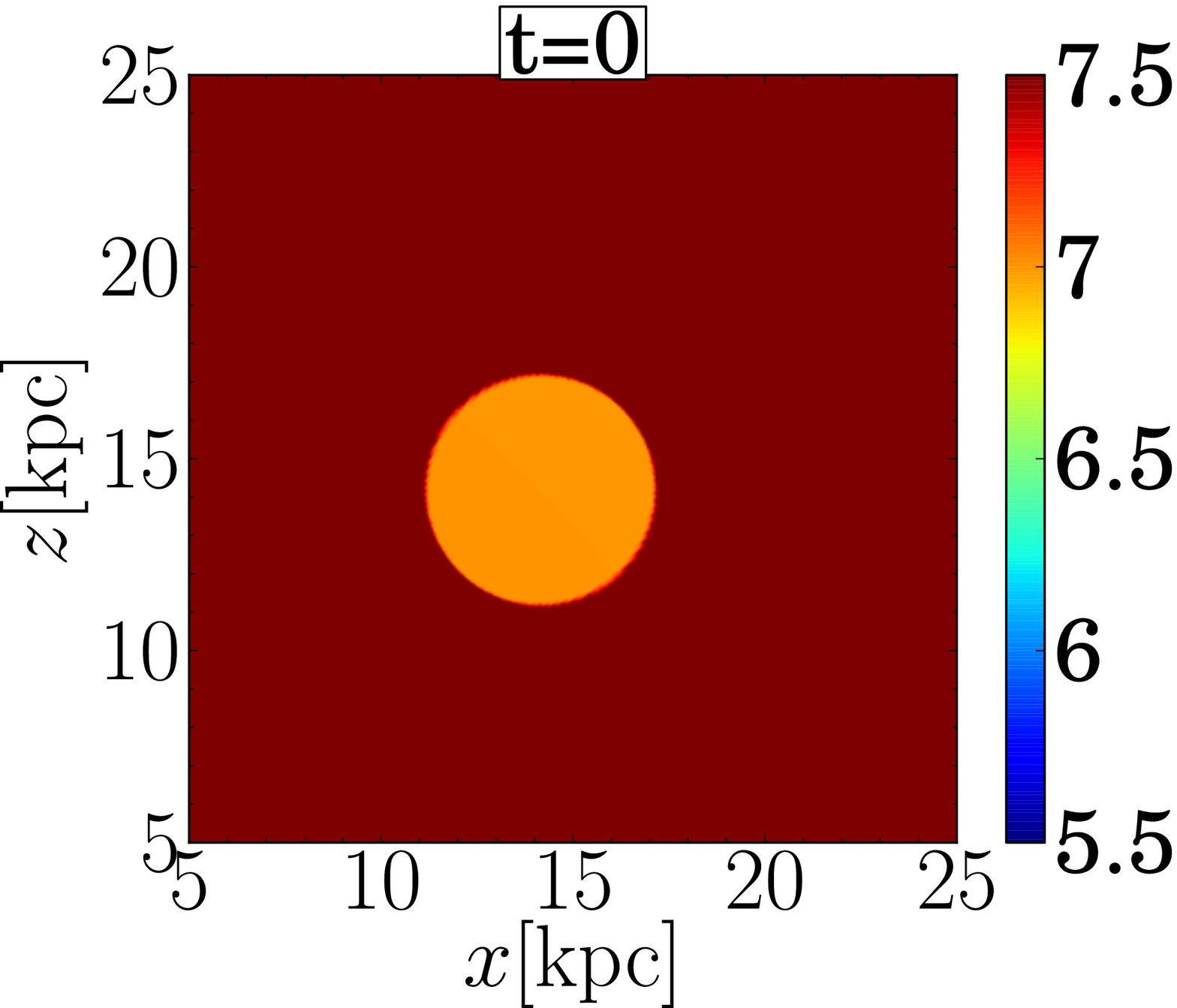}} \\
\subfigure{\includegraphics[width=0.32\textwidth]{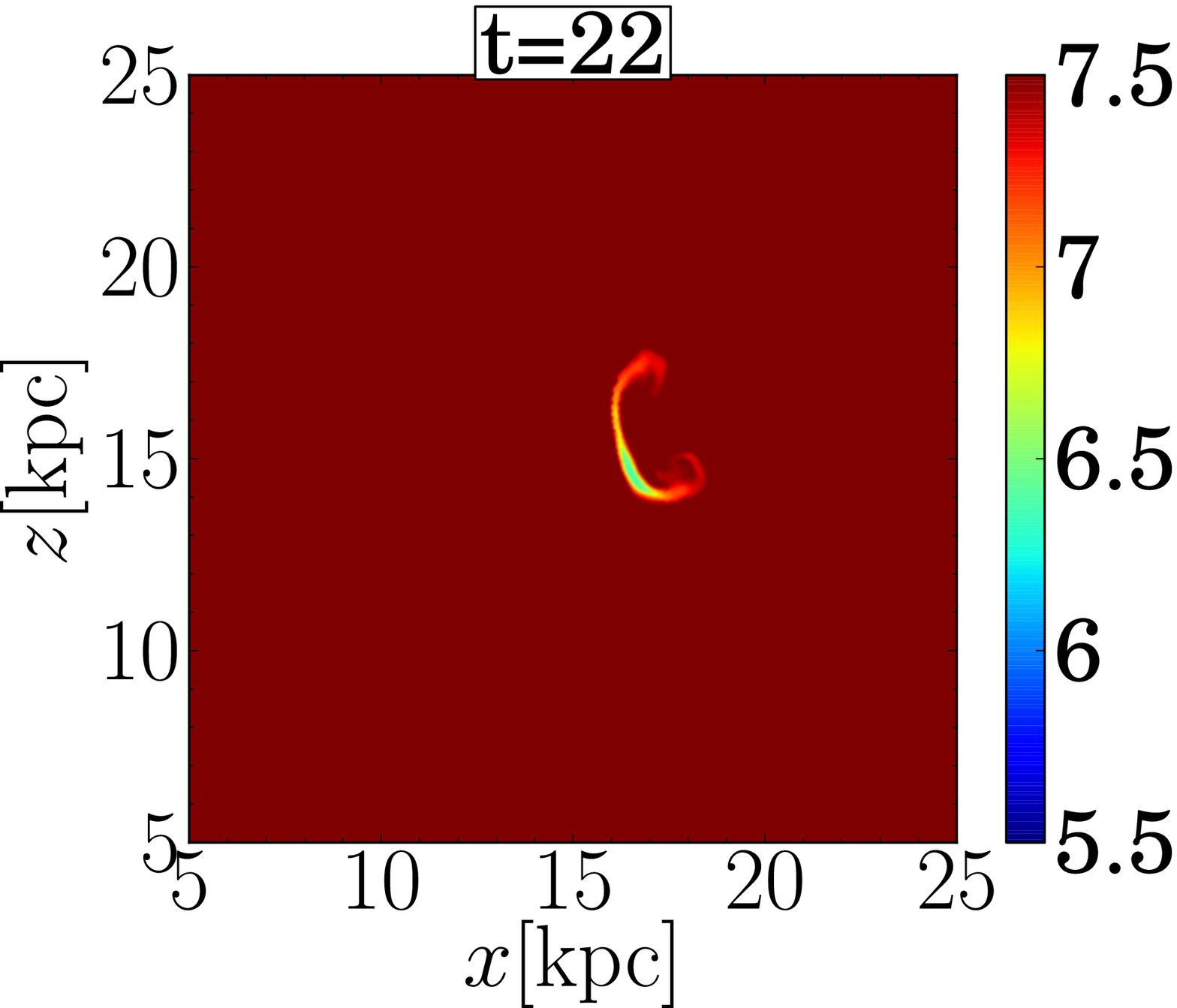}}
\subfigure{\includegraphics[width=0.32\textwidth]{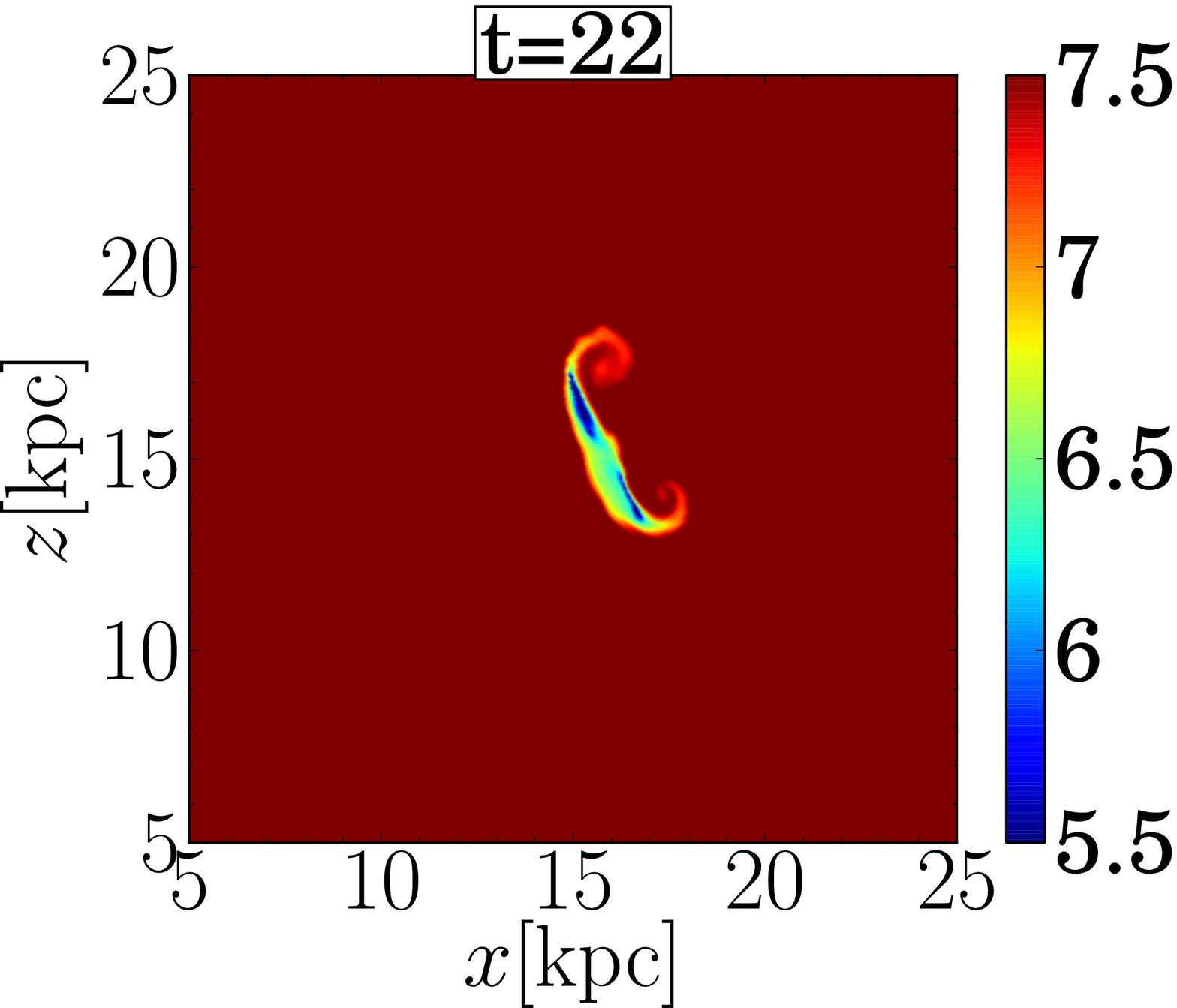}}
\subfigure{\includegraphics[width=0.32\textwidth]{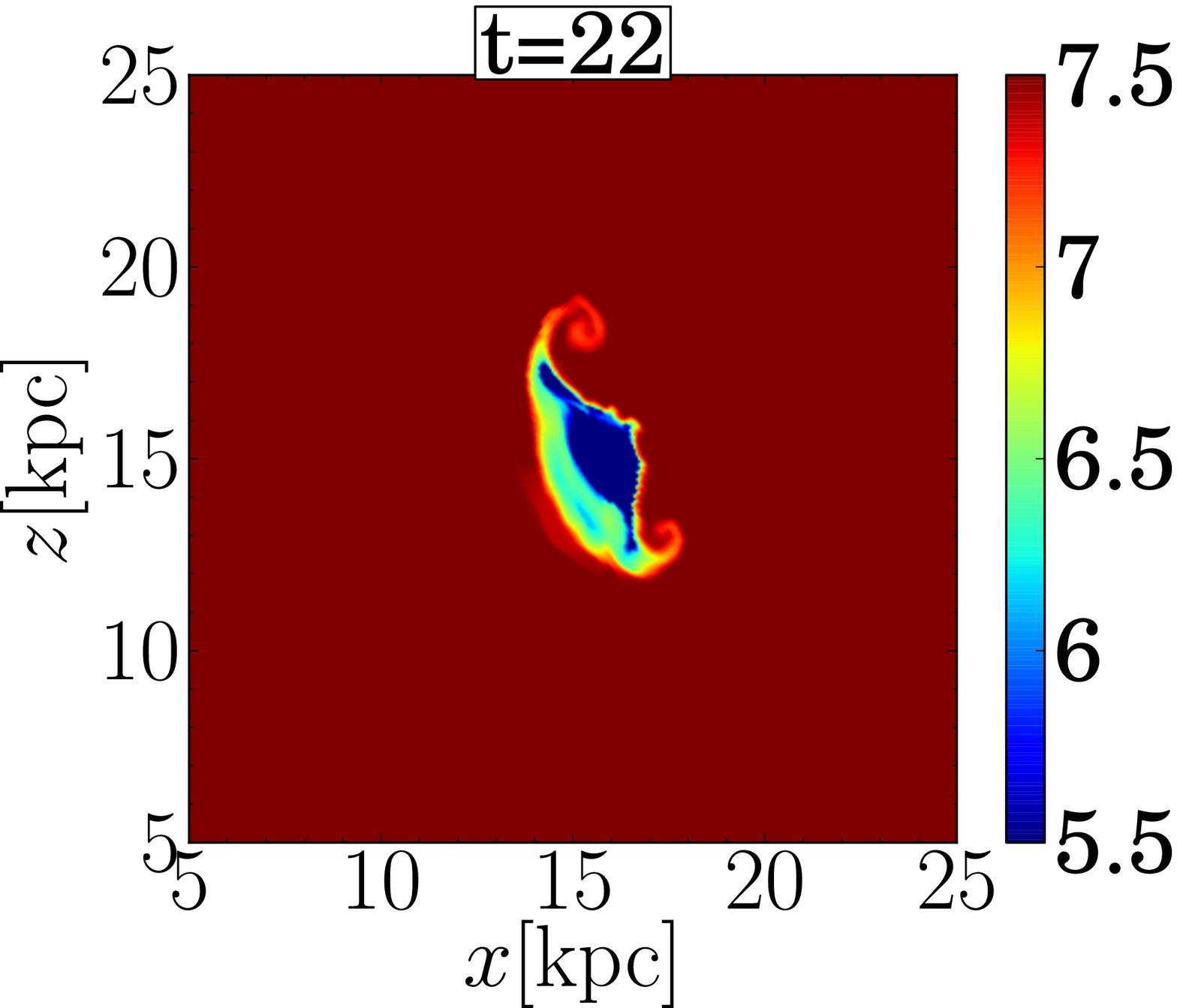}} \\
\subfigure{\includegraphics[width=0.32\textwidth]{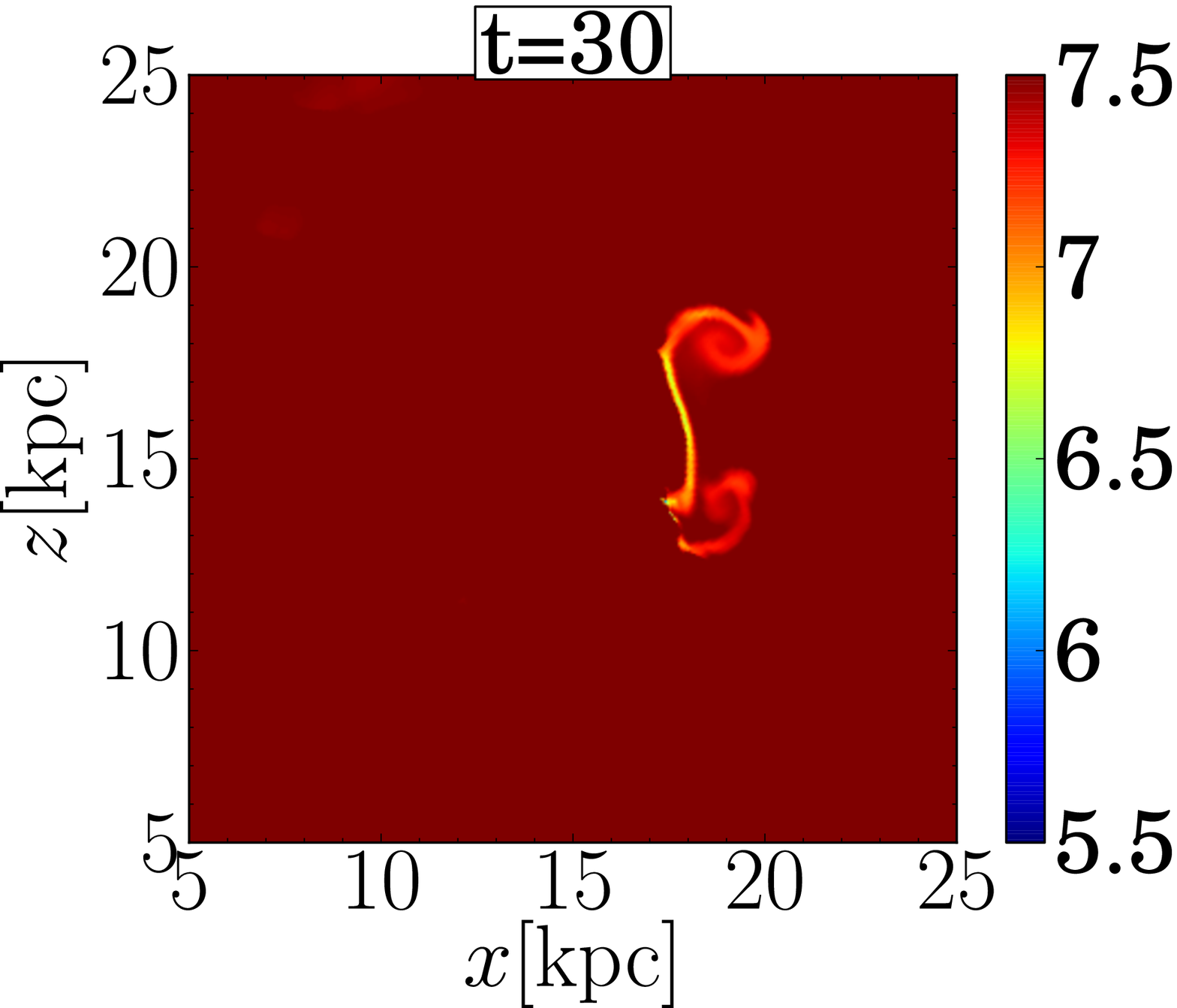}}
\subfigure{\includegraphics[width=0.32\textwidth]{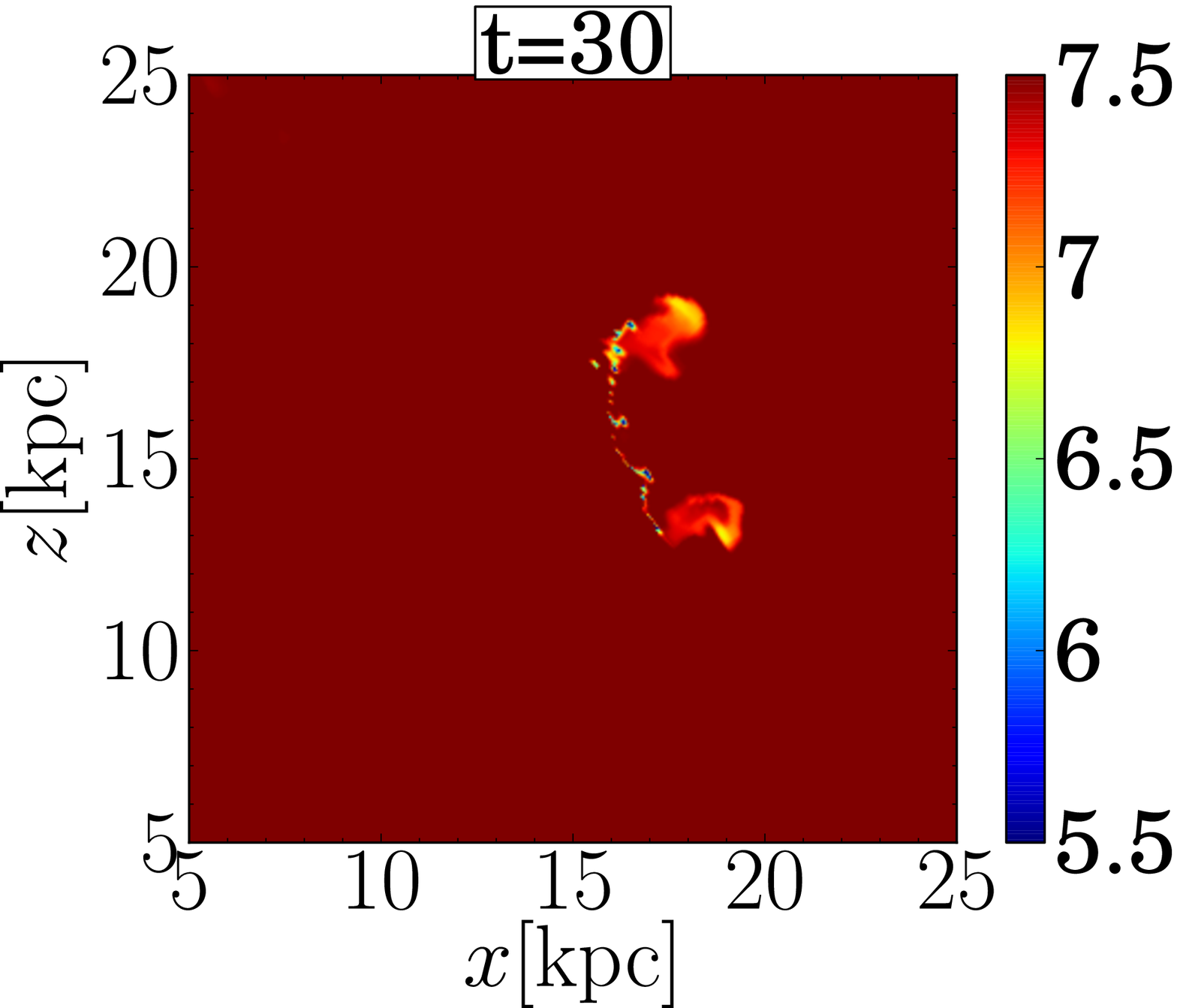}}
\subfigure{\includegraphics[width=0.32\textwidth]{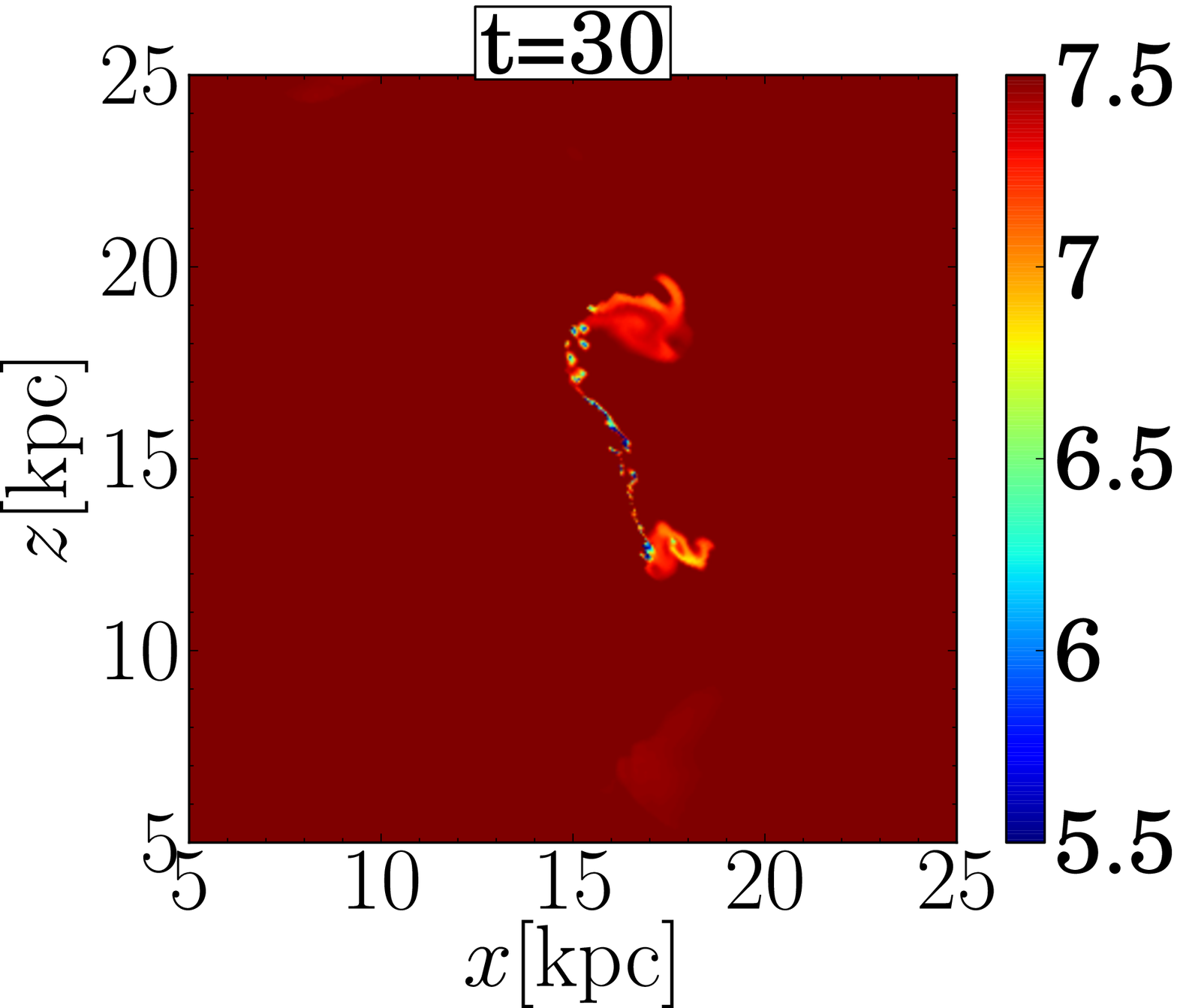}}
\caption{
Temperature maps in log scale and units of $\K$  of three simulations differ in the initial cross section radii of the clumps,
$R = 1$ (left panels), $R=2$ (middle panels), $R=3 \kpc$ (right panels).
All clumps start with a density contrast of $\delta = 3$, a distance of $r = 20 \kpc$, and at an angle of $45^\circ$ with respect to the jet axis.
Plots are shown at times of $t = 0, 22, 30 \Myr$ as indicated.
Note the catastrophic cooling of part of the clump material in the middle row,
and the compression of this gas to very low volume (not resolved) in the bottom row.
The numerical temperature floor is $10^4 \K$.
}
\label{figure: rhofactor4_size123}
\end{figure}
%FFFFFFFFFFFFFFFFFFFFFFFFFFFFFFFFFFFFFFFFFFFFFFFFFFF

% ==========================================================
\subsection{Filamentary structures of clumps}
\label{s-morphology}
% ==========================================================

We here show that the interaction of some clumps with the bubbles and vortices formed by the jets lead to the
formation of dense filaments. These can later cool and form optical filaments.
The morphological study here is brief because of the limitation of the {{{{2D}}}} grid to explore morphology properly.
At $t=0$ we place five clumps as shown in Fig. \ref{figure: confA_rhofactor2}, all having
an initial density contrast of $\delta=1$.
One jet is launched for $20 \Myr$ with properties as in section \ref{s-mixing} and as described in section \ref{s-numerical-setup}.
This simulated case is termed {{{{Run~S20$\delta$1C5}}}}.
%FFFFFFFFFFFFFFFFFFFFFFFFFFFFFFFFFFFFFFFFFFFFFFFFFFF
\begin{figure}[htb]
\centering
\subfigure{\includegraphics[width=0.49\textwidth]{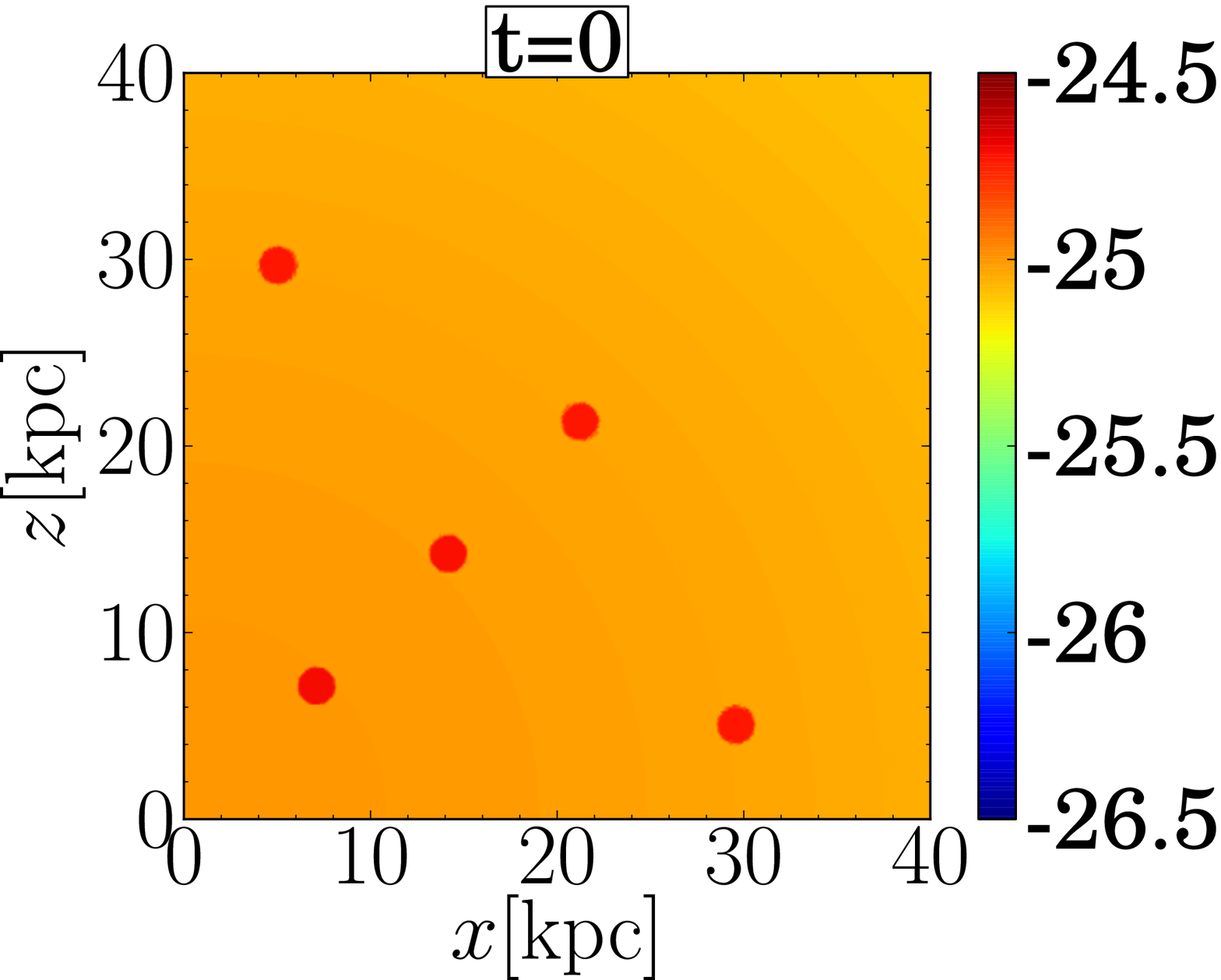}}
\subfigure{\includegraphics[width=0.49\textwidth]{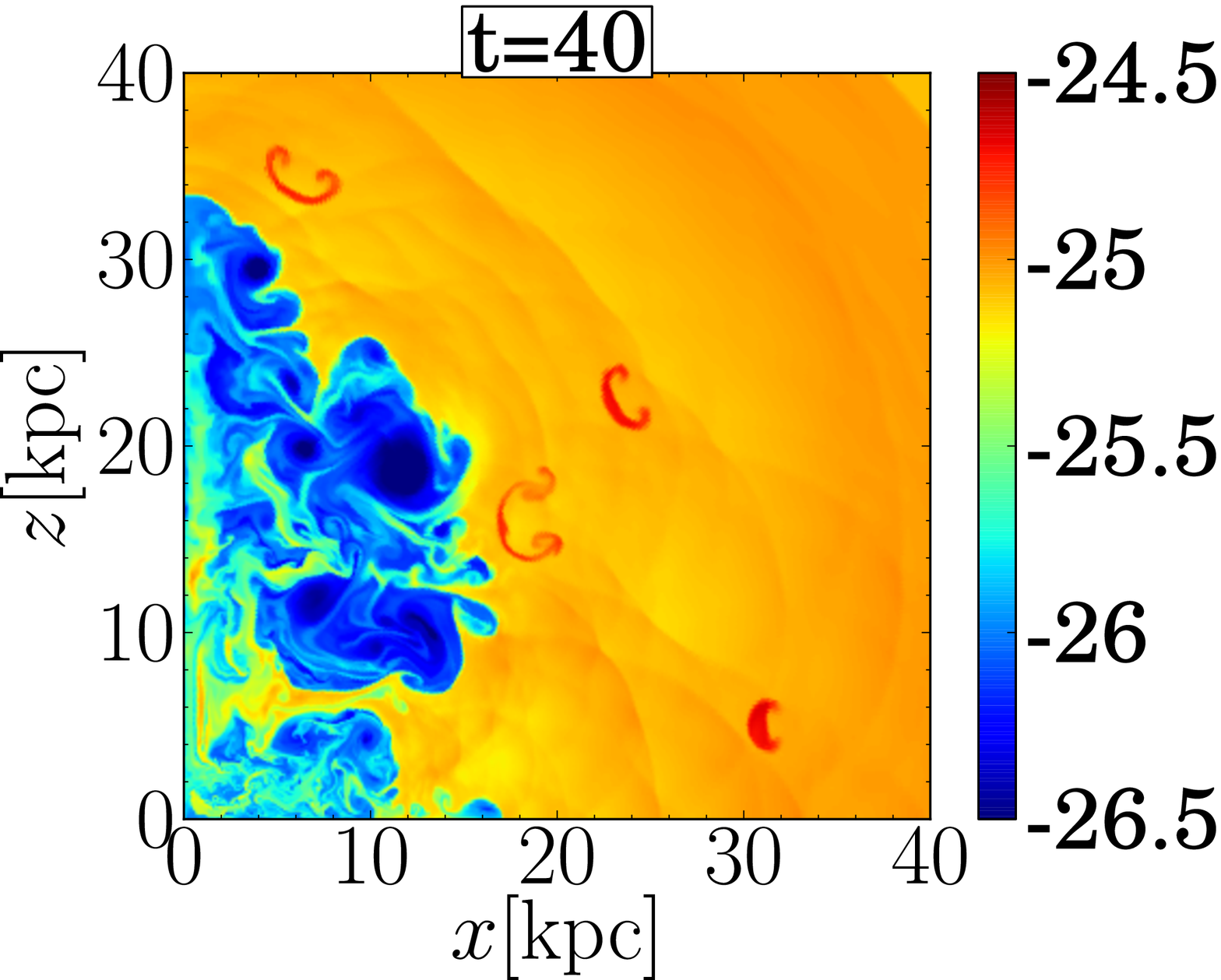}} \\
\subfigure{\includegraphics[width=0.49\textwidth]{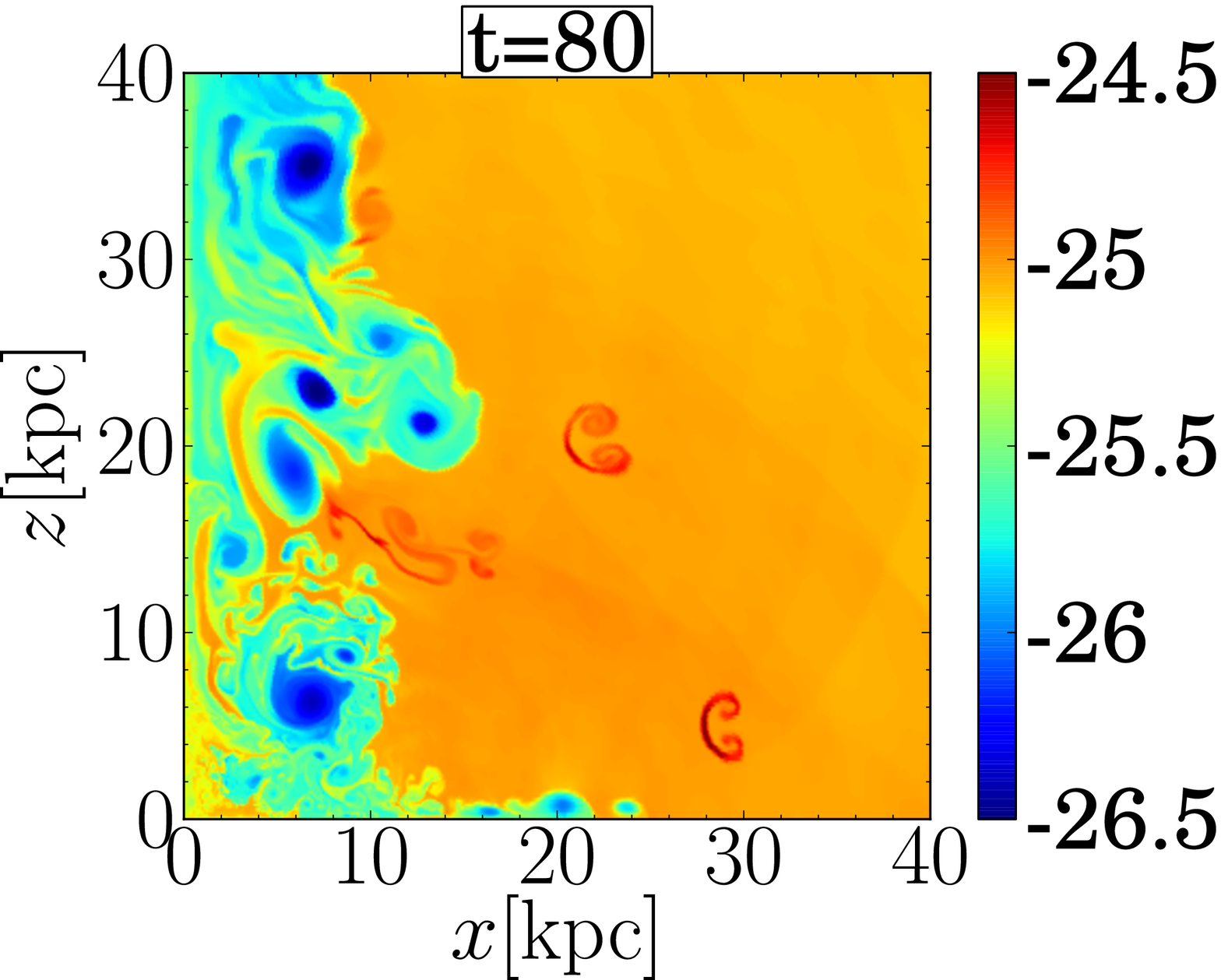}}
\subfigure{\includegraphics[width=0.49\textwidth]{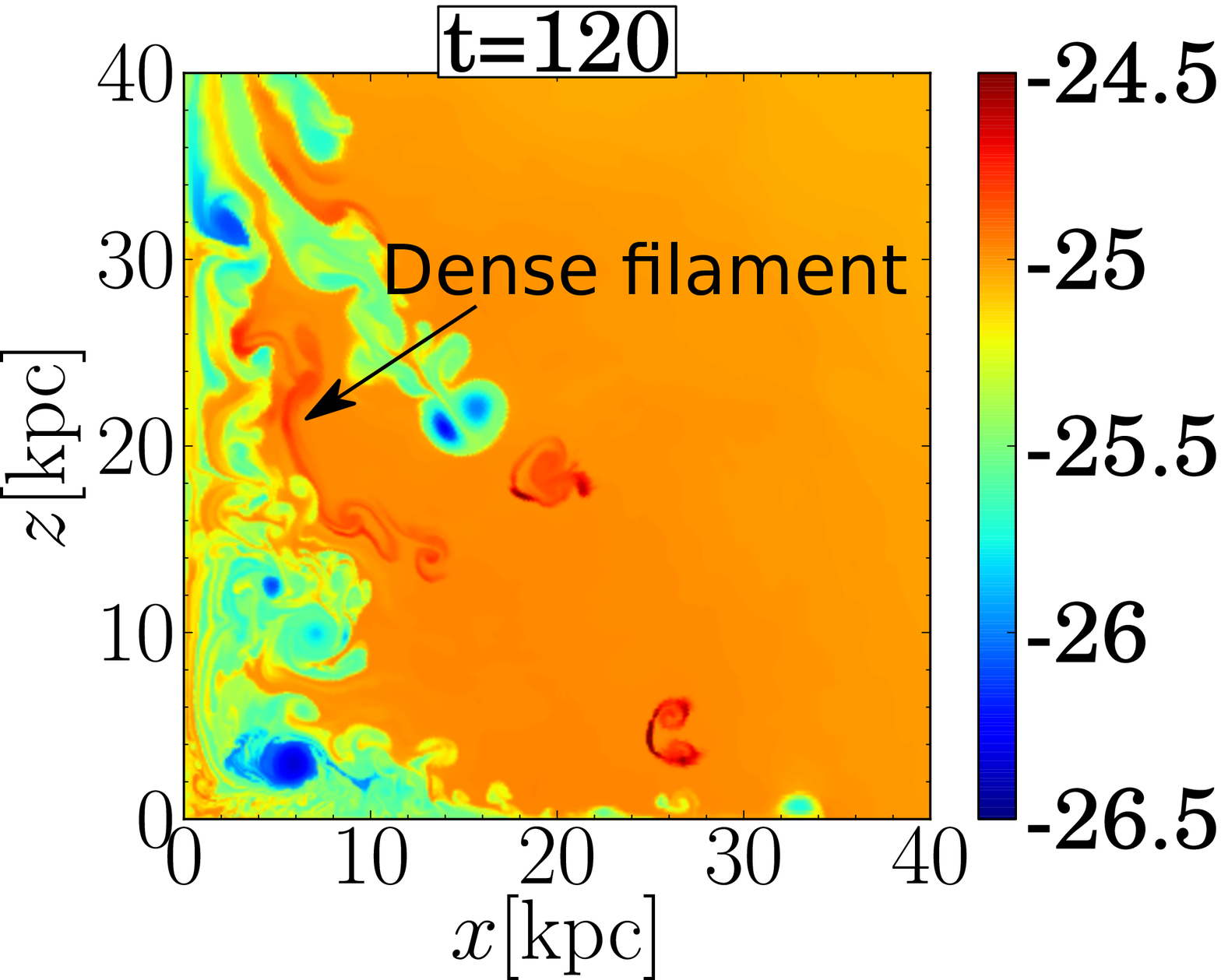}}
\caption{
Five dense clumps with $\delta = 1$ are inserted at various locations with respect to the jet axis in this {{{{Run~S20$\delta$1C5}}}}.
Times are indicated in $\Myr$.
The color code is the density in log scale.
}
\label{figure: confA_rhofactor2}
\end{figure}
%FFFFFFFFFFFFFFFFFFFFFFFFFFFFFFFFFFFFFFFFFFFFFFFFFFF

{{{{Run~S20$\delta$1C5}}}} emphasizes the different evolution of clumps resulting from different locations.
Those that are close to the jet are mixed and heated first, and
can be dragged efficiently outward.
We concentrate on the middle clump that evolves to a filamentary structure,
and show its `tracer' map in Fig.~\ref{figure: confA_rhofactor2_tracer}. The tracer indicates the concentration of the initial clumps' gas at each point.
A filamentary morphology is clearly seen at $t=120 \Myr$.
%FFFFFFFFFFFFFFFFFFFFFFFFFFFFFFFFFFFFFFFFFFFFFFFFFFF
\begin{figure}[htb]
\centering
\subfigure{\includegraphics[width=0.49\textwidth]{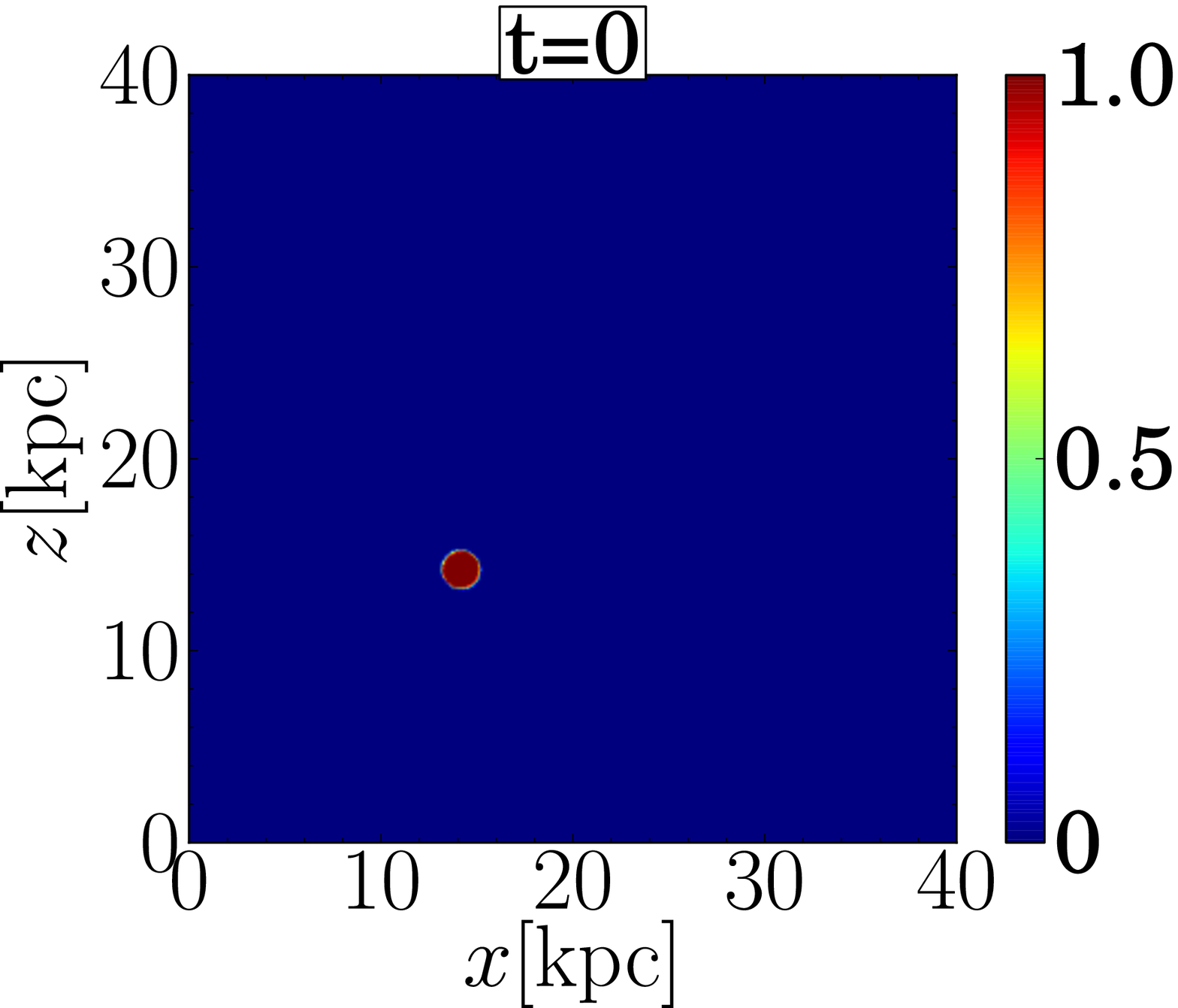}}
\subfigure{\includegraphics[width=0.49\textwidth]{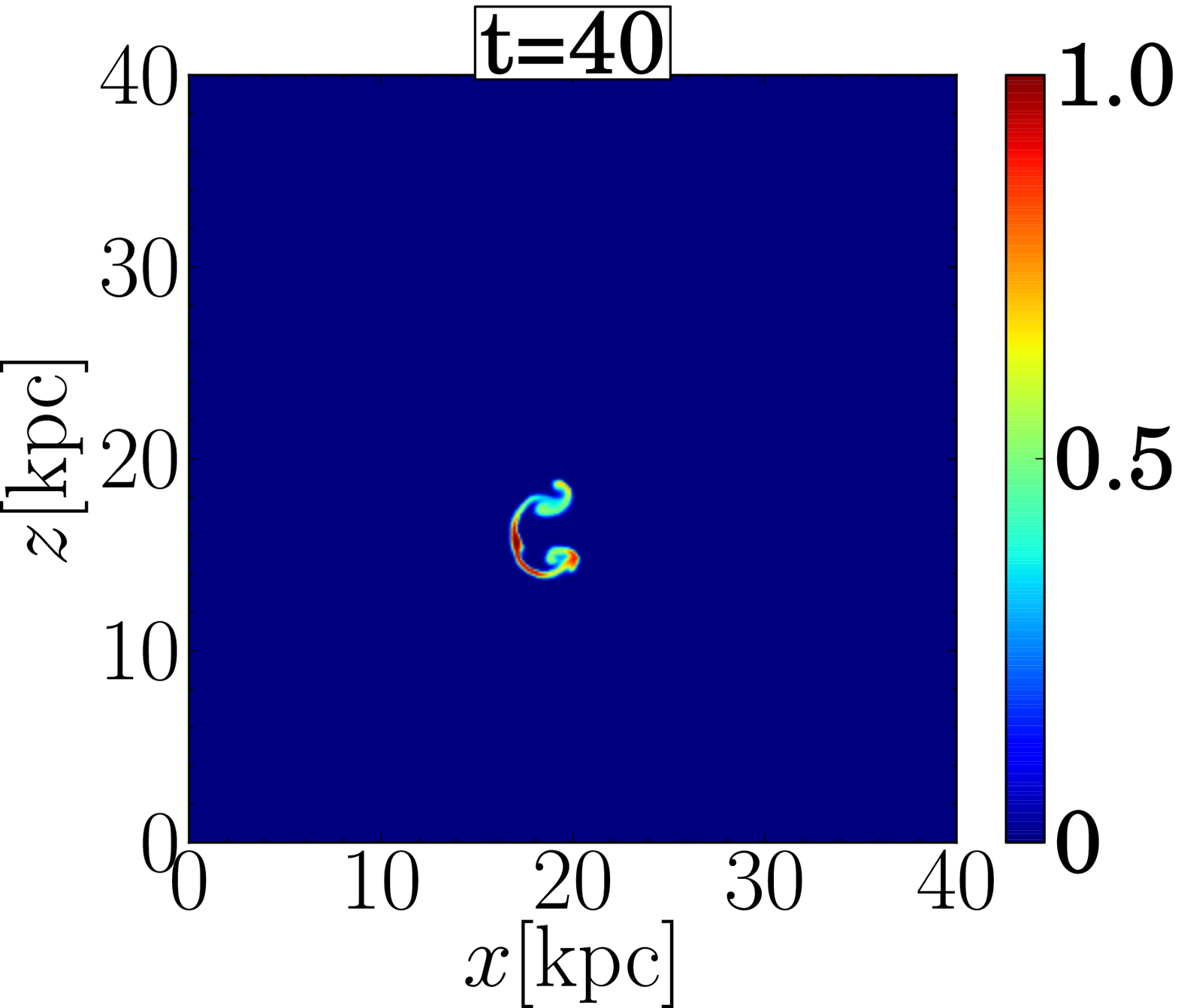}} \\
\subfigure{\includegraphics[width=0.49\textwidth]{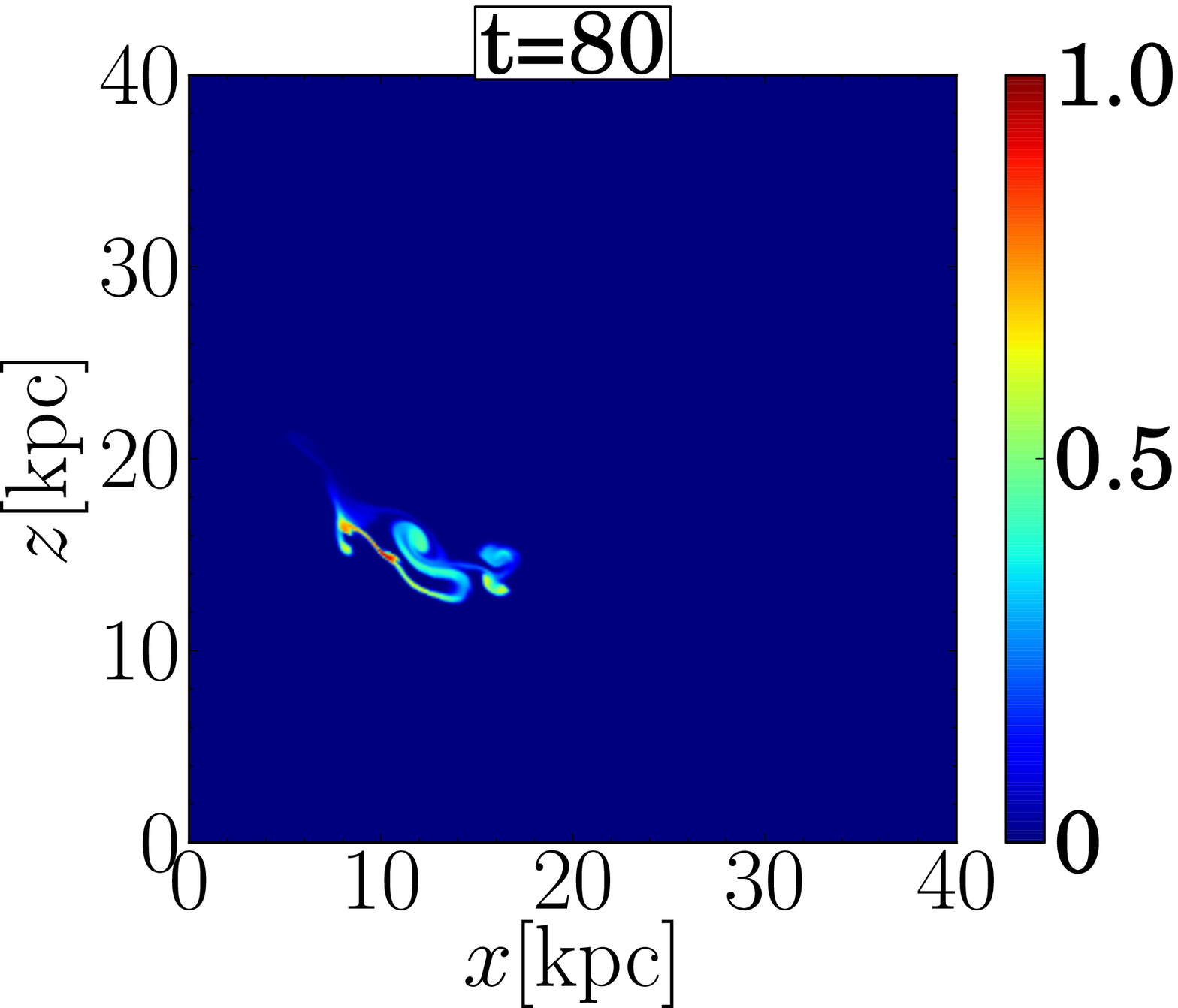}}
\subfigure{\includegraphics[width=0.49\textwidth]{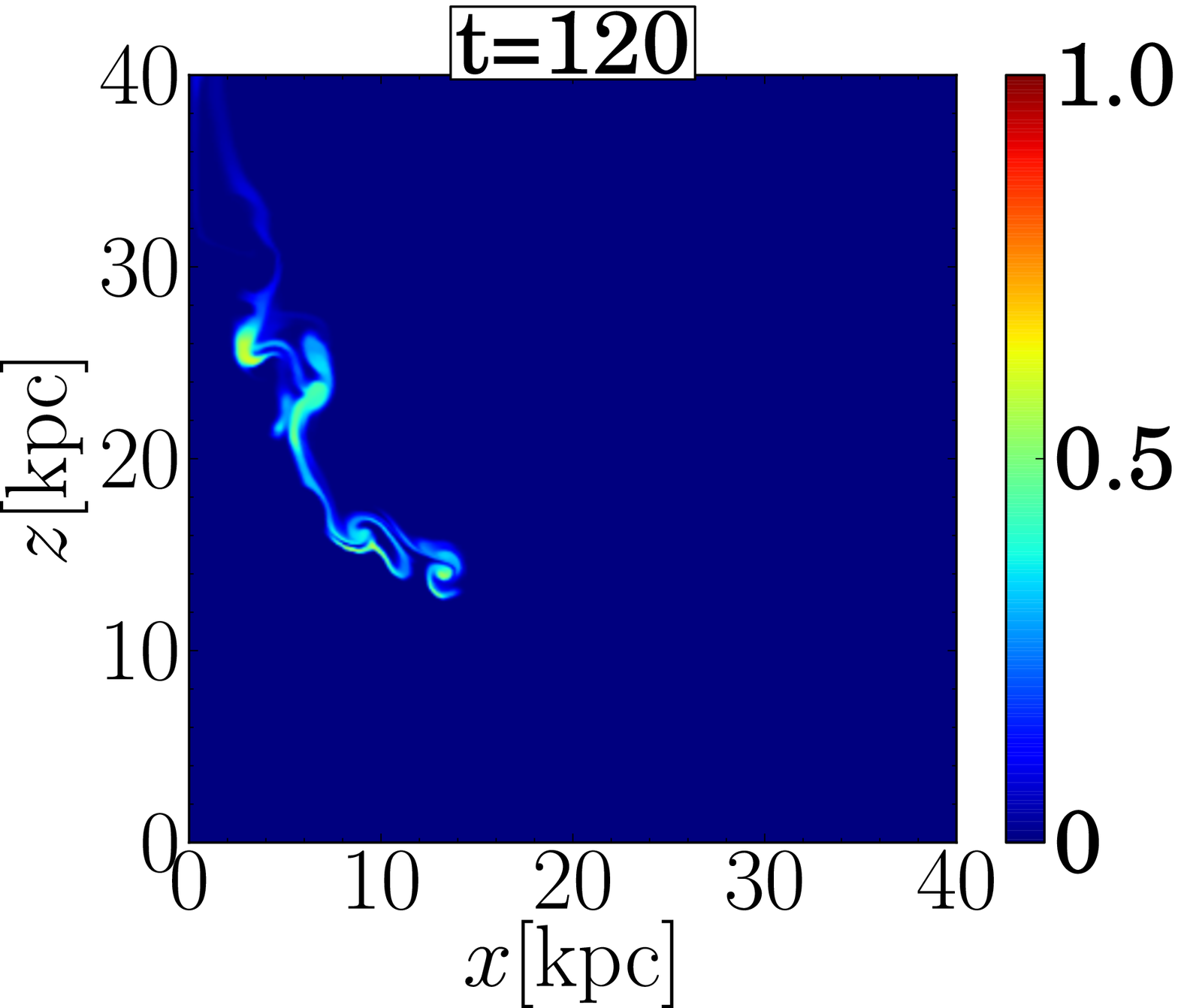}}
\caption{
The tracer of the middle clump, that starts at a distance of $r = 20 \kpc$ at $45^\circ$ with respect to the jet axis,
of {{{{Run~S20$\delta$1C5}}}} shown in Fig. \ref{figure: confA_rhofactor2}.
Color coding is the fraction of the initial clump's material at each point (the tracer).
Times are indicated in $\Myr$.
}
\label{figure: confA_rhofactor2_tracer}
\end{figure}
%FFFFFFFFFFFFFFFFFFFFFFFFFFFFFFFFFFFFFFFFFFFFFFFFFFF

% =======================================
\section{MULTIPLE JET-LAUNCHING EPISODES}
\label{multiepisodes}
% =======================================
% =======================================
\subsection{Flow structure}
\label{multiICM}
% =======================================

We follow the evolution of three dense clumps having a density contrast of $\delta  = 0.3$
through 25 jets launching episodes for a total time of $500 \Myr$. Each episode was active for $10 \Myr$ and followed by
a quiescence period of $10 \Myr$. We term it {{{{Run~M20$\delta$0.3}}}}, and its global flow structures were presented in Figs.~\ref{figure: 3clumps_r45_rhofactor1.3_t50} and \ref{figure: 3clumps_r45_rhofactor1.3_t305}.
We place the three clumps at azimuthal angles, measured from the $z$ axis, of  $\theta = 36.9^\circ, 45^\circ, 53.1^\circ$,
and all at an equal distance of $r = 45 \kpc$ from the source of the jet (center).
All have cross section radius of $1 \kpc$. Again, in our {{{{2D}}}} grid the initial clumps are torii.
The evolution time of $500 \Myr$ is chosen for two reasons:
(a) To examine many shocks running through the clumps and the ICM, and
(b) this is about half the initial radiative cooling time of the ICM at the initial location of the clumps.
Hence we can study the heating by many shocks.
The temperature and clumps' tracers of this run are presented in Figs. \ref{RunM20D0.3Temp} and \ref{RunM20D0.3Trace}, respectively.
%FFFFFFFFFFFFFFFFFFFFFFFFFFFFFFFFFFFFFFFFFFFFFFFFFFF
\begin{figure}[htb]
\centering
\subfigure{\includegraphics[width=0.32\textwidth]{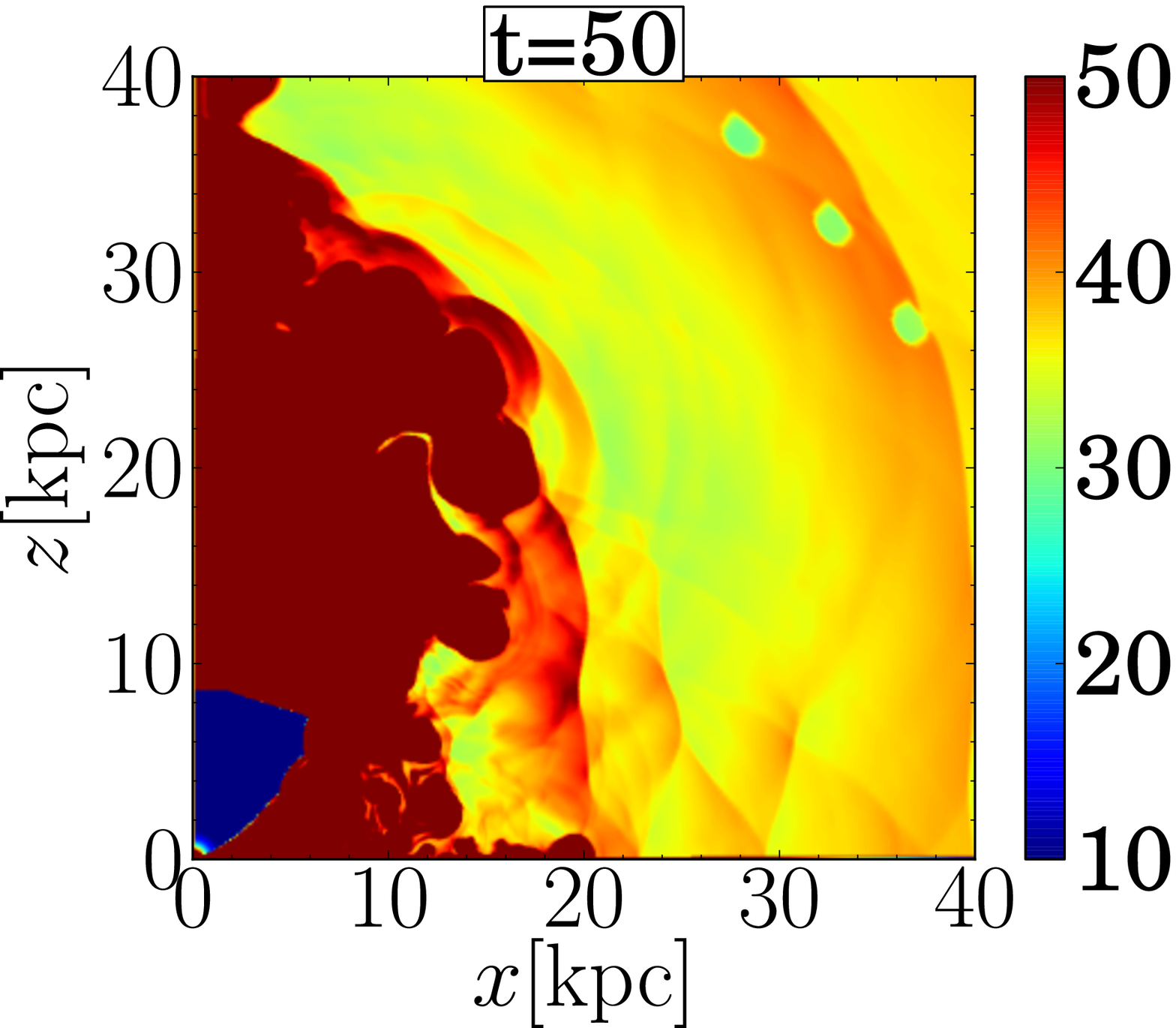}}
\subfigure{\includegraphics[width=0.32\textwidth]{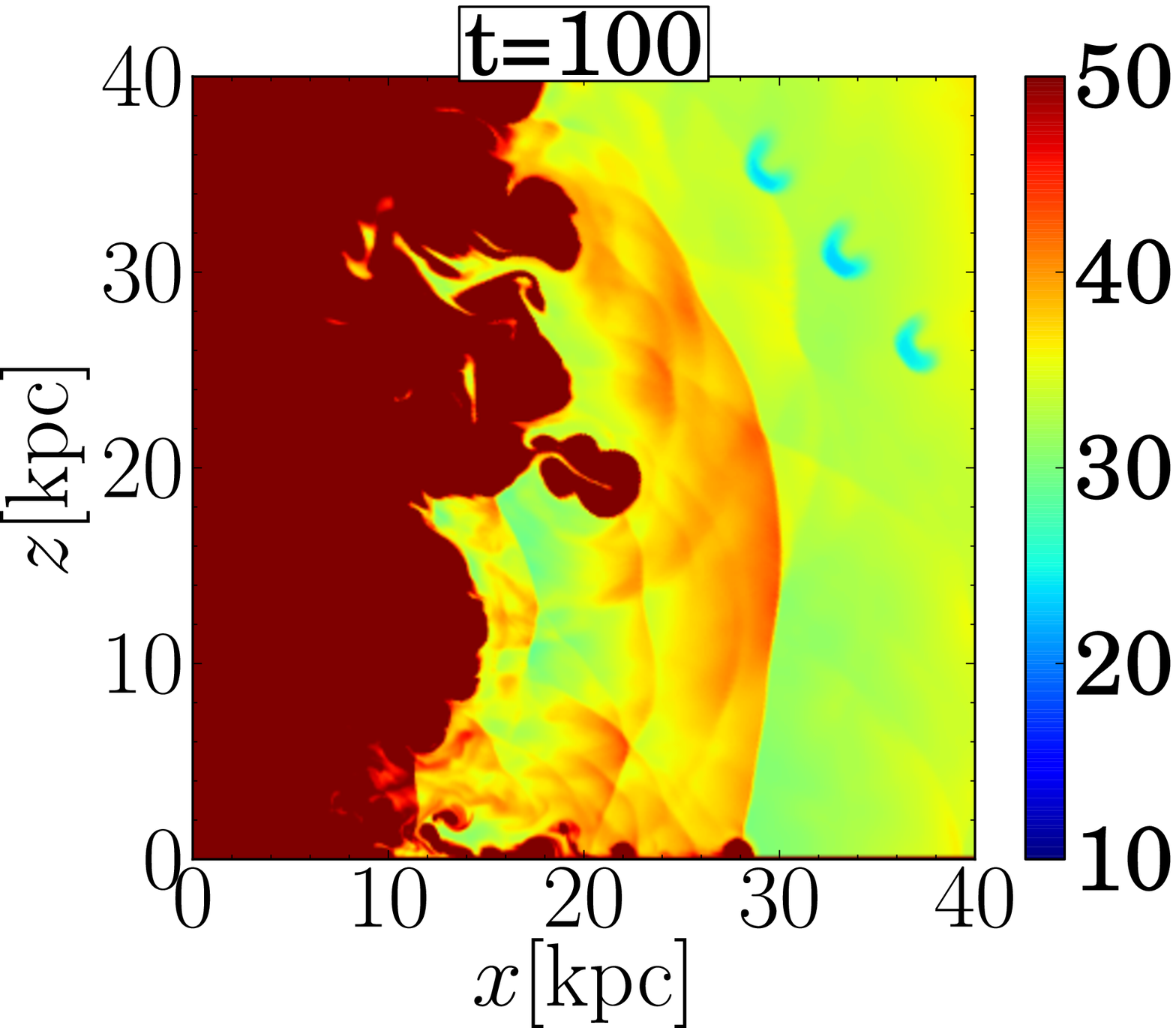}}
\subfigure{\includegraphics[width=0.32\textwidth]{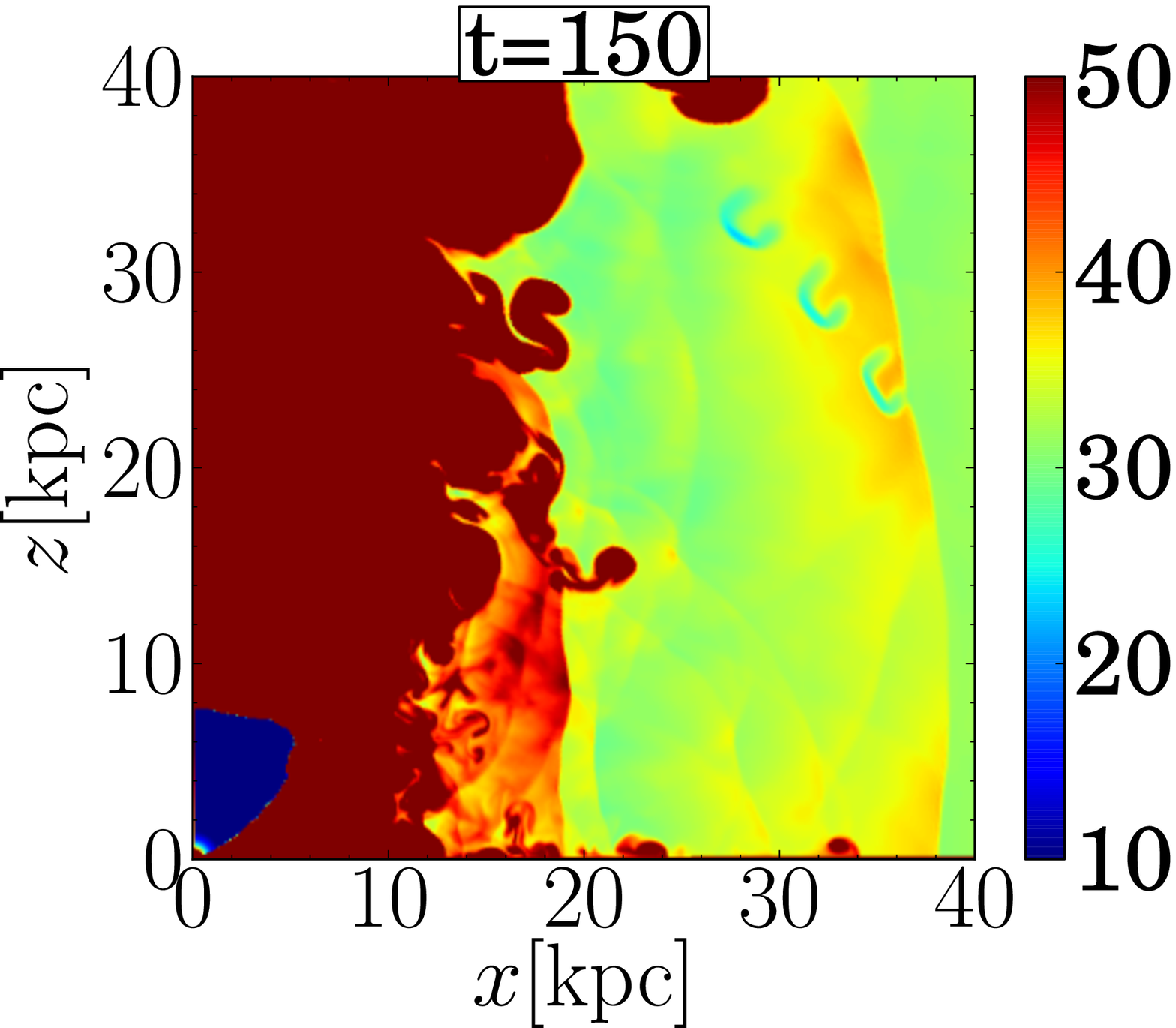}} \\
\subfigure{\includegraphics[width=0.32\textwidth]{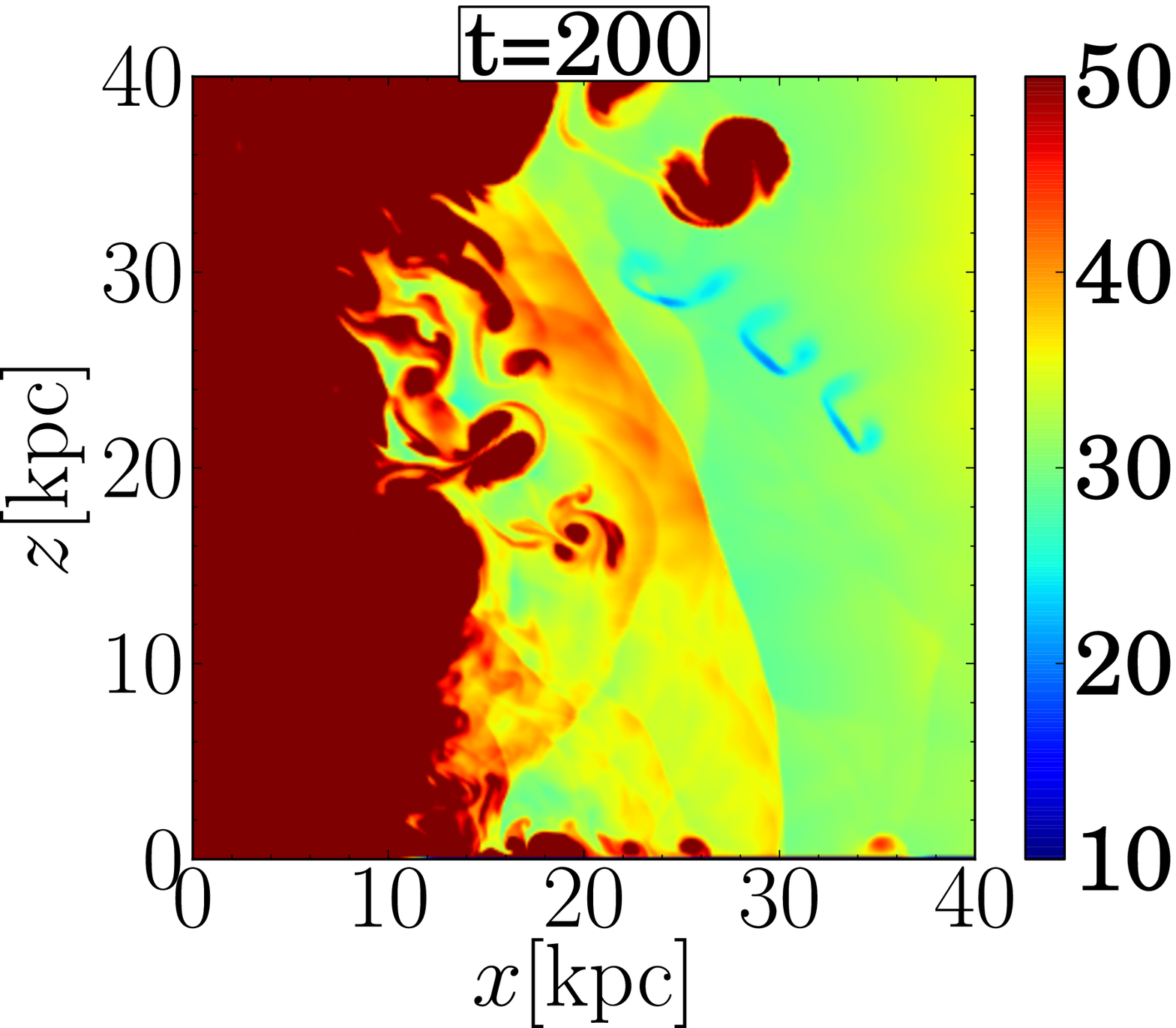}}
\subfigure{\includegraphics[width=0.32\textwidth]{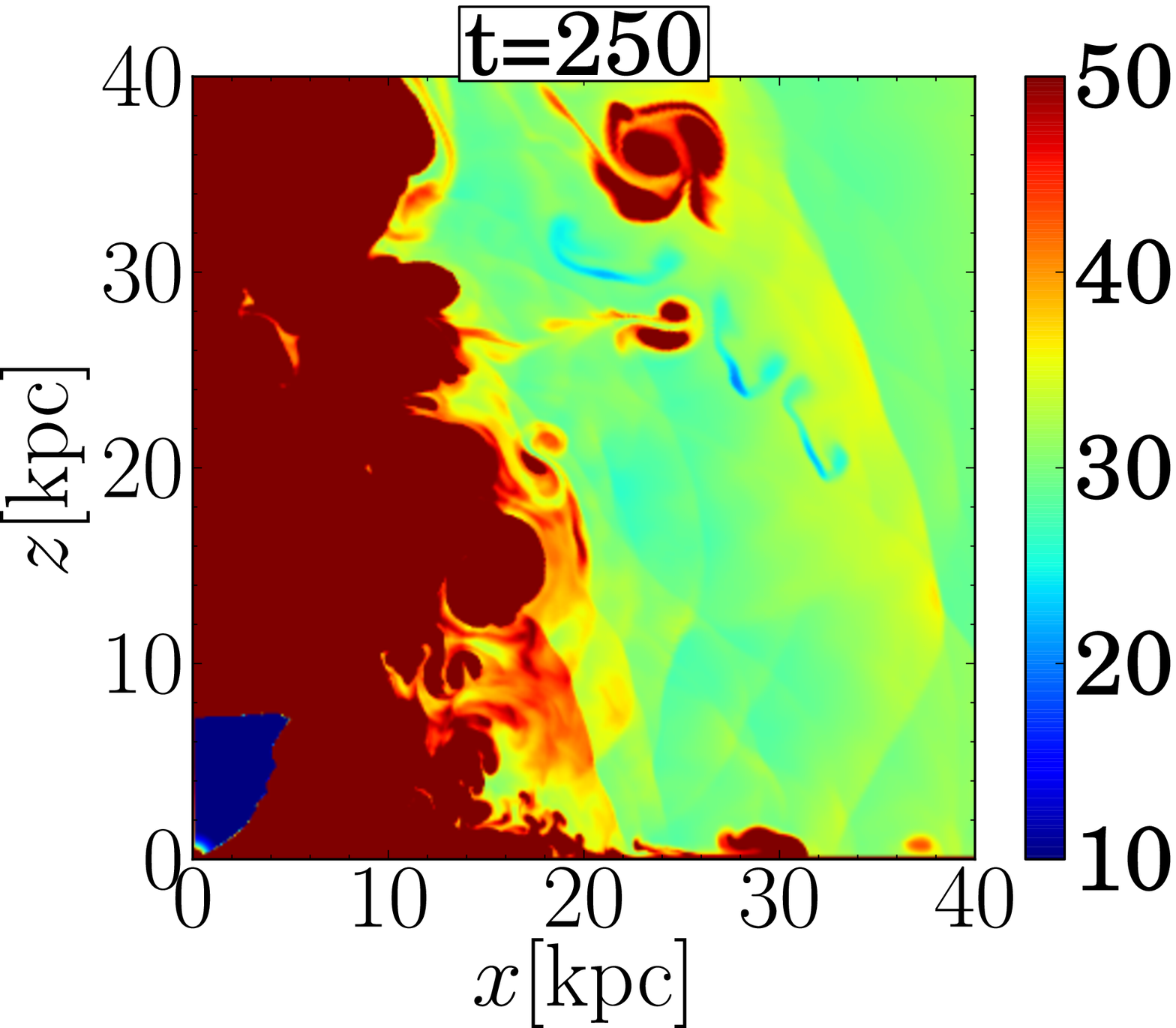}}
\subfigure{\includegraphics[width=0.32\textwidth]{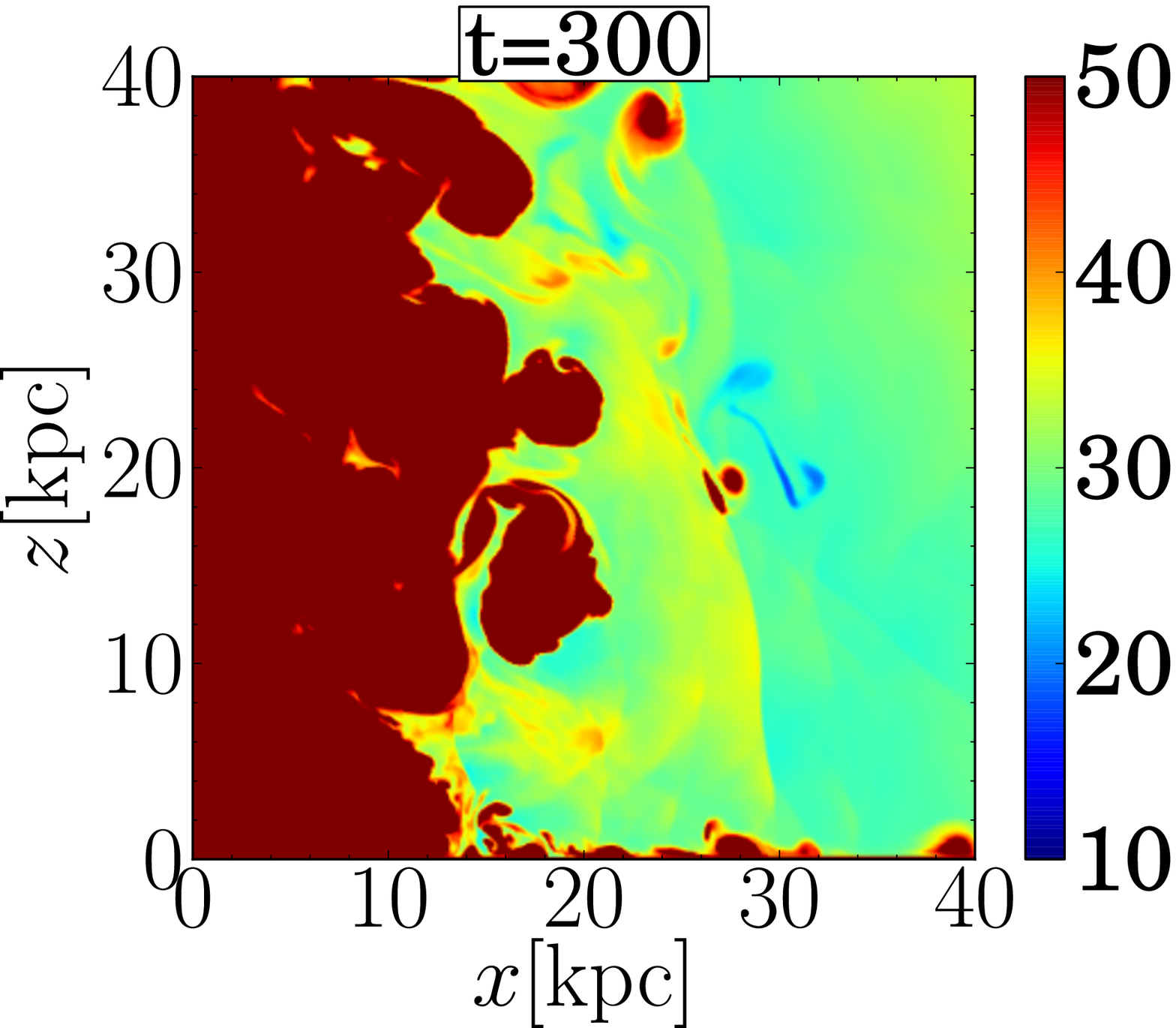}} \\
\subfigure{\includegraphics[width=0.32\textwidth]{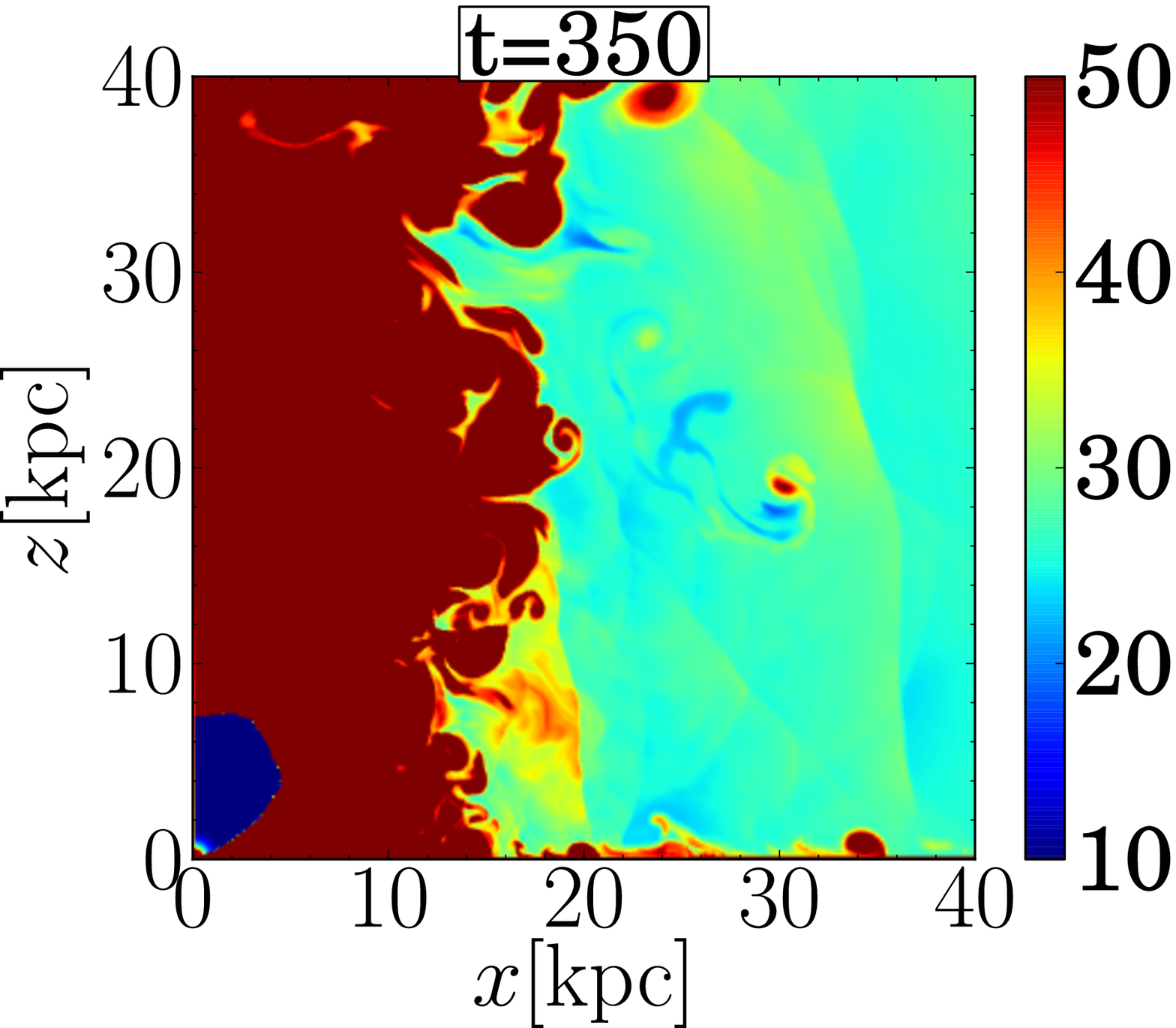}}
\subfigure{\includegraphics[width=0.32\textwidth]{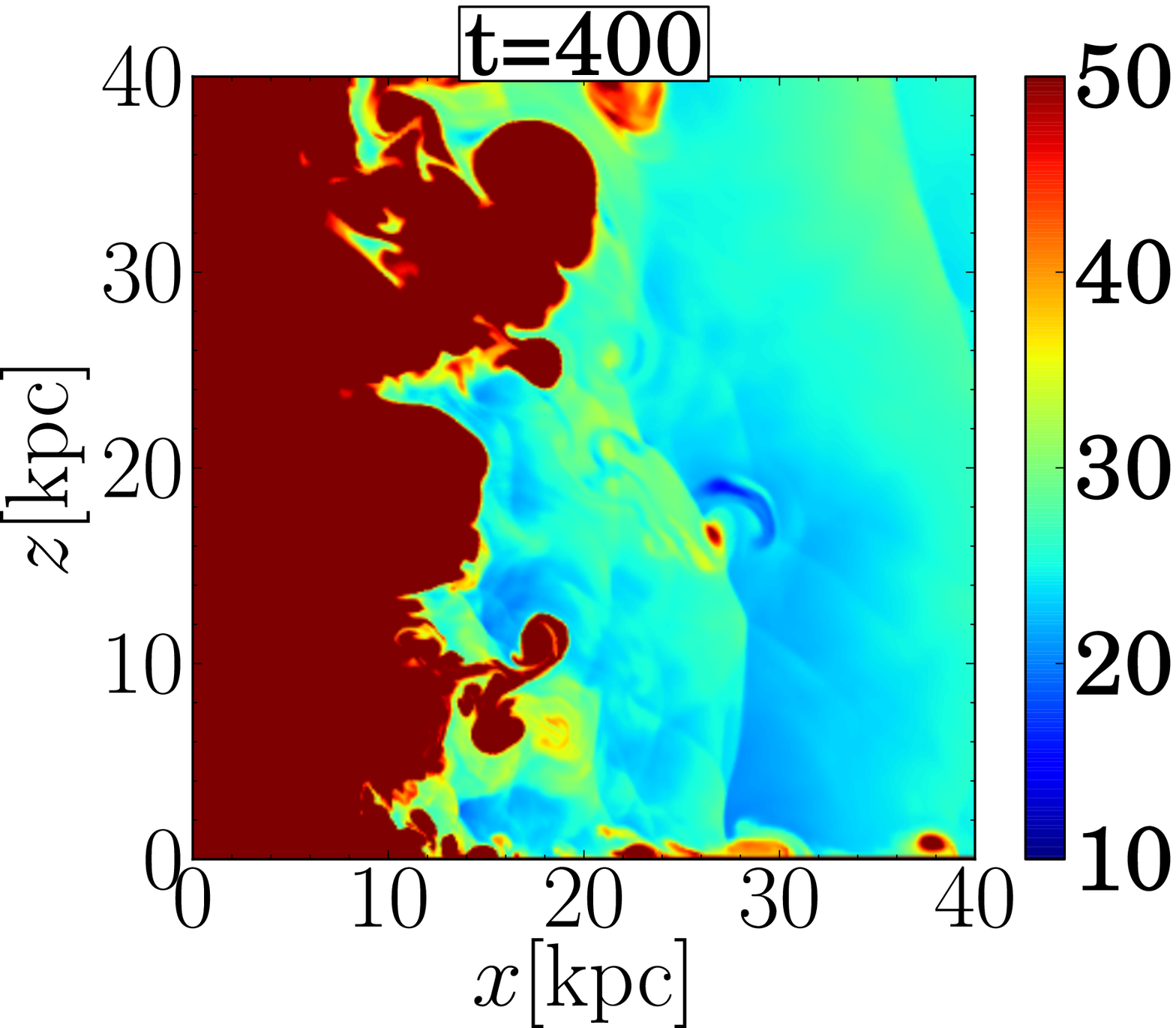}}
\subfigure{\includegraphics[width=0.32\textwidth]{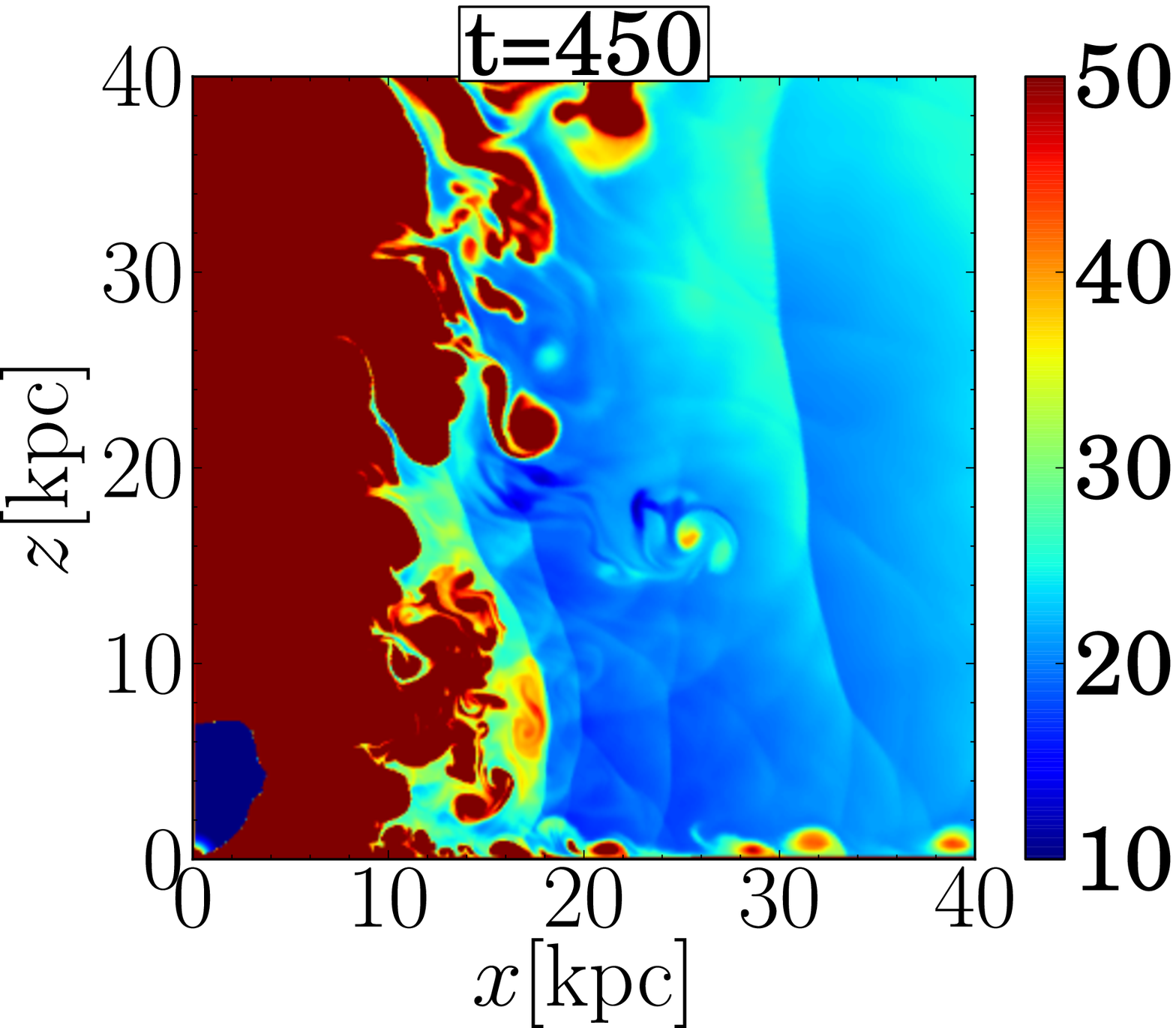}}
\caption{
The evolution of three dense clumps with a density contrast of $\delta = 0.3$ that start at a distance of $r = 45 \kpc$ from the jet's source
({{{{Run~M20$\delta$0.3}}}}; see also Figs.~\ref{figure: 3clumps_r45_rhofactor1.3_t50} and \ref{figure: 3clumps_r45_rhofactor1.3_t305}).
Times are indicated in $\Myr$.
The color coding is the temperature in millions of degrees $\K$ and in a linear scale.
Shocks propagating through the ICM are clearly seen at different times, but they do not manage to prevent catastrophic cooling
of the ICM near the equatorial plane and of the clumps that do not suffer mixing, as seen by the light-blue color in the last 3 panels.
}
\label{RunM20D0.3Temp}
\end{figure}
%FFFFFFFFFFFFFFFFFFFFFFFFFFFFFFFFFFFFFFFFFFFFFFFFFFF
%FFFFFFFFFFFFFFFFFFFFFFFFFFFFFFFFFFFFFFFFFFFFFFFFFFF
\begin{figure}[htb]
\centering
\subfigure{\includegraphics[width=0.32\textwidth]{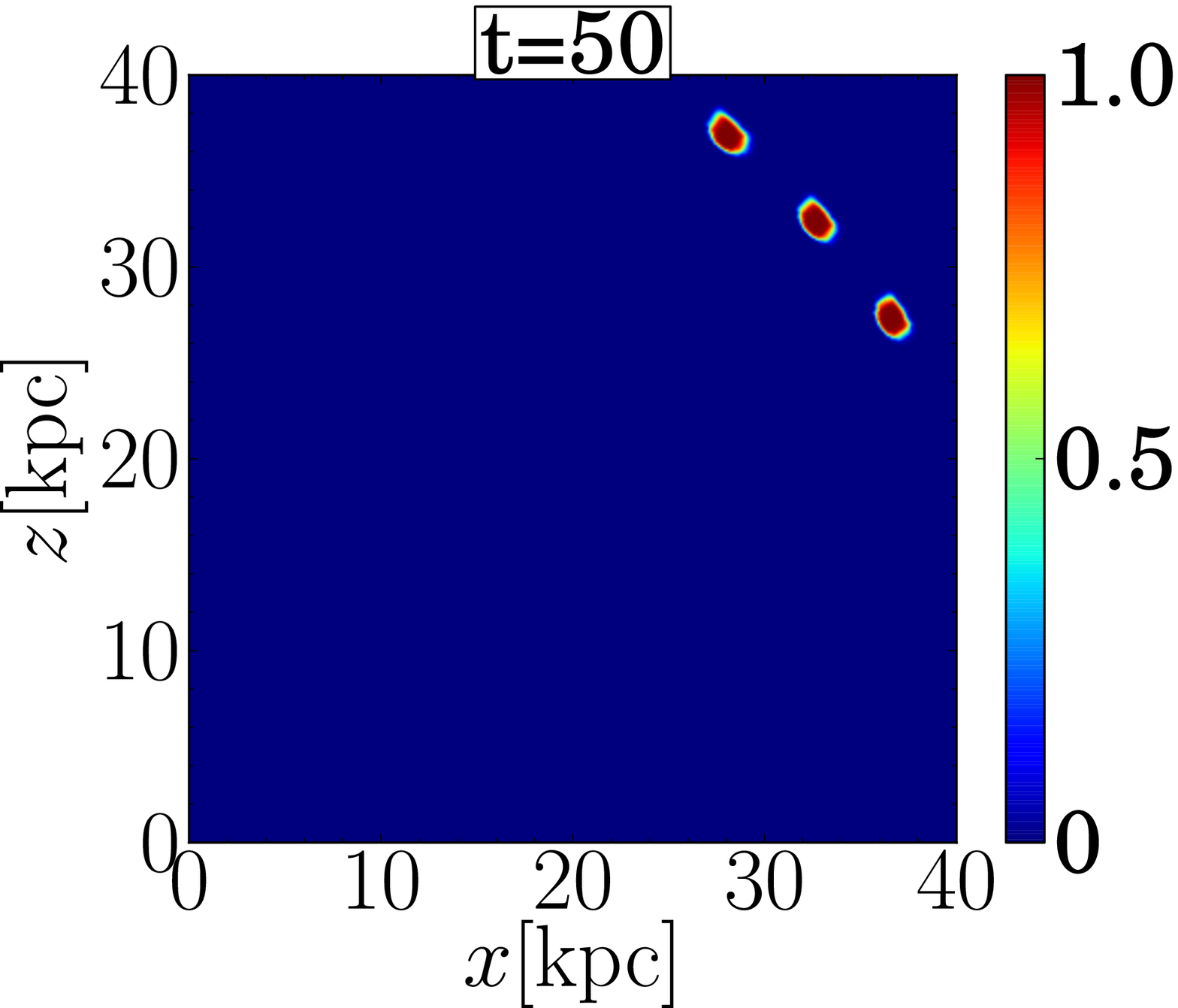}}
\subfigure{\includegraphics[width=0.32\textwidth]{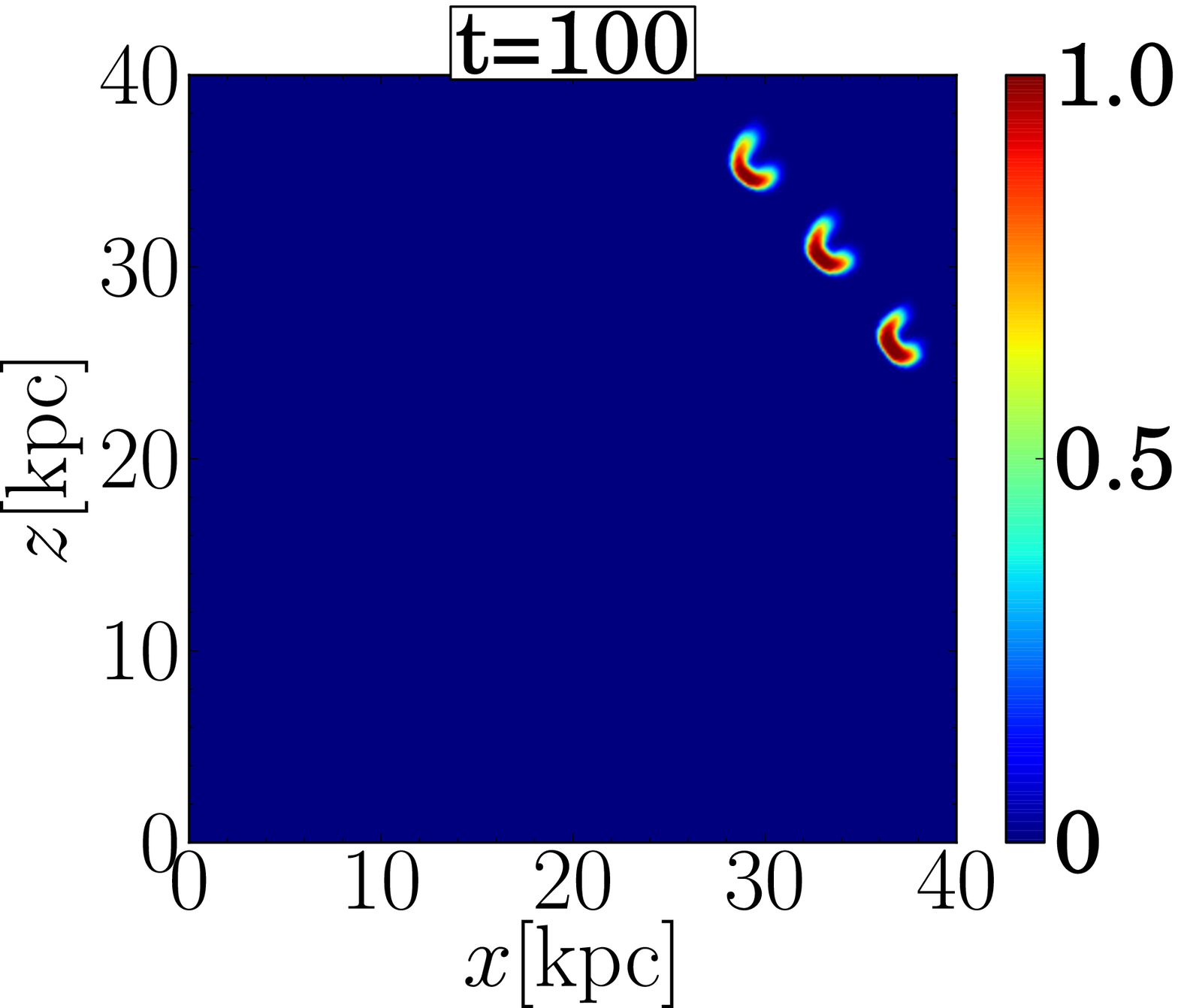}}
\subfigure{\includegraphics[width=0.32\textwidth]{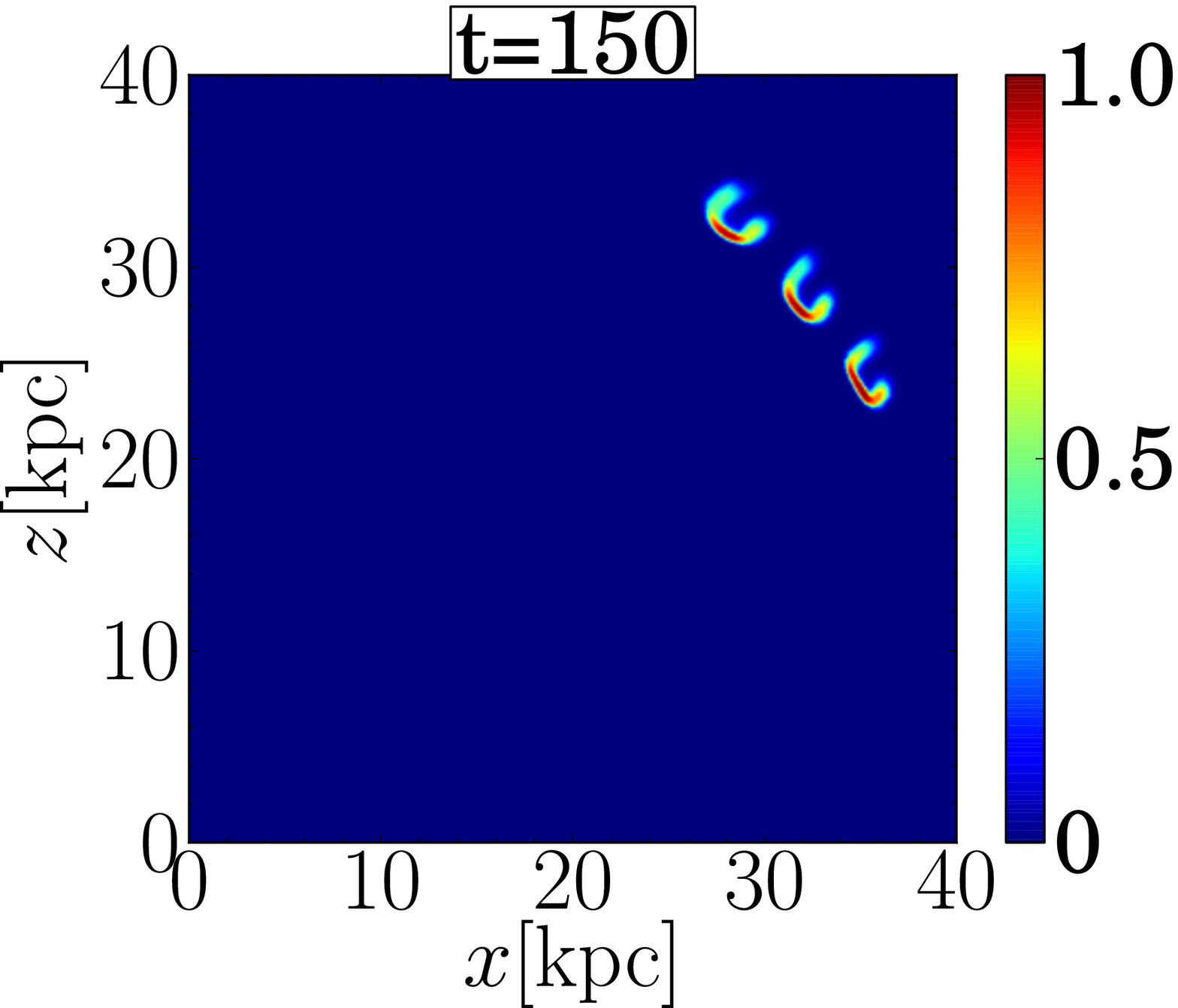}} \\
\subfigure{\includegraphics[width=0.32\textwidth]{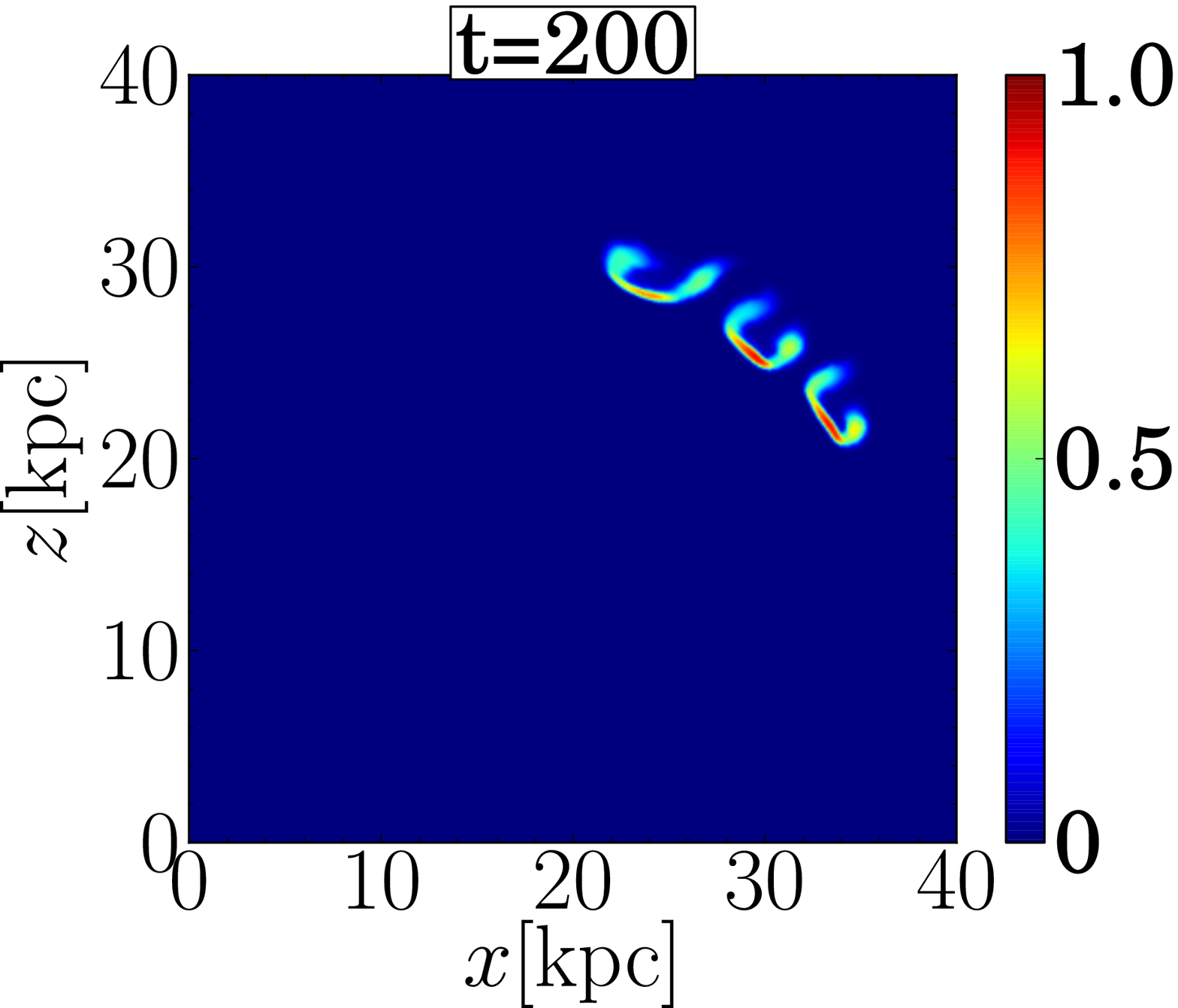}}
\subfigure{\includegraphics[width=0.32\textwidth]{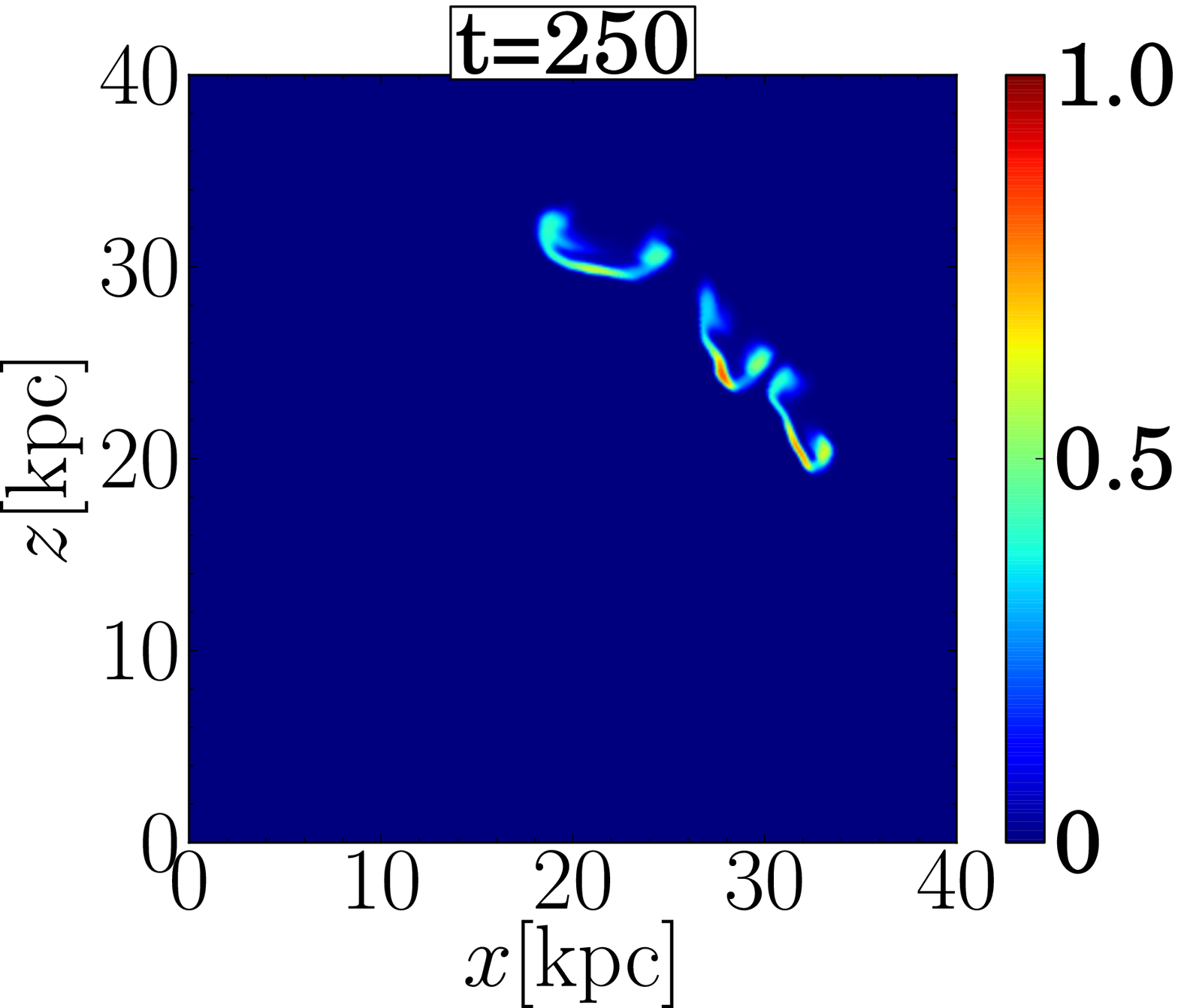}}
\subfigure{\includegraphics[width=0.32\textwidth]{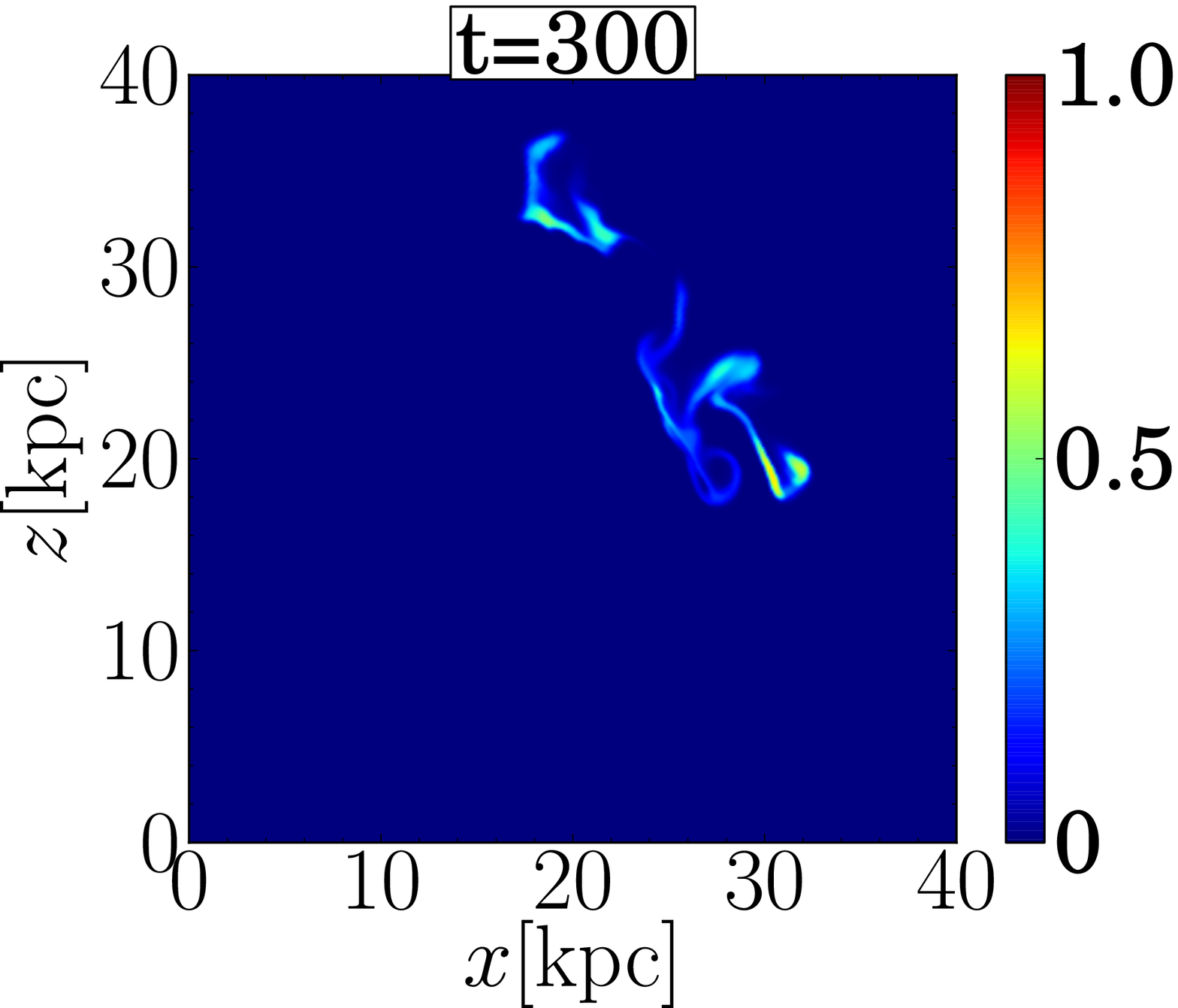}} \\
\subfigure{\includegraphics[width=0.32\textwidth]{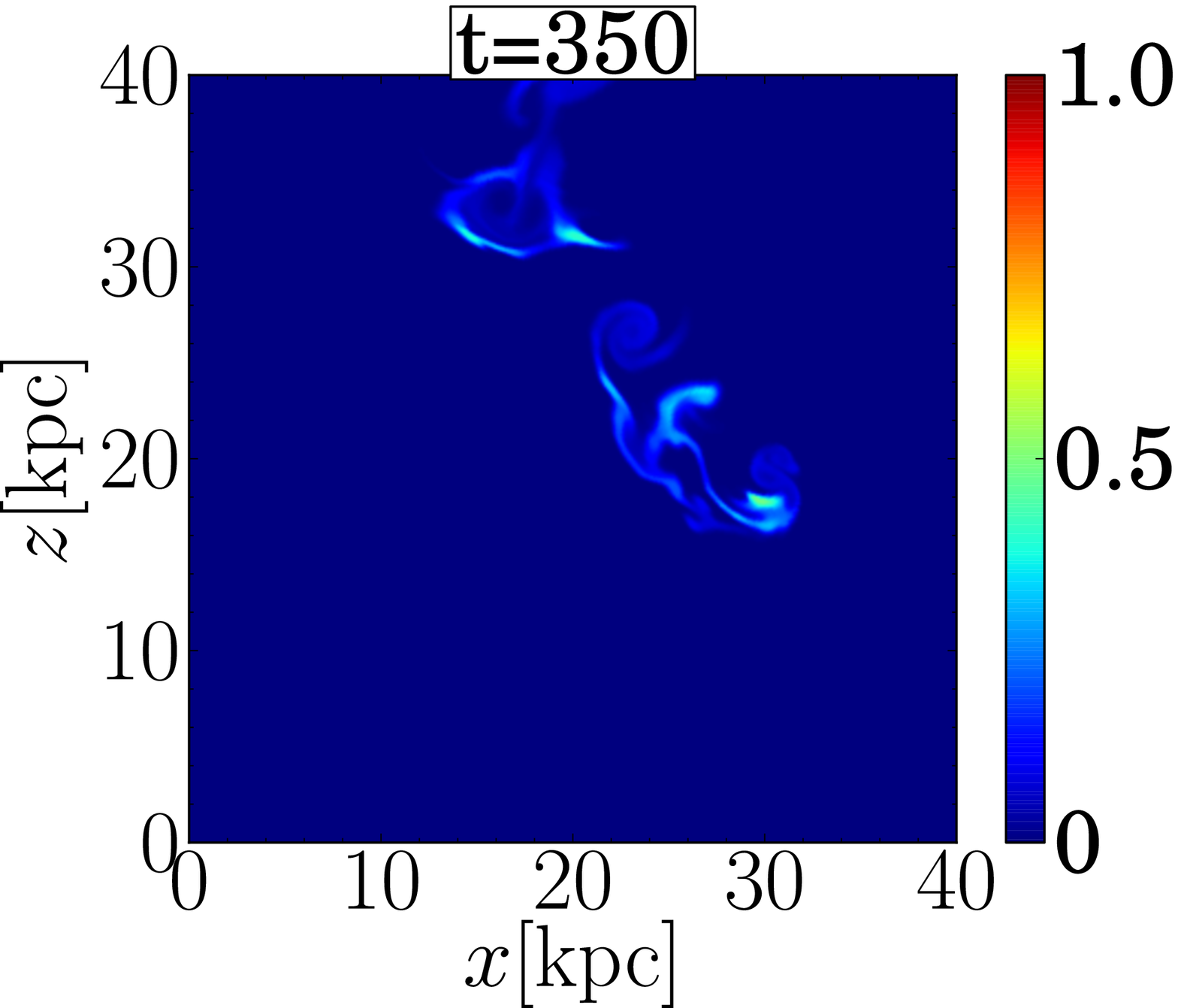}}
\subfigure{\includegraphics[width=0.32\textwidth]{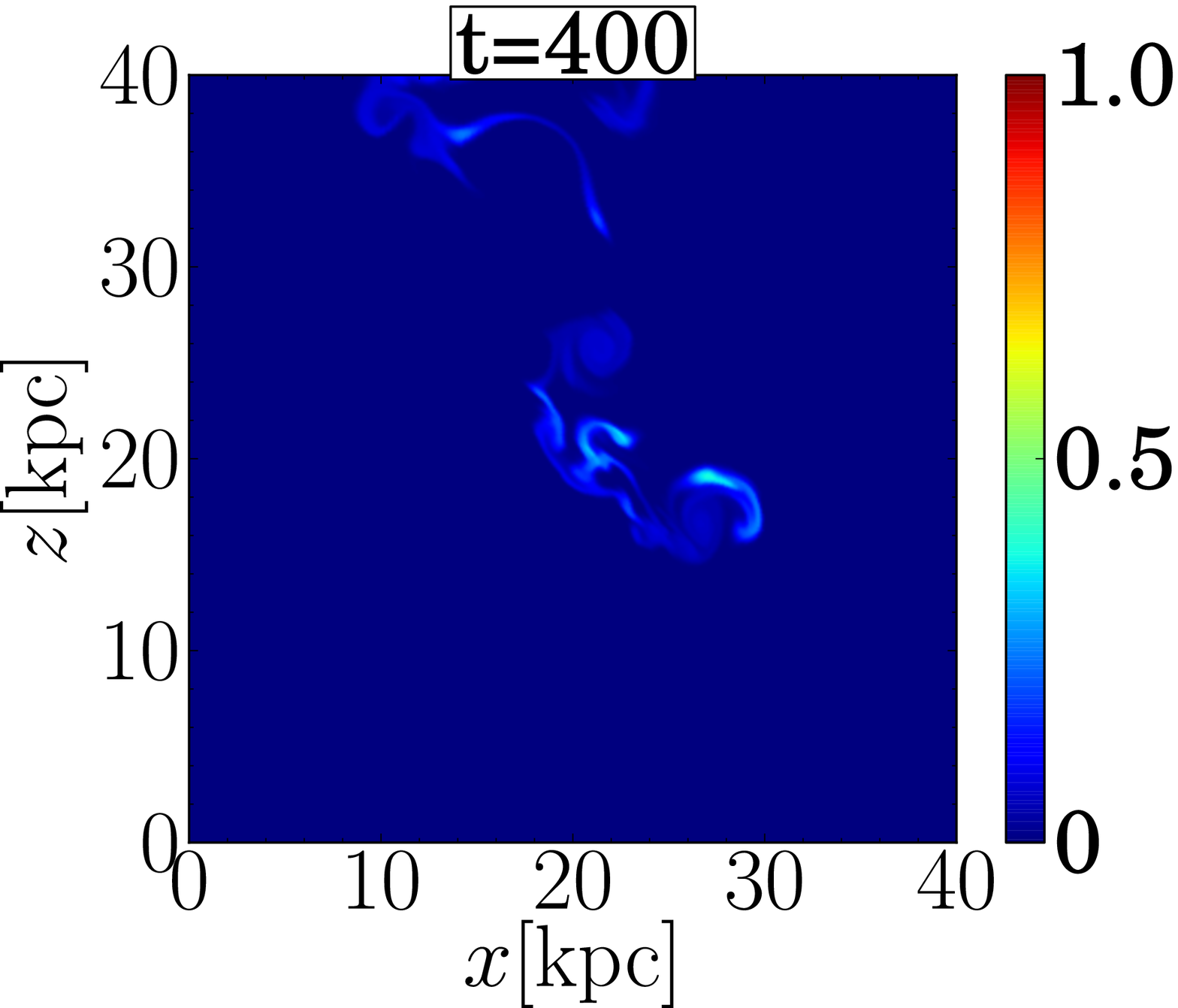}}
\subfigure{\includegraphics[width=0.32\textwidth]{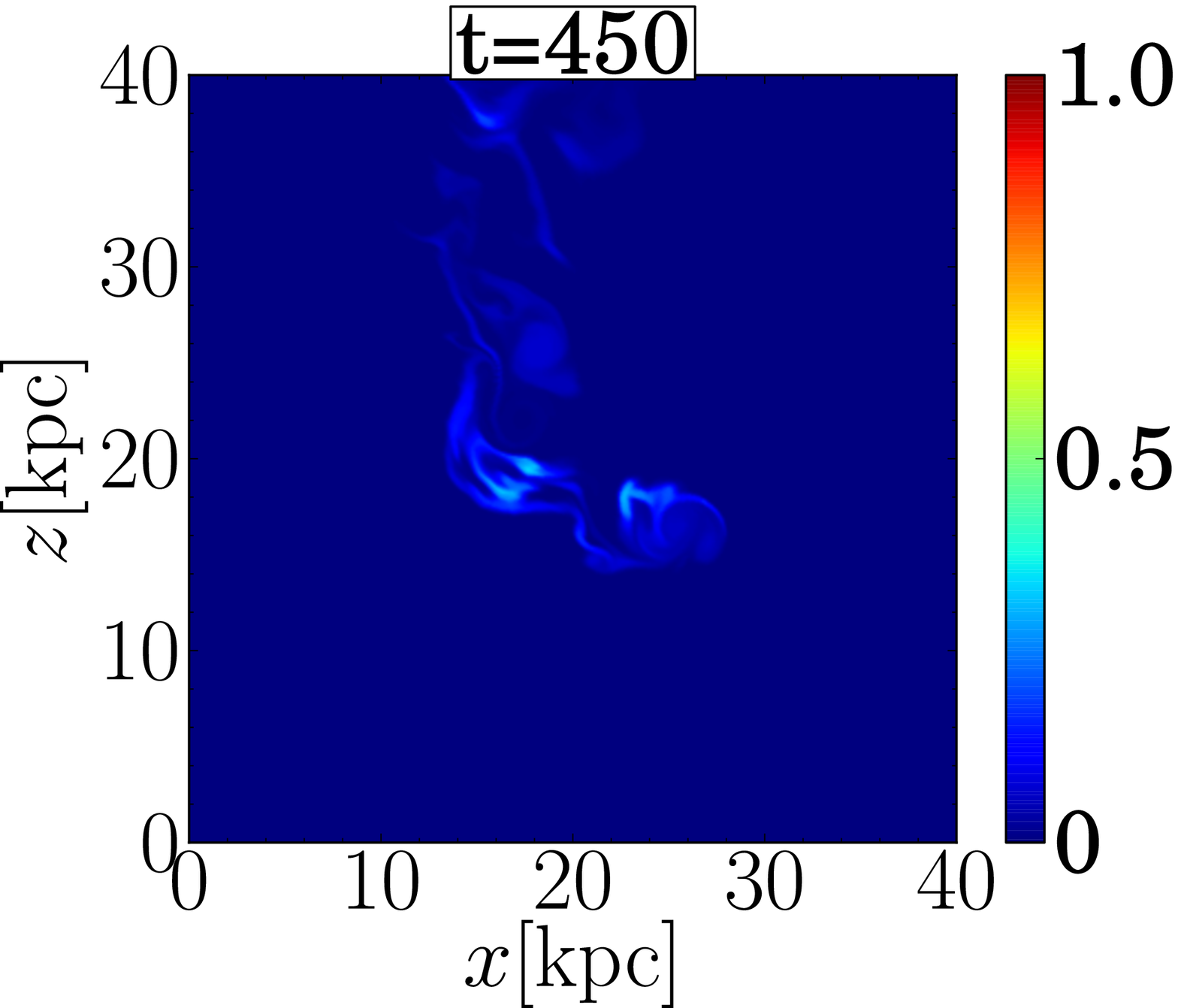}}
\caption{
Like Fig. \ref{RunM20D0.3Temp} but showing the tracers of the clumps. Color coding is the fraction of the initial clumps' material at each point.
The mixing of the clumps, with the ICM and shocked jets' gas, is evident.
}
\label{RunM20D0.3Trace}
\end{figure}
%FFFFFFFFFFFFFFFFFFFFFFFFFFFFFFFFFFFFFFFFFFFFFFFFFFF

We present the temperature and tracers, rather than density and velocity, as we are now turning to study the thermal evolution of the clumps.
Consecutive shock waves pass through the ICM gas every $20 \Myr$.
The cooling time at the initial density and temperature is $\tau_{\rm cool} \simeq 10^9 \yr$.
At about half that time, $t \sim 500 \Myr$, we expect the ICM gas to cool substantially, as we indeed see in the lower
panels of Fig. \ref{RunM20D0.3Temp}, where light blue areas ($T\simeq 20 \times 10^6 \K$) start to appear, about half the initial temperature.
This shows that the shocks do not manage to heat the ICM and the clumps.
As can be seen in the latest panel, the upper clump, that starts closest to the jet's axis, is mixed with the shocked
jet's material and is heated up (this is further discussed in section \ref{s-heating-the-cold-gas} below).

As stated above, our numerical code is limited to one jet axis. We expect that in real cluster environments during
a time period of $0.5 \Gyr$, either different jets' launching directions or relative motion of the ICM and the AGN will take place.
Such a displacement can be seen in misalignment of the ghost bubbles in Perseus from the axis of the inner bubble-pair \citep{Fabianetal2006}.
Any of these transverse (to the initial jets' axis) ICM-jets motion will lead to efficient mixing of gas that is
initially near the equatorial plane. Over a long time our code is limited in handling the mixing of such gas.

% ==========================================================
\subsection{Heating the clumps}
\label{s-heating-the-cold-gas}
% ==========================================================

In the previous sections we showed the general flow structure of the system, emphasizing mixing.
We now turn to quantitatively examine the thermal evolution of cold clumps.
We define artificial flow quantities called 'tracers'
that are frozen-in to the flow (see section \ref{s-numerical-setup}).
The initial value of a tracer is set to $\xi=1$ in the clump and $\xi=0$ elsewhere.
Mixing of the traced gas with the ICM or the jet's material changes the tracer to a value of $0 < \xi < 1$.
Since this quantity is advected with mass, the sum $\Sigma \xi_i M_i$ is constant with time.
Using these tracers we define the average property $Q$ of a clump as
\begin{equation}
Q_c \equiv \frac{\Sigma_i \xi_i M_i Q_i}{\Sigma_i \xi_i M_i},
\label{eq:tracers1}
\end{equation}
where here $Q_c$ stands for the temperature or specific entropy of the clump \citep{GilkisSoker2012}.

Before moving to the case of multiple jet episodes, in Fig.~\ref{figure: clump temperature and entropy} we present the influence of one jet
from {{{{Run~S20$\delta$1}}}}. Shown are the temperature $T_c$ and specific entropy $s_c$ histories of the cold clump whose evolution is presented
in Figs. \ref{figure: rhofactor2_6panels}, \ref{figure: rhofactor2_zoomed}, and the left column of Fig. \ref{Tracers}.
The initial drop in both $T_c$ and $s_c$ is due to radiative cooling before the shock hits the clump.
When the outward propagating shock wave reaches the clump it increases both $T_c$ and $s_c$ of the clump.
However, after the passage of the shock the clumps material re-expands and the temperature drops.
The entropy increase brings it more or less back to its initial value.
More temperature variations occur as sound waves cross the clumps.
Substantial heating starts only when the clump starts mixing with the hot bubble material at $t \simeq 35 \Myr$.
This corresponds to the period between the middle-left panel and middle-right panel of Fig. \ref{figure: rhofactor2_6panels}.
From there on the temperature and the entropy increase till $t \simeq 70 \Myr$.
The major heating of the clump during the period of $t \simeq 35-70 \Myr$ takes place when the clump suffers vigourous mixing
with the hot bubble material, as seen in the middle-right and bottom-left panels of Fig.~\ref{figure: rhofactor2_6panels}.
Overall, the shock can heat the clump, but mixing is much more efficient in doing this.
{{{{This is to be expected in light of the analysis in \cite{Sokeretal2013}.}}}}
Later in the history of the clump, radiative cooling becomes dominant again, and a drop in temperature and specific entropy is seen.
%FFFFFFFFFFFFFFFFFFFFFFFFFFFFFFFFFFFFFFFFFFFFFFFFFFF
\begin{figure}[htb]
\centering
\subfigure{\includegraphics[width=0.477\textwidth]{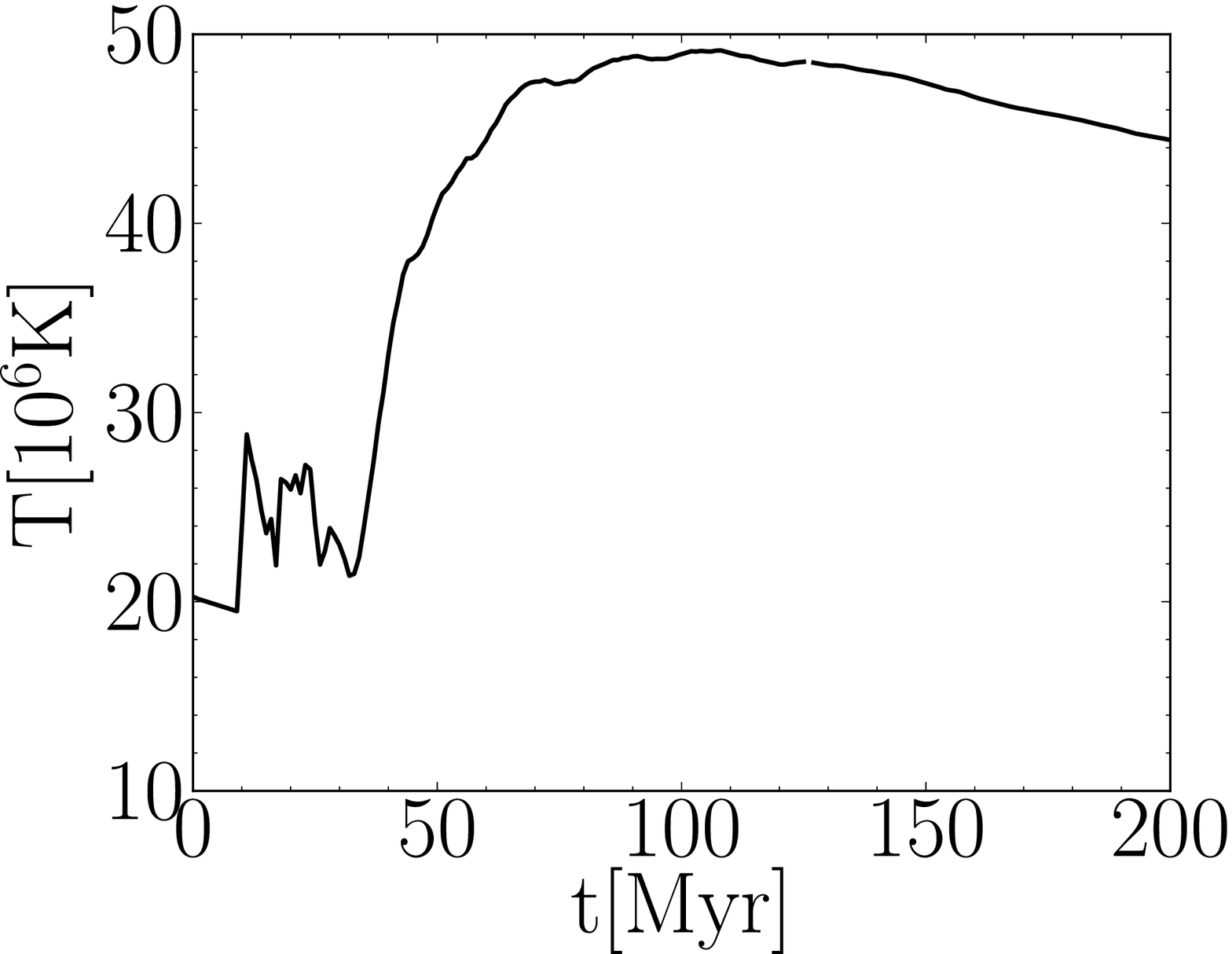}}
\subfigure{\includegraphics[width=0.503\textwidth]{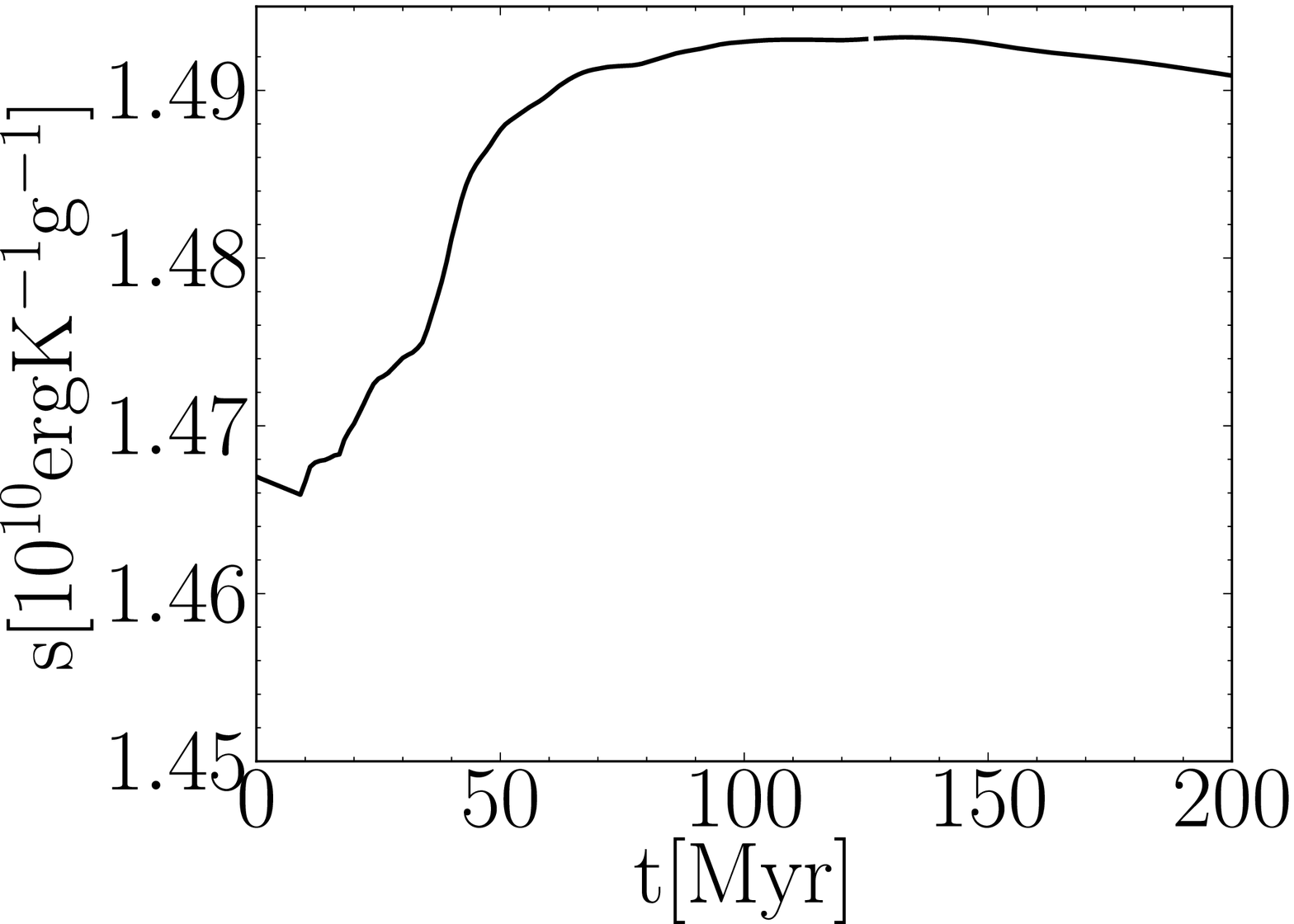}}
\caption{
The mean temperature (left panel) and {{{{specific}}}} entropy (right panel) history of a cold clump having an initial density contrast of $\delta = 1$
in the presence of one jet launching episode lasting $t_{\rm jets} = 20 \Myr$ and starting at $t = 0$ ({{{{Run~S20$\delta$1}}}}).
The initial radius of the clump is $R = 1 \kpc$, and its initial location and evolution are presented
in figures Figs.~\ref{figure: rhofactor2_6panels}, \ref{figure: rhofactor2_zoomed}, and the left column of Fig.~\ref{Tracers}.
The first temperature jump occurs as the shock hits the clump.
Subsequently the clump expands and adiabatically cools.
Further temperature variations occur due to sound waves.
The substantial heating starts only at $t \simeq 35 \Myr$ when very hot shocked jet material mixes with the clump's gas
(Fig.~\ref{figure: rhofactor2_6panels}).
}
\label{figure: clump temperature and entropy}
\end{figure}
%FFFFFFFFFFFFFFFFFFFFFFFFFFFFFFFFFFFFFFFFFFFFFFFFFFF

In Fig.~\ref{figure: temperature and entropy history, 1.3, periodic jet} we follow the thermal evolution of the three dense clumps of {{{{Run~M20$\delta$0.3}}}},
all starting with a density contrast of $\delta = 0.3$.
The initial location and evolution of the clumps are presented in Figs.~\ref{figure: 3clumps_r45_rhofactor1.3_t50},
\ref{figure: 3clumps_r45_rhofactor1.3_t305}, \ref{RunM20D0.3Temp}, and \ref{RunM20D0.3Trace}.
These three clumps start at a distance of $r = 45 \kpc$ from the center, rather than $20 \kpc$ for the clumps whose thermal evolution is presented in
Fig.~\ref{figure: clump temperature and entropy}, and are less influenced by sound waves.
The almost periodic variation in temperature and {{{{specific}}}} entropy seen in the first $\sim 300 \Myr$ are due to the periodically jet-excited
shock waves that propagate through the ICM (see Fig.~\ref{figure: 3clumps_r45_rhofactor1.3_t50}).
The shock waves heat the clumps and increase their entropy, but only by a small amount that does not compete with radiative cooling for
the parameters used here.
The central clump, presented by the black line, suffers mixing at $t \simeq 260 \Myr$ and its temperature and entropy increase.
However, this heating process lasts for only $\sim 40 \Myr$, after which the average temperature and entropy of the clump decrease
due to radiative cooling.
We point again that these are average quantities, as the clumps by this stage are highly deformed and spread.
Namely, some parts of the clumps might be heated while other cooling.
At $t \simeq 320 \Myr$ the clump closest to the jet's axis (the top-left clump) suffers vigorous mixing and its
entropy and temperature increase by a substantial amount (upper blue line in the figure).
The bottom-right clump has too large a distance from the jet's axis, and before it comes close and mixes with hot gas,
it experiences catastrophic radiative cooling.
The evolution presented in Fig. \ref{figure: temperature and entropy history, 1.3, periodic jet} shows that even tens of shocks
excited by multiple jet-activity episodes are far less efficient than mixing in heating the cooling ICM.
%FFFFFFFFFFFFFFFFFFFFFFFFFFFFFFFFFFFFFFFFFFFFFFFFFFF
\begin{figure}[htb]
\centering
\subfigure{\includegraphics[width=0.477\textwidth]{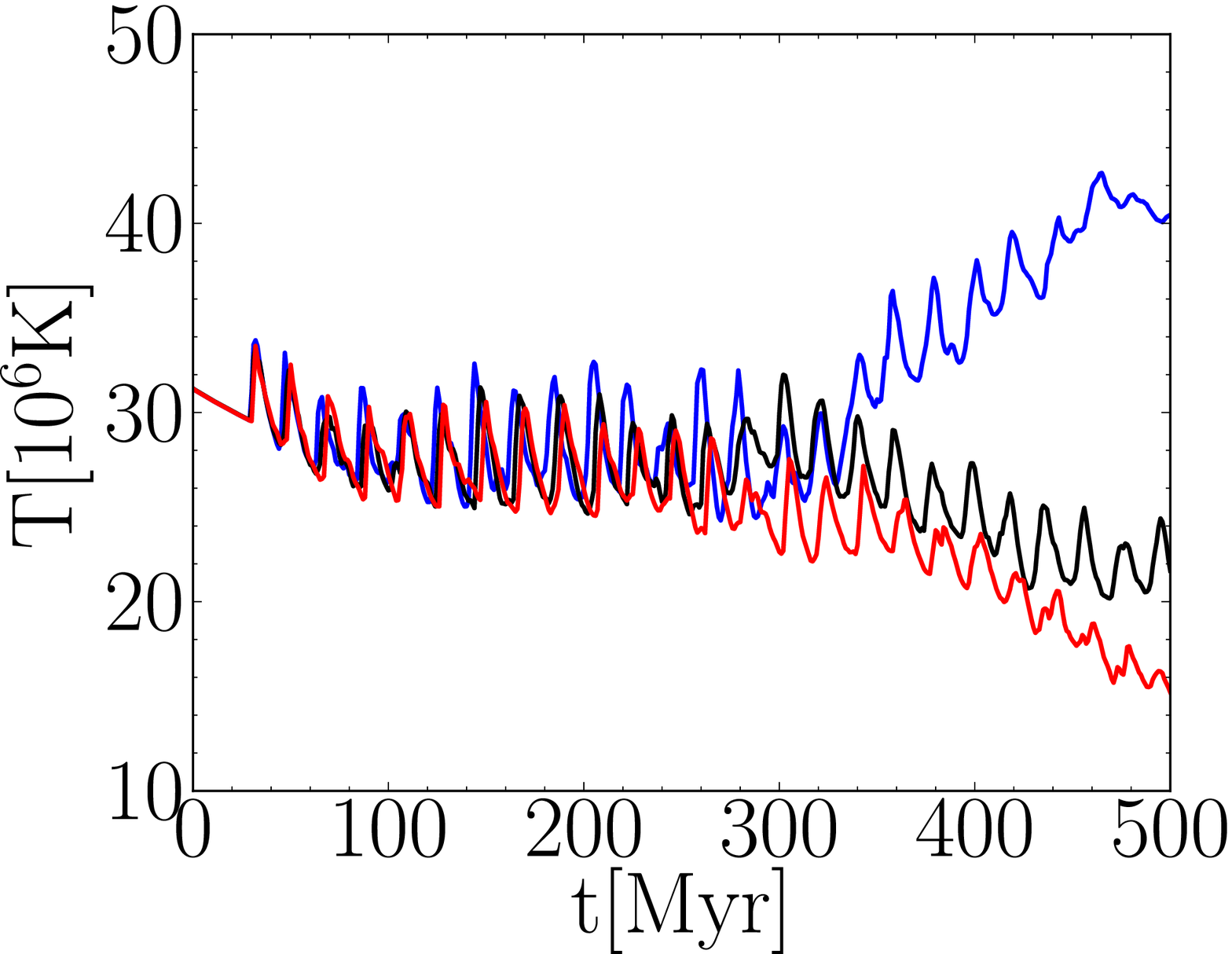}}
\subfigure{\includegraphics[width=0.503\textwidth]{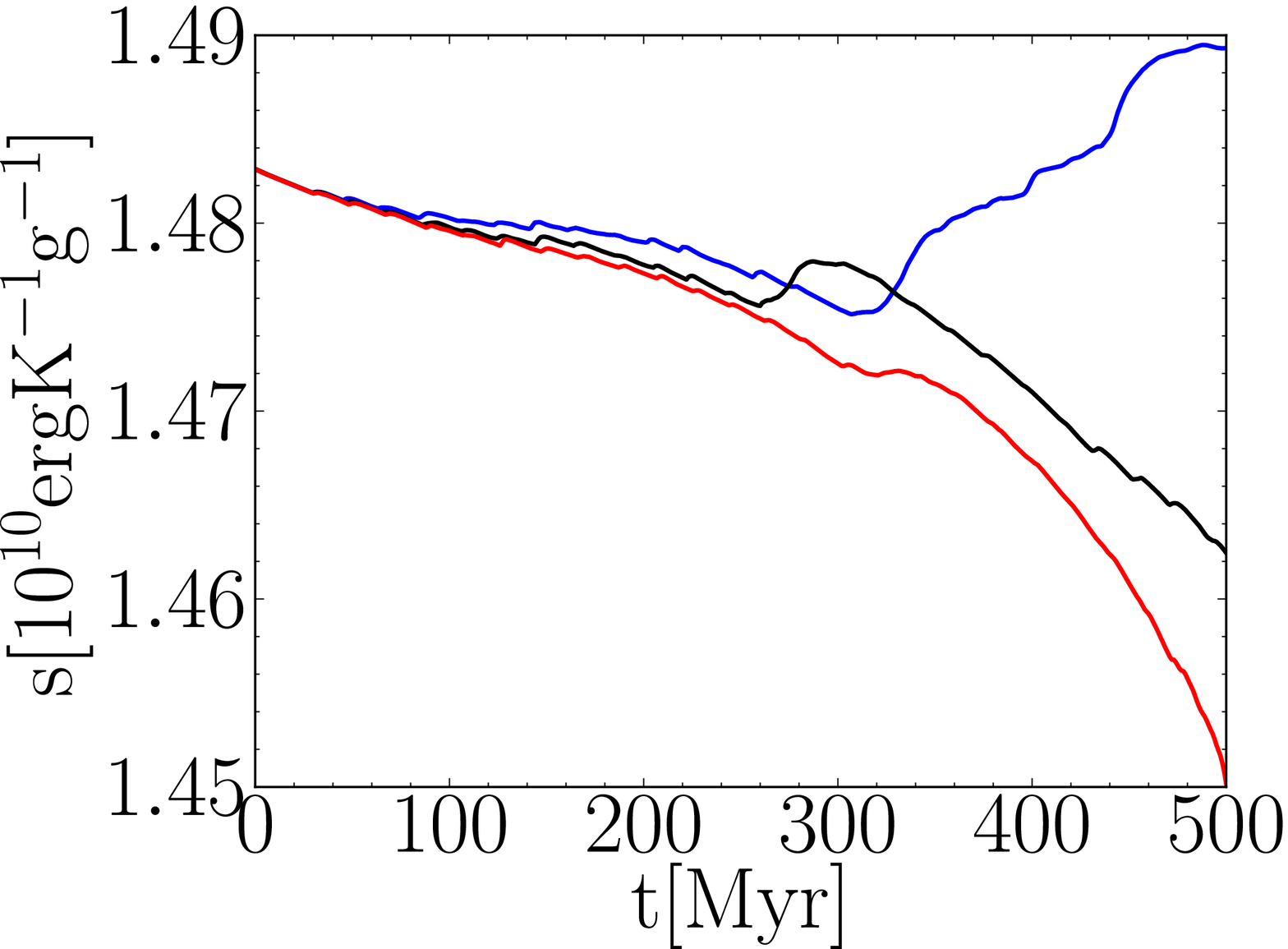}}
\caption{
The mean temperature (left panel) and {{{{specific}}}} entropy (right panel) history of three cold clumps in the presence of a periodic jet of $t_{\rm jet} = 10 \Myr$ and quiescence time of
$t_{\rm q}=10 \Myr$, i.e., a period $ t_{\rm jp} = 20 \Myr$ ({{{{Run~M20$\delta$0.3}}}}).
The initial density contrast of the three clumps is $\delta = 0.3$,
and their initial location and evolution are presented in Figs.~\ref{figure: 3clumps_r45_rhofactor1.3_t50},
\ref{figure: 3clumps_r45_rhofactor1.3_t305}, \ref{RunM20D0.3Temp}, and \ref{RunM20D0.3Trace}.
}
\label{figure: temperature and entropy history, 1.3, periodic jet}
\end{figure}
%FFFFFFFFFFFFFFFFFFFFFFFFFFFFFFFFFFFFFFFFFFFFFFFFFFF

% ==========================================================
\section{SUMMARY}
\label{s-summary}
% ==========================================================

We used the PLUTO hydrodynamic code \citep{Mignone2007} to study the heating of dense clumps embedded in the intra-cluster medium (ICM) of cooling
flow clusters of galaxies.
We conducted {{{{2D axisymmetric}}}} hydrodynamic simulations, i.e., the flow is 3D but with an imposed azimuthal symmetry around the $z$ axis, to study the
influence of multiple jets-activity cycles on the thermal evolution of the dense clumps.
The initial cross section of each clump is a circle in the meridional plane of the {{{{2D}}}} numerical grid, which implies a torus in 3D.
Only one side of the equatorial plane was simulated.
In some cases only one jet-launching episode was simulated, and in others we run 25 activity episodes, with an off period of $10 \Myr$ between $10 \Myr$ long active phases.

We reproduced (Fig. \ref{figure: 3clumps_r45_rhofactor1.3_t50}) the formation of a fat bubble by a slow massive wide (SMW) jet \citep{Sternberg2007, GilkisSoker2012},
and the formation of multiple sound waves with a single jet-activity episode \citep{Sternberg2009, GilkisSoker2012}.
We strengthened the finding of \cite{GilkisSoker2012} and \cite{Sokeretal2013} that vorticity plays major roles in the structure and evolution of bubbles
and their interaction with the ICM.
Our addition here is the study of dense clumps and the simulation of many jet-activity episodes.

We considered two main heating mechanisms in AGN feedback: heating by shock waves initiated by jets and mixing of cold gas with shocked hot jet material.
The thermal evolution of dense clumps is summarized in Figs.~\ref{figure: clump temperature and entropy} and \ref{figure: temperature and entropy history, 1.3, periodic jet}.
For the parameters used in our study (see section \ref{s-numerical-setup}) we found that heating by shock waves cannot compete with radiative cooling over a long time
(for an opposite view see \citealt{Randall2011}).
Shocks increase the temperature of the clumps and compress the gas, but after the clumps re-expand the temperature drops back to almost its initial value.
Shocks also increase the clumps' entropy, but the compression shorten the radiative cooling time of the gas.
Even in simulations with multiple frequent jet episodes, shock waves did not nearly offset radiative cooling.
The inefficiency of shock heating was derived analytically in a previous paper \citep{Sokeretal2013}, and was shown to be much less efficient than mixing.
On the other hand, we found, like \cite{GilkisSoker2012}, that once mixing with the jets' shocked material (the hot bubbles) begins, it is very efficient in heating the
cold clump's material and increasing its entropy.
The mixing process studied by \cite{GilkisSoker2012} and explored here, can go much beyond the direct mixing of shocked jets' material with the ICM and cold clumps,
and continue with turbulence in a larger volume of the ICM in cooling flows \citep{BanerjeeSharma2014}.
{{{{ Based on a 2D hydrodynamical study, \cite{Peruchoetal2014} argued recently that heating by shocks is the main heating process.
However, they inflate bubbles on scales of $>500 \kpc$, larger by an order of magnitude than typical bubbles in cooling flow clusters.
Their bubbles occupy a huge fraction of the ICM volume, hundreds of times the typical volume in observed cases.
We attribute their conclusion about shock heating to their unrealistically large bubbles.
In any case, they also note the importance of mixing. }}}}

Our {{{{2D}}}} numerical code is constrained to launch jets along a constant direction, and the mixing is not efficient in directions at large angles to this direction.
Observations, however, show that bubbles of different episodes are not exactly aligned with each other, and even two opposite bubbles inflated
together lose alignment over time.
These misalignments result from a relative motion of the central AGN and the ICM, and from jets' precession.
{{{{Thus}}}}, mixing is expected to be efficient in all directions, and so to be the major heating mechanism in cooling flows in galaxies and clusters of galaxies,
as well as in the process of galaxy formation during which cooling flow could have taken place \citep{Soker2010a}.

The numerical constraint of a constant jet axis has another effect.
In the simulations we conducted we did not get clear fat bubbles at late times of multiple-episodes simulations.
Rather an elongated bubble was formed, a shape which is mostly inconsistent with the observations.
The elongated shape was formed since the jets we simulated were always along the same axis, the rotational-symmetry axis of our {{{{2D}}}} simulations.
In reality, different jet episodes are often directed at different directions, and in such cases two opposite bubbles are formed \citep{Sternberg2007, GilkisSoker2012}.

The complicated flow structure induced by bubbles' inflation (Figs. \ref{figure: 3clumps_r45_rhofactor1.3_t50}, \ref{figure: rhofactor2_zoomed}, \ref{figure: rhofactor3})
has some further implications for the thermal evolution and feedback mechanism.
(1) Vortices on all scales entangle magnetic field lines in the ICM. This suppresses any global heat conduction in the ICM near the center.
(2) The same entanglement process mixes the magnetic fields of the ICM and the shocked jets' material.
This leads to reconnection of the magnetic field lines, hence allowing for local heat conduction between
the mixed ICM and jets' gas.
We emphasize the efficiency of local heat conduction (scales of $\la 0.1 \kpc$) as opposed to the inefficiency of global
(scales of $\ga 1 \kpc$) heat conduction (see review by \citealt{Soker2010b}).
{{{{The typical grid size at $10 \kpc$ from the center of our numerical code is $0.06 \kpc$.
Therefore, with our resolution and if the claim of \cite{Soker2010b} holds, there is no need to include heat conduction
in the inner region of $r \la 50 \kpc$. The outer regions are of less interest to us here.}}}}
(3) Heating by mixing, while being very efficient, is not $100 \%$ efficient.
Our results show that some cold clumps do indeed cool to low temperatures. These will form very dense clumps that,
if not heated by another jet within a short time, fall inward and feed the AGN.
Our results support the cold feedback mechanism as suggested by \cite{Pizzolato2005}, and that has gained considerable support
by recent observations of cold gas and by more detailed studies (see section \ref{s-intro}).

We thank an anonymous referee for very helpful and detailed comments.

\end{document}